%% file: main.tex
\documentclass[aps,prx,twocolumn,psfig,superscriptaddress,longbibliography]{revtex4-2}

\usepackage{times}
\usepackage{graphicx}
\usepackage{float}
\usepackage{latexsym,bm,euscript}
\usepackage{color}
\usepackage{subfigure}
\usepackage{epstopdf}
\usepackage[colorlinks=true,linkcolor=blue,citecolor=blue]{hyperref}
\usepackage{hyperref}
\usepackage[normalem]{ulem}
\usepackage{mathrsfs}
\usepackage{amsmath,amssymb}
\usepackage{xspace}
\usepackage{amsthm}
\usepackage{amsbsy}

\input{shortcuts.tex}

\begin{document}

\title{RKKY to Kondo Crossover in Helical Edge of a Topological Insulator}

\author{Pol Alonso-Cuevillas Ferrer}
\affiliation{Physics Department,
Arnold Sommerfeld Center for Theoretical Physics, and Center for NanoScience,
Ludwig-Maximilians-Universit\"at, Theresienstrasse 37, 80333 Munich, Germany}

\author{Oleg M. Yevtushenko}
\affiliation{Institut f\"ur Theorie der Kondensierten Materie,
Karlsruhe Institute of Technology, 76128 Karlsruhe, Germany}

\author{Andreas Weichselbaum}
\email{weichselbaum@bnl.gov}
\affiliation{Department of Condensed Matter Physics and Materials
Science, Brookhaven National Laboratory, Upton, NY 11973-5000, USA}

\begin{abstract}
Two spatially separated magnetic impurities coupled to itinerant
electrons give rise to a dynamically generated exchange (RKKY)
inter-impurity interaction that competes with the individual
Kondo screening of the impurities. It has been recently shown
by \citex{Yevt_2018}, that the RKKY interaction and the RKKY
vs. Kondo competition become nontrivial on helical edges of
two-dimensional topological insulators where there is lock-in
relation between the electron spin and its direction of motion.
Kondo screening always takes over and dominates at large
inter-impurity distances and it can also dominate all the way to
short distances if the Kondo coupling is sufficiently large and
anisotropic.
In the present paper, we study the Kondo-RKKY competition in
detail on a qualitative and quantitative level. For this we
employ the numerically exact numerical renormalization group
(NRG) for a broad parameter scan of two Kondo coupled impurities
vs. magnetic anisotropy, impurity distance and temperature, and
comment on the role of finite bandwidth. We give a pedagogical
introduction on the the setup of the two-impurity setting within
the NRG in the helical context. Overall we establish a plain
crossover from RKKY to Kondo with increasing impurity distance
which permits an intuitive physical picture by simply comparing
length scales set by the Kondo screening cloud vs. the thermal
length scale vs. the impurity distance.
\end{abstract}

\date{\today}

\maketitle

\section{Introduction}

Physics of magnetic impurities (MIs) coupled to helical
electrons on one-dimensional (1D) edges of two dimensional (2D)
time-reversal invariant topological insulators (TIs) attracted
the attention of researchers soon after the experimental
discovery of the TIs \cite{HasanKane,QiZhang,TI-Shen,hsu_2021}.
This interest resulted from a search of possible backscattering
mechanisms which could make the virtually protected helical
conductance sub-ballistic in relatively long samples
\cite{Molenkamp-2007,konig_2013,EdgeTransport-Exp1,EdgeTransport-Exp2}.
Since the helical electrons possess a lock-in relation between the
spin projection on the quantization axis and the direction of
propagation (the so-called chirality), their backscattering is
expected to involve some nontrivial spin processes, e.g. the
spin flip. The MI can provide such an inelastic backscattering
of the individual helical electrons.
However, the helical conductance can be suppressed only
if the spin conservation on the edge is violated, see papers
\cite{OYeVYu_2021,OYeVYu_2022} and references therein.
If the edge is spin-conserving and the MI does not break the
spin U(1) symmetry, it backscatters the helical electrons but
cannot influence the dc conductance \cite{FurusakiMatveev}.
The anisotropic MI is able to suppress the helical conductance
only if it breaks the spin conservation and is not Kondo
screened \cite{kurilovich_2017,TI-Anisotr-KI,vinkler_2020}.
The latter requires either the temperature being larger than the
Kondo temperature, $T > \TK$, or a large value of the MI spin,
$S > 1/2$. This points out the importance of understanding the
Kondo effect in TIs which is substantially different from that
in usual (non-helical) 1D wires in the presence of the
electron-electron interaction or the magnetic anisotropy of the
XXZ type \cite{MaciejkoOregZhang}. Here as well as throughout
this paper, $\TK \equiv \TK^{(1)}$ represents the Kondo scale of
a single, possibly anisotropic spin-half impurity coupled to a
helical edge \cite{MaciejkoOregZhang}. Its value is identical
to the plain non-helical Kondo model given that that helicity in
the non-interacting bath is irrelevant from the point of view of
a single impurity.

The Kondo effect can be suppressed by the indirect exchange MI
interaction, the Ruderman--Kittel--Kasuya--Yosida (RKKY)
interaction \cite{Kittel}, if the helical edge is coupled to a
dense array of the MIs. According to the simple picture of the
Doniach criterion \cite{doniach_1977}, the ``winner of the RKKY-Kondo
competition'' can be found by comparing $ T_{K} $ with the
characteristic RKKY energy scale \ER. The latter means the
energy gap which opens after the RKKY correlations lift a
degeneracy in the energy of the uncorrelated MIs. If $T \to 0$
but $\ER \gg \TK$, the RKKY correlations overwhelm the Kondo
screening which may lead to many nontrivial effects, including
Anderson localization of the helical electrons caused by the
random magnetic anisotropy \cite{AAY,YeWYuA}, and
magnetically correlated phases
\cite{MaciejkoLattice,Hsu_2017,TI-SupSol,Hsu_2018}. The
RKKY-induced magnetic order in helical 1D systems is nontrivial
and qualitatively different from that in their non-helical counterparts,
cf. Refs. \cite{TsvelikOYe_2019,TsvelikOYe_2020} and references therein.

Since \TK does not depend on the MI density while \ER typically
decays with increasing the inter-impurity distance, $ R $ --
see \Eq{H-RKKY} below, one can surmise that there is a
characteristic distance defined by the equality $\TK \simeq
\ER(x_c) $, which separates the RKKY- and Kondo-dominated
phases, $ R < x_c $ and $ R > x_c $ respectively.
This conclusion is inspired by a misleading analogy with the
physics of non-helical wires \cite{LeeToner,FuruNaga,egger_1996,
hallberg_1997,egger_1998}.
One of us (in collaboration with V.\,I.\,Yudson \cite{Yevt_2018})
have recently considered two MIs coupled to the helical edge and
shown that, if either the electron-electron interaction or the
magnetic XXY anisotropy or both are strong, $ x_c $ shrinks and
the Kondo effect overwhelms the RKKY interaction over all
macroscopic inter-impurity distances. This unexpected conclusion
has been drawn based on phenomenological arguments and on the
analytical consideration of limiting cases. This theory reveals
many quantitative features but is far from being complete. In
particular, it cannot predict whether the above mentioned phases
with finite and vanishing $ x_c $ are separated by a crossover
or by a phase transition and whether the MIs remain somehow
correlated even in the Kondo-dominated phase.

In the present paper, we expand and complete the theory of
Ref.~\cite{Yevt_2018}. We present an analytical theory of the
RKKY correlation between the two MIs coupled to the helical
electrons but, as the main working tool, we have chosen the
numerical renormalization group (NRG; \cite{Wilson75,Bulla08,Wb07}).
This powerful and well-established method has allowed us to
answer the aforementioned open questions. In particular we will
show that 1) the different phases are separated by the
crossover, and 2) the MIs are uncorrelated, i.e. independently
screened, in the Kondo dominated phase.

The paper is organized as follows: We introduce the model in
\Sec{sec:model}, followed by analytical considerations in
\Sec{sec:FT-RKKY}. The remainder of the paper then is dedicated
to a detailed analysis and discussion of the model based on the
NRG in \Sec{sec:NRG:results}, concluded by summary and
outlook. In appendix \App{sec:NRG:setup}, we give a detailed
pedagogical derivation of how the helical 2-impurity system is
setup and mapped into the standard NRG machinery. In
particular, this highlights the possibility to map the system
onto a Wilson ladder, with complex coefficients only within the
coupling to the impurity. Furthermore, we included in the
appendix a brief reminder on the poor-mans scaling of the
anisotropic Kondo model, as well as a plain second order
perturbative derivation of the RKKY effective Hamiltonian and
RKKY energy that is complementary to \Sec{sec:FT-RKKY}.

\begin{figure}[b!]
\begin{center}
\includegraphics[width=.55\linewidth]{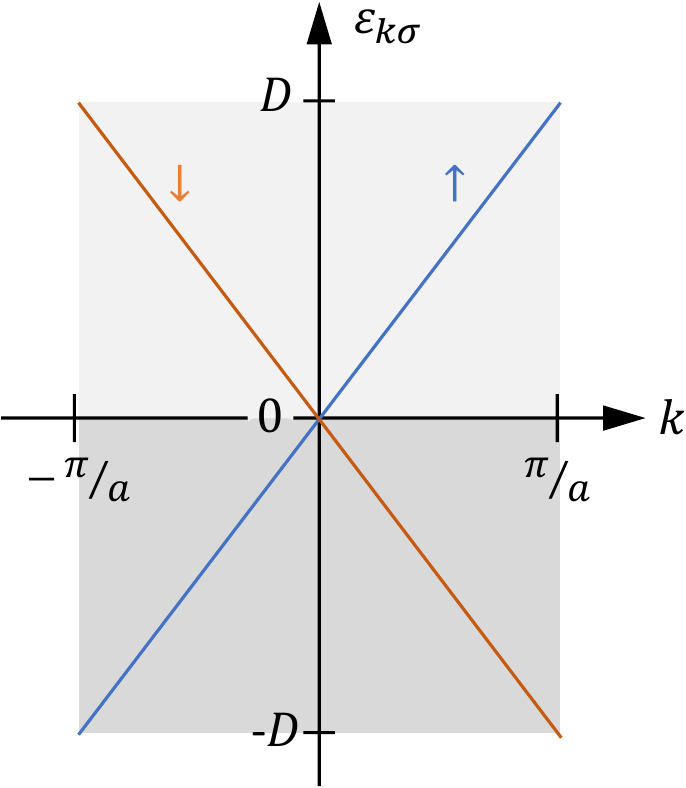}
\end{center}
\caption{
   Helical edge mode of the one-dimensional edge in a 2D
   topological insulator, e.g., as it occurs in the Kane-Mele
   model \cite{KaneMeleQSH}. Focusing on low energies,
   the dispersion vs. momentum $k$ is given by
   $\varepsilon_{k\sigma} = \sigma v k$ with spin $\sigma \in
   \{\uparrow,\downarrow\} \equiv \{+1, -1\}$ indicated by the
   blue (red) line, respectively, where $v$ denotes Fermi
   velocity. We assume a finite half-bandwidth $D$, throughout.
   This gives rise to an effective Brillouin zone with lattice
   spacing $a$ [cf. \Eq{eq:lattice:a}].
}
\label{fig:helical}
\end{figure}

\section{The model \label{sec:model}}

\subsection{Hamiltonian of the helical edge \label{sec:H0}}

We study spins coupled to a helical edge mode in a 2D
topological insulator. This may be approached in various
ways. The edge mode can be simulated (i) by fully modeling an
underlying 2D lattice model in real space, such as the Kane-Mele
model with the Dresselhaus spin-orbit interaction \cite{KaneMeleQSH}.
This naturally introduces a cutoff in terms of
bandwidth of the helical edge mode  which is located
{\it inside} the gap of the continuum of extended bulk states.
Together with interacting correlated
impurities, this may be simulated numerically, for example,
using the density matrix renormalization group (DMRG,
\cite{White92,Schollwoeck11}). There the non-interacting 2D
lattice without the impurities can be conveniently mapped to an
effective 1D impurity setting via Lanczos tridiagonalization
\cite{Allerdt18}. However, this approach bears significant
overhead in terms of the precise choice of the underlying 2D
lattice model and its parameters, the details of which are
considered irrelevant for the low-energy physics.
Conversely, (ii) the model can be considerably simplified by
focusing on a single pure effective 1D helical edge described
in energy-momentum space at low energies as depicted in
\Fig{fig:helical} and described by \Eq{eq:H0:k} below.
The latter representation of the ballistic non-interacting
edge modes may be (iii) exactly transformed into a 1D real-space
lattice realization which, however, involves long-range hoppings
[cf. \Sec{sec:ftilde} and in particular \Eq{eq:tb:helical}
for more on this]. Working in energy-momentum space, instead,
is appealing from an analytical point of view \cite{Yevt_2018},
but is also perfectly well-suited for the numerical treatment
via the NRG. With the additional goal to scan many orders
of energy scales with Kondo physics in mind, the energy-momentum
representation is preferred over real-space lattice descriptions.
Therefore we follow approach (ii) throughout this work.

The standard model Hamiltonian for a single helical edge mode
can be written in discrete form
in momentum space as follows [cf. \Fig{fig:helical}],
\begin{eqnarray}
   \hat{H}_0
   = \sum_{k\sigma}
   \underbrace{\sigma \, v k
   }_{\equiv \varepsilon_{k\sigma}} \cdot\,
   \hat{c}^\dagger_{k\sigma} \hat{c}^{\phantom{\dagger}}_{k\sigma}
\ \equiv\ \sum_k vk \cdot
  \hat{\bf c}^\dagger_{k} \tau_z \hat{\bf c}^{\ }_{k}
\text{ ,}\label{eq:H0:k}
\end{eqnarray}
where $\hat{c}_{k\sigma}$ are the annihilation operators of the
helical fermions with spin $\sigma \in \{\uparrow,\downarrow\}
\equiv \{+1, -1\}$, and $\tau_\alpha$ with $\alpha \in x,y,z$
the Pauli matrices. The fermions are described by a linearized
dispersion relation with the Fermi velocity $v$. Below we
consider a finite half-bandwidth $D$ (UV cutoff) such that
$\varepsilon_{k\sigma} \in [-D,D]$. This is required in the
numerical context yet and also for the sake of regularization.
We assume that the TI edge is oriented along the $x$-axis, with
$xy$ being the TI plane, and the $z$-direction is the
quantization axis for the spins of the helical edge modes.
This convention corresponds to the experimentally relevant
situation where the quantization axis is often fixed being
perpendicular to the TI plane \cite{Molenkamp-2007, konig_2013,
EdgeTransport-Exp1, EdgeTransport-Exp2}.

The helical system in \Eq{eq:H0:k} respects time-reversal
symmetry (TRS), in that $\varepsilon_{k\sigma} =
\varepsilon_{-k,-\sigma}$. The crossing of the spin selective
dispersions, i.e., the Dirac point in \Fig{fig:helical} is
chosen for simplicity at $k_0=0$ without restricting the case.
The precise choice of $k_0$ is irrelevant for our purposes, as
it can be absorbed into the definition of the basis states, and
hence can be gauged away.
\footnote{
   For this note that while inversion symmetry is evidently
   broken on a single edge, in the actual 2D case with another
   open edge at a far other end of a sample, exactly the same
   dispersion is duplicated there, yet with $k \to -k$. Then
   given that on a large but finite plain 2D sample only a
   single helical edge mode exists around the entire sample,
   this pins the position of the Dirac cone to either $k_0=0$
   (as in \Fig{fig:helical}) or $k_0=\pi$ (cf. armchair vs.
   zigzag edges in graphene in the Kane-Mele model
   \cite{KaneMeleQSH}).
}

By assuming a finite half-bandwidth $D$ and having translational
invariance by working in momentum space, this directly implies
an effective Brillouin zone (BZ) with momentum range $ | k | \le
k_{\rm max} $, see \Fig{fig:helical}, with a discontinuous
dispersion across the Brillouin zone (BZ) boundary. Conversely,
this defines an effective lattice constant, $ a = \pi / k_{\rm
max} $, via the one-particle dispersion, having
$|\varepsilon(\tfrac{\pi}{a})| = D$, i.e.,
\begin{eqnarray}
   a \equiv \tfrac{\pi v}{D}
\text{ .}\label{eq:lattice:a}
\end{eqnarray}
Below, we use the local density of states $\rho_0 = a \rho_{1D}$
as experienced by an impurity where $ \rho_{1D} = 1 / 2 \pi v $
is the constant one-particle density of states of a 1D system with
linear dispersion relation \cite{Giamarchi} (thus units are $[\rho_0]
= 1/$energy, whereas $[\rho_{1D}] = 1/$(energy\,$\cdot$\,distance)).
Using \Eq{eq:lattice:a}, the local density of states becomes
$\rho_0 = \frac{1}{2 D}$, which is consistent with standard NRG
conventions. For more on the effects of finite bandwidth and an
effective 1D lattice defined by \Eq{eq:lattice:a}, see
\Sec{sec:ftilde} [cf. \Eq{eq:helical:xgrid}]).

In numerical simulations, furthermore, adhering to standard NRG
conventions, we choose the unit of energy $D:=1$. By also
setting the unit of distance $a:=1$, this fixes the velocity to
$v=1/\pi$. We also set $\hbar = k_B=1$.

The last expression in \Eq{eq:H0:k} provides a more compact
notation using the spinor $\hat{\bf c}_{k} \equiv
(\hat{c}_{k\uparrow}; \hat{c}_{k\downarrow})$, where the
semicolon denotes a column vector. This explicitly shows that
the SU(2) spin symmetry is broken in the helical setting. It
reduces to the abelian U(1) symmetry with preserved component of
the total spin $S_z^\mathrm{tot}$ [this is in contrast to a
{\it chiral} system where both spins move in the same direction
\cite{Giamarchi,Lotem22}, which thus preserves SU(2) spin
symmetry]. In the helical case, we thus only explore the
combination of abelian symmetries U(1)$_\mathrm{charge}
\otimes$ U(1)$_\mathrm{spin}$ [see \Sec{sec:symmetries}
for further comments on symmetries].

The Hamiltonian (\ref{eq:H0:k}) can be written in real-space in
the standard continuous form
\begin{equation}
       \hat{H}_0 = - \iu v \int
       {\rm d} x \
       \hat{\Psi}^{\dagger}(x) \tau_z \partial_x \hat{\Psi}(x)
\text{ ,}\label{eq:H0:cont}
\end{equation}
where $\hat{\Psi} \equiv ( \hat{\Psi}_R; \hat{\Psi}_L )$ is the
spinor constructed from the slow helical fields of the right and
left moving electrons $ \hat{\Psi}_{R,L} $. For the free
electrons in the helical edge mode, i.e., the bath to which the
spin impurities are coupled, spins up and down and, respectively,
chirality (the direction of propagation) of the right/left
movers can be denoted in more general fashion by the index
\begin{eqnarray}
   \etaB \in
   \{\uparrow, \downarrow\} \equiv \{ R, L \} \equiv {\{+1, -1\}} ;
\label{def:etaB}
\end{eqnarray}
in particular, $ \hat{\Psi}_{R,L} \equiv \hat{\Psi}_{\uparrow,
\downarrow}$. When used as variable in equations below, the
particular meaning of this index is always clear in context.
The Hamiltonian \Eq{eq:H0:cont} may be defined on a finite-length
system with periodic boundary conditions (BCs) and hence
discrete momenta, or in the thermodynamic limit with a continuous
energy-momentum space. \Sec{sec:H0} represents the simplest
effective model that describes a {\it single} helical edge, for
example, in HgTe/CdTe quantum-well heterostructures that
possess axial and inversion symmetry around the growth axis.

\subsection{Coupling between the helical fermions and MIs}
\label{sec:H:cpl}

Let us introduce two MI spins $\hat{S}^{\alpha}_\etaI$ separated
at distance $x$ and located symmetrically around the origin at
positions $x_\etaI = \tfrac{\etaI x}{2}$ with
\begin{eqnarray}
   \etaI \in \{ \iR, \iL \} \equiv \{+1, -1\}
\label{def:etaI}
\end{eqnarray}
for left and right impurity, respectively [to be differentiated
from right / left movers denoted by $\etaB \in \{R,L\}$ in
\Eq{def:etaB}].
By working in energy-momentum space, the bath operators at
the location of the impurity are obtained via Fourier transform
[e.g., cf. \App{sec:NRG:setup}].  With this, one can focus both,
analytically and numerically, on the two impurities being
located along a single edge without having to worry about
periodic boundary conditions. This is possible, when the actual
(2D) sample is always considered much larger than the impurity
distance $x$. Thus without restricting the case, one is free to
think of the two impurities as being symmetrically located at
$\pm x/2$ along a straight edge around some arbitrary but fixed
origin.

The exchange interaction between the helical
electrons and these two MIs is described by the Hamiltonian
\begin{eqnarray}
  \hat{H}_{\mathrm{int}}
  &=& 2 \pi v \sum_{\etaI=\iR,\iL}
  \bigl[
    j_0
    \bigl(
       \hat{S}^{+}_\etaI
       \hat{\sigma}^{-}_\etaI
    + \mathrm{H.c.}
    \bigr)
    + j_z
      \hat{S}^{z}_\etaI
       \hat{\sigma}^{z}_\etaI
   \bigr]
\text{ .}\label{eq:Hint}
\end{eqnarray}
Here $ j_0 \equiv \rho_0 J$ and $j_z \equiv \rho_0 J_z$ are the
constant dimensionless exchange couplings, such that, for example,
$2\pi v\, j_0 = a J$ with $J$ the coupling strength of the
impurity in units of energy and $[a \hat{\sigma}_\eta^\alpha]=1$
dimensionless as typically used within the NRG, having
\begin{eqnarray}
  \hat{\sigma}^{-}_\etaI &\equiv&
  [\hat{\Psi}^{\dag} \, \tau^-\, \hat{\Psi}](x_\etaI)
= \Psi^{\dag}_{\downarrow}(x_\etaI)
  \Psi_{\uparrow}^{\,}(x_\etaI)
\notag \\
  \hat{\sigma}^{z}_\etaI &\equiv&
  [\hat{\Psi}^{\dag} \, \tau^z\, \hat{\Psi}](x_\etaI)
= [\Psi^{\dag}_{\uparrow}\Psi_{\uparrow}^{\,}
- \Psi^{\dag}_{\downarrow}\Psi_{\downarrow}^{\,}](x_\etaI)  \\
  S^{\pm}_\etaI &=& S^{x}_\etaI \pm S^{y}_\etaI
\text{ .} \notag
\end{eqnarray}
with $\tau^{\pm} \equiv \tfrac{1}{2} (\tau^x \pm \iu \tau^y)$.
The exchange interaction \eqref{eq:Hint} may be anisotropic,
having $J \ne J_z$, while it always conserves the $z$-projection
of the total (electron and MIs) spin.

In the next section, we adhere to the standard notations of the
literature devoted to the analytical study of Kondo impurities
coupled to helical electrons, where the Kondo coupling is
measured in units of the Fermi velocity, $\tilde{J} = (2 \pi v) j$,
and similarly for $\tilde{J}_z$.

\section{Analytical theory of the RKKY regime \label{sec:FT-RKKY}}

Let us briefly review the RKKY theory for the helical edge mode
coupled to two MIs \cite{Yevt_2018} using field theoretical
machinery.

\subsection{Non-interacting fermions}

To describe degrees of freedom of the MIs, one should approximately
integrate out the helical fermionic edge modes. A natural way to do
this is to exploit the formalism of functional integrals. As a
result, one arrives at an effective action for the impurity spins:
\begin{eqnarray}
   e^{-\delta \mathcal{S}_{\rm imp}} =
   \tfrac{1}{Z_0} \int D[\Psi]\ e^{-[\mathcal{S}_0 + \mathcal{S}_\mathrm{int}]}
\text{ .}\label{Spin-effective-action}
\end{eqnarray}
Here $\mathcal{S}_0$ and $\mathcal{S}_\mathrm{int}$ are the
action of the free electron system, i.e., the bath, and of the
electron-impurity interaction (\ref{eq:Hint}), respectively;
$Z_0$ is the statistical sum of the electron system without spin
impurities. The functional integration is performed over fermionic
(Grassmann) variables. The spin degrees of freedom in the action
are described by one of the known approaches (e.g. coherent
spin representation or a set of Majorana Grassmann variables,
etc., see the book \cite{ATsBook}). A particular choice of the
spin variable is not important for our current purposes.

Let us start from the simplest case of the non-interacting
fermions. The electron action $\mathcal{S}_0$ in the Matsubara
representation reads
\begin{eqnarray}\label{free-e-action}
   \mathcal{S}_{0} &\overset{\eqref{eq:H0:cont}}{=}&
   \int\limits^{\beta}_{0}\!\! d\tau\!\! \int\!\! dx \
      \hat{\Psi}^{\dagger}(\tau, x) \underbrace{
         \bigl( \partial_{\tau} - \iu \tau^z v \partial_x \bigr)
      }_{\equiv -\hat{\mathcal{G}}^{-1}_{0} }
      \hat{\Psi}(\tau, x)
\end{eqnarray}
where
\begin{eqnarray*}
\hat{G}_0^{-1} &=&
    \begin{pmatrix}
      G^{-1}_{0 R}   & 0 \\
      0  & G^{-1}_{0 L}  \\
    \end{pmatrix} ;
\ \quad \
  G^{-1}_{0 \etaB}(\tau,x)
  \equiv -\partial_{\tau} + \iu \etaB v \partial_x
\text{ .}% \label{eq:G}
\end{eqnarray*}
with $\etaB = \pm 1 $ as in \Eq{def:etaB}. The Matsubara Green's
functions of the helical electrons in the momentum-frequency and
space-frequency representation are given by
\begin{eqnarray}
\label{Green-functions-1}
   G_{0\etaB}(\omega_n,k) &=& \tfrac{1}{\iu\omega_n - \etaB v k} \\
\label{Green-functions-2}
   G_{0\etaB}(\omega_n,x) &=& -\iu\etaB\,
     \tfrac{\mathrm{sgn}(x)\theta(\etaB x\omega_n)}{v}\, e^{-\frac{|x \omega_n|}{v}}
\text{ .}
\end{eqnarray}
The combined action reads
\begin{eqnarray}
   \mathcal{S}_0 + \mathcal{S}_\mathrm{int} &=&
   \int\limits^{\beta}_{0} \! d\tau \! \int \!dx\
   \hat{\Psi}^{\dagger}
      \bigl[ -\hat{G}^{-1}_{0} + \hat{V} \bigr]
   \hat{\Psi} ,
\label{e+S-action}
\end{eqnarray}
where
\begin{eqnarray}\label{V}
   \hat{V}(x) & = & \sum_{\etaI} \delta(x-x_\etaI)
   \begin{pmatrix}
      \tilde{J}_z S^{z}_{\etaI}  &  \tilde{J}   S^{-}_{\etaI} \\
      \tilde{J}   S^{+}_{\etaI}  & -\tilde{J}_z S^{z}_{\etaI} \\
   \end{pmatrix}
\text{ .}\label{eq:V:def}
\end{eqnarray}
Calculating the Gaussian integral over the Grassmann variables,
we obtain the contribution to the spin action,
\begin{eqnarray}\label{Spin-action-formal}
   e^{-\delta \mathcal{S}_{\rm imp}}
   &=& \tfrac{1}{Z_0}\det\bigl(
      -\hat{G}^{-1}_{0} + \hat{V}
   \bigr)
   = \det\bigl(
   \hat{I} - \hat{G}_{0}\hat{V}
   \bigr) \notag\\
      & = & e^{\mathrm{Tr}\ln(\hat{I} - \hat{G}_{0}\hat{V}) }
\text{ .}
\end{eqnarray}
This expression is formally exact and, being properly regularized,
describes all effects of the electron coupling to the spin impurities.
Following the standard RKKY scheme, let us now focus on the weak
coupling regime by restricting ourselves to terms up to second
order in $J$ in the action. This yields
\begin{eqnarray}\label{Spin-action-expanded}
  \delta\mathcal{S}_{\rm imp} &=& J^2
       \int^{\beta}_{0} \!\! d\tau_1
       \int^{\beta}_{0} \!\! d\tau_2\ \bigl[ \\
 &\phantom{+}&
      S^{+}_{1}(\tau_1)\, G_{0R}({\cal X}_1{;}{\cal X}_2) \,
      S^{-}_{2}(\tau_2)\, G_{0L}({\cal X}_2{;}{\cal X}_1) \notag\\
 &+&
      S^{-}_{1}(\tau_1)\, G_{0L}({\cal X}_1{;}{\cal X}_2) \,
      S^{+}_{2}(\tau_2)\, G_{0R}({\cal X}_2{;}{\cal X}_1)
   \ \bigr]
\notag
\end{eqnarray}
where ${\cal X}_j \equiv (\tau_j,x_j)$. As argued below, other
second order combinations do not contribute. After Fourier
transform to the Matsubara frequencies, the first term in
(\ref{Spin-action-expanded}) takes the form
\begin{subequations}\label{eq:Simp:F}
\begin{eqnarray}\label{Spin-action-Matsubara}
  &&
  \tilde{J}^2 T\sum_{n}
  S^{+}_{1}(\Omega_n) \,
  S^{-}_{2}(-\Omega_n)\,
  \mathcal{F}(\Omega_n) \, , \\
\label{F-definition}
   && \mathcal{F}(\Omega_n) = T\sum_{m}
   G_{0R}(-x, \omega_m) \,
   G_{0L}( x, \omega_m{+}\Omega_n)
\,,\quad
\end{eqnarray}
with $x \equiv x_2 - x_1 > 0$ being the inter-impurity distance.
The most interesting is the low temperature regime, $x \ll L_T \equiv v/T$,
where the summation over frequencies in (\ref{F-definition})
can be replaced by the integration over $\frac{d\omega}{2\pi T}$,
resulting in
\begin{eqnarray}\label{F-answer}
  \mathcal{F}(\Omega_n) =
  -\tfrac{1}{4\pi v x} e^{-|\Omega_n| \frac{x}{v}}
\text{ .}
\end{eqnarray}
\end{subequations}
Expressions similar to \Eqs{eq:Simp:F} are governed also by the
second term in (\ref{Spin-action-expanded}). Combining all terms
together, we obtain
\begin{eqnarray}\label{Spin-action-with-kernel}
   \delta\mathcal{S}_{\rm imp}
&=&
   - \tfrac{\tilde{J}^2 T}{4\pi v x} \sum_{\Omega_n}
   e^{-|\Omega_n| \frac{x}{v}} \bigl[\
   S^{+}_{1}(\Omega_n)S^{-}_{2}(-\Omega_n) + \quad \\[-2ex]
&& \hspace{1.17in}
   S^{-}_{1}(\Omega_n)S^{+}_{2}(-\Omega_n)\
\bigr]
\, .
\nonumber
\end{eqnarray}
We are interested in slow motion of the MI spins with
characteristic frequencies being much smaller than the inverse
time-of-flight of the electron between the MIs, $ |\Omega_n| \,
\frac{x}{v} \ll 1$. In this case, $e^{-\frac{x |\Omega_n|}{v}}
\simeq 1 $ and, returning back to the imaginary time, we arrive
at the expression
\begin{eqnarray}\label{Spin-action-local}
   \delta\mathcal{S}_{\rm imp} &=&
    - \tfrac{\tilde{J}^2}{4 \pi v x}
      \int\limits^{\beta}_{0}d \tau \ \HR(\tau) \\
    \HR(\tau) & \equiv &
        S^{+}_{1}(\tau)S^{-}_{2}(\tau) + S^{-}_{1}(\tau)S^{+}_{2}(\tau)
\text{ .}
\end{eqnarray}
This is the action of a system described by an effective
RKKY-like Hamiltonian of the MI spins:
\begin{subequations}\label{H-RKKY}
\begin{eqnarray}\label{H-RKKY:1}
   \HR & = & - \ER
   \left(S^{+}_{1}S^{-}_{2} + S^{-}_{1}S^{+}_{2}\right)
\end{eqnarray}
with the RKKY energy scale
\begin{eqnarray}
   \ER & \equiv & \tfrac{\tilde{J}^2}{4\pi v x}
   = \tfrac{\pi v j_0^2
   }{x}
   \ \overset{\eqref{eq:lattice:a}}{\equiv}\
   \tfrac{j_0^2}{x/a} D \ \geq 0
\text{ .} \label{E-RKKY}
\end{eqnarray}
\end{subequations}
This anisotropic spin coupling is ferromagnetic
for all distances. Note that the RKKY coupling
in a normal metal also starts out with a ferromagnetic
sign at short distances \cite{Yafet87}.
The first two expressions in \Eq{E-RKKY}
are valid in the wide-band limit.
In the presence of a finite but large bandwidth, this RKKY scale
can also be rewritten as in the last expression where the RKKY
coupling in units of the bandwidth ($D$) is simply given by the
dimensionless $j_0^2$ [\Eq{eq:Hint}] divided by the distance of
the impurities in units of the lattice spacing [\Eq{eq:lattice:a}].
The wideband limit in the analytical approach therefore
implies two assumptions: based on the second order approach used
to derive \Eq{E-RKKY}, this implies (i) $j_0 \equiv \rho_0 J \ll
1$, i.e., $J\ll D$.  Yet via \Eq{eq:lattice:a} [see also
\Eq{eq:HRKKY:I1}], the wideband limit also implies (ii) $x \gg a$.

\begin{figure}[tb]
\begin{center}
\includegraphics[width=\linewidth]{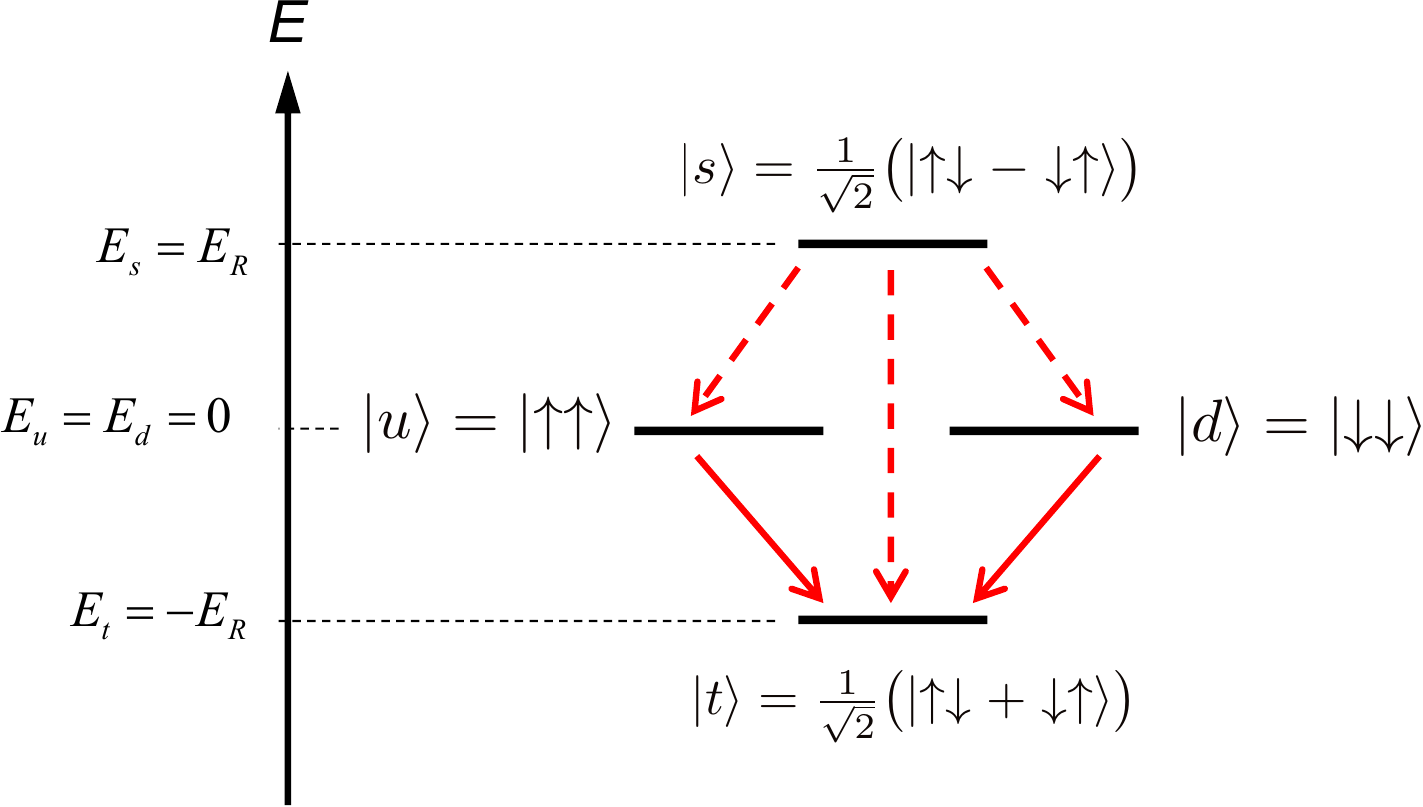}
\end{center}
\caption{
   Eigenstates and eigenlevels of the RKKY Hamiltonian
   (\ref{H-RKKY}). Red lines and arrows show various decay
   channels of the singlet state (dashed lines) and up/down
   states (solid lines). These include spin-flips (outer paths)
   or phase flip (center path), as probed by the transverse
   $\langle S_{\iL}^+ \Vert S_{\iR}^-\rangle_\omega$ etc. and the
   longitudinal $\langle S_{\iL}^z \Vert S_{\iR}^z\rangle_\omega$
   dynamical spin correlation function, and which in the helical
   context reflect backward and forward scattering,
   respectively.
}
\label{fig:Spectrum}
\end{figure}

The ground state of this MI spin Hamiltonian is the triplet
state with $S_z{=}0$ (cf. \Fig{fig:Spectrum}):
\begin{eqnarray}\label{t-state}
   |t\rangle = \tfrac{1}{\sqrt{2}}
   \bigl(|{\uparrow\downarrow}\rangle +
         |{\downarrow\uparrow}\rangle
   \bigr)_\mathrm{imp}
\text{ .}
\qquad (E_{t}= - \ER)
\end{eqnarray}
Hence despite the ferromagnetic coupling in \Eq{H-RKKY:1},
the spins are antialigned. Yet by not being in a singlet state,
they are still also exposed to Kondo screening.

The forward-scattering (${\sim} J_z$) parts of the electron--MI
coupling do not give a contribution $\sim S^z_1S^z_2$ to the
second-order effective spin action. This is because the
contribution of the $ S^{z}_1 S^{z}_2 $-term to
Eqs.(\ref{Spin-action-expanded}--\ref{F-definition}) contains
the product of two Green's functions of the same chirality,
$G_{\etaB}(x, i\omega_n) G_\etaB(-x, i\omega_n)$, which vanishes
in Eq.(\ref{F-definition}) due to Eq.(\ref{Green-functions-2}).
Cross-scattering of the type $S^z S^{\pm}$ is also absent in the
effective second-order spin action in \Eq{Spin-action-local}
because, in equilibrium, the electrons of different chiralities
are not correlated. Namely, the averaging of the corresponding
combinations of fermion operators contain three operators of the
same chirality, e.g. $\langle \Psi^{\dag}_R(X_1)\Psi_L(X_1)
\Psi^{\dag}_R(X_2) \Psi_R(X_2)\rangle$ which vanishes due to the
property $\langle \Psi_L(X_1) \Psi^{\dag}_R(X_2) \rangle =0$.

In addition to the ground state $|t\rangle$, the effective spin
Hamiltonian (\ref{H-RKKY}) possesses three other eigenstates
with higher energies (cf. \Fig{fig:Spectrum}): the remaining
triplet states denoted as the degenerate doublet (`up' and
`down') states,
\begin{eqnarray}
  |u\rangle = |{\uparrow\uparrow}\rangle ,
  |d\rangle = |{\downarrow\downarrow}\rangle
\qquad (E_{u/d}=0)
\end{eqnarray}
and the singlet state,
\begin{eqnarray}\label{s-state}
   |s\rangle = \tfrac{1}{\sqrt{2}}\left(
   |{\uparrow\downarrow}\rangle - |{\downarrow\uparrow}\rangle
   \right)
\text{ .}\qquad (E_s = \ER)
\end{eqnarray}
The excitation energies of the doublet and the singlet relative to the
ground state are $\ER$ and $2\ER$, respectively (cf. \Fig{fig:Spectrum}).

There is an important physical difference between the ground and
all excited states of the effective spin Hamiltonian. The first
one is stable and corresponds to the ground state of the total
many-particle (electrons + spins) system projected onto the spin
sector. To be specific, in the wideband limit $D\to\infty$ (i.e.,
$j_0 \to 0$) and zero temperature, the pair of impurities live
in an exact non-correlated product ground state with the helical
edge channel, i.e., $|g\rangle \equiv |t\rangle \otimes
|0\rangle_\mathrm{edge}$. In contrast, the excited states of
the spins remain connected to the many-electron ``reservoir''
and as such cannot be simply written like product states for
true eigenstates of the entire system. Instead, they represent
resonant states, in the sense that they give rise to (narrow)
resonances in the dynamical spin response functions.

By using the Fermi Golden rule, one can estimate decay rates
of the excited spin states: the decay rate of the states $ | u
\rangle $ and $ | d \rangle $ is $ \propto \tilde{J}_0^2 \ER $
while the state $ | s \rangle $ has the parametrically smaller
decay rate $\tilde{J}_z^2 \tilde{J}_0^4 \ER $ (see
[\onlinecite{VIYu_2022}] for more details).

\subsection{Taking into account non-perturbative effects of \\
electron interactions or finite $ J_z $}

Non-perturbative effects of the electron interactions and of a
finite $ J_z $ on the indirect exchange interaction of two MIs
attached to the helical edge can be described by using the
bosonized theory. The standard free (without impurities) action
reads as:
\begin{equation}
  S_b = \tfrac{1}{2 \pi K u} \int {\rm d} \tau \, {\rm d} x \,
      \left(
        \partial^2_\tau \phi + u^2 \partial^2_x \phi
      \right) \, .
\end{equation}
The single bosonic field $ \phi $ describes both the spin and
chiral degrees of freedom \cite{MaciejkoOregZhang,MaciejkoLattice}.
$ K $ and $ u $ are the Luttinger parameter and speed of the
helical plasmons, respectively. $ K $ incorporates effects of
the electron interaction.
The boson-spin exchange interaction is described by actions:
\begin{eqnarray}
  S_{\rm fs} & = & \iu \tfrac{\tilde{J}_z}{\pi u K} \int {\rm d} \tau \, {\rm d} x \,
                   \sum_{j=1,2} \delta(x-x_j) \, S^z_j \, \partial_\tau \phi \, ; \\
  S_{\rm bs} & = & \tfrac{\tilde{J}}{2 \pi \xi} \int {\rm d} \tau \, {\rm d} x \,
                   \sum_{j=1,2} \delta(x-x_j) \times \\
             &   & \times \left[ S^+_j e^{-2 \iu \phi} + c.c. \right] \, .
             \nonumber
\end{eqnarray}
Here, subscripts $ \mbox{fs} $ and $ \mbox{bs} $ stand for
forward-/backward scattering, and $ \xi $ is the lattice constant
which is usually needed to make the bosonized theory regular.

By using the Emery-Kivelson gauge transformation \cite{EmKivRes},
one can completely reduce the effect of $ J_z $ to changing the
dimension of the backscattering, which can be described by the
effective dependence of $ K $ on $ J_z $:
\begin{equation}
   K_{\rm eff} = K \left( 1 -
     \tfrac{\rho_{1D} \tilde{J}_z}{2 K}
   \right)^2
\text{ .}\label{eq:Keff}
\end{equation}
Below, we assume that $\tilde{J}_z$ is included in $ K $.

The bosonic theory is not quadratic and a
functional integral over the bosonic fields
\begin{equation}
  \label{Bos-FI}
  \int {\cal D} \{ \phi \} e^{-(S_b + S_{\rm bs})}
\end{equation}
cannot be calculated exactly, even formally. However, the
effective action of the indirect spin interaction can be
obtained for small $J$ by calculating the integral over $\phi$
as the first cumulant. This is similar to the renormalization
group (RG) treatment of the sine-Gordon theory \cite{Giamarchi}:
one Taylor expands the exponential in Eq.(\ref{Bos-FI}) in $S_{\rm bs}$
up to the second order, calculates the integral over $\phi$ and
re-exponentiates the answer:
\begin{eqnarray}
  \int {\cal D} \{ \phi \} e^{-(S_b + S_{\rm bs})}
  &\simeq& 1 + \tfrac{1}{2} \langle S_{\rm bs} S_{\rm bs} \rangle_{S_b} \\
  &\simeq& \exp \left( \tfrac{1}{2} \langle S_{\rm bs} S_{\rm bs} \rangle_{S_b} \right)
  \nonumber
\end{eqnarray}
where $ \langle A \rangle_{S_b} \equiv \int {\cal D} \{ \phi \}
\, A e^{-S_b} $. Using the well-known expression for the bosonic
correlation function
\begin{eqnarray*}
  \Pi(\tau) & = & \left( \tfrac{\pi \xi}{\beta u} \right)^{2K}
              \tfrac{1}{\left( \sin^2(\pi \tau T) + \sinh^2( \frac{R}{L_T}) \right)^K} \\
  \beta & \equiv & \tfrac{1}{T} \, , \ L_T = \tfrac{\beta u}{\pi} \, ;
\end{eqnarray*}
we arrive at the perturbative (in $ J $) expression for the
effective spin action
\begin{equation}
   \label{RKKY-act}
   {\cal S} = - \tfrac{\tilde{J}^2}{(2 \pi \xi)^2} \sum_{j,j'} \int \, {\rm d} \tau_{1,2} \,
   S^+_j(\tau_1) \Pi(\tau_1 - \tau_2) S^-_{j'}(\tau_2)
\, .
\end{equation}
This theory is non-local in time and, thus, takes into account
retardation effects which are beyond the Hamiltonian formulation
of the RKKY theory. The action can be reduced to a local one if
$ \xi \ll | x_1 - x_2 | \ll L_T $ and $ 1/2 < K \le 1 $.
In this case, \Eq{RKKY-act} reduces to
\begin{eqnarray}
  \SR & = & - \ER \int {\rm d} \tau \HR (\tau) \, , \\
  \ER & \equiv & \tfrac{2 \tilde{J}^2}{(2 \pi \xi)^2} \int {\rm d} \tau \Pi(\tau) \, .
\end{eqnarray}
These expressions were analyzed in Ref.\cite{Yevt_2018}. They
coincide with answers derived in the previous section for the
noninteracting case, $ K = 1 $.

The application of the bosonized theory has several advantages.
Firstly, it takes into account non-perturbative effects of the
electron interaction and of $ J_z $. Besides, it is straightforward
to go beyond the quadratic approximation in $J$ and derive
renormalization of this Kondo coupling constant
\cite{MaciejkoOregZhang,MaciejkoLattice}.

\begin{figure*}[p!]
\begin{center}
\includegraphics[width=.8\linewidth]{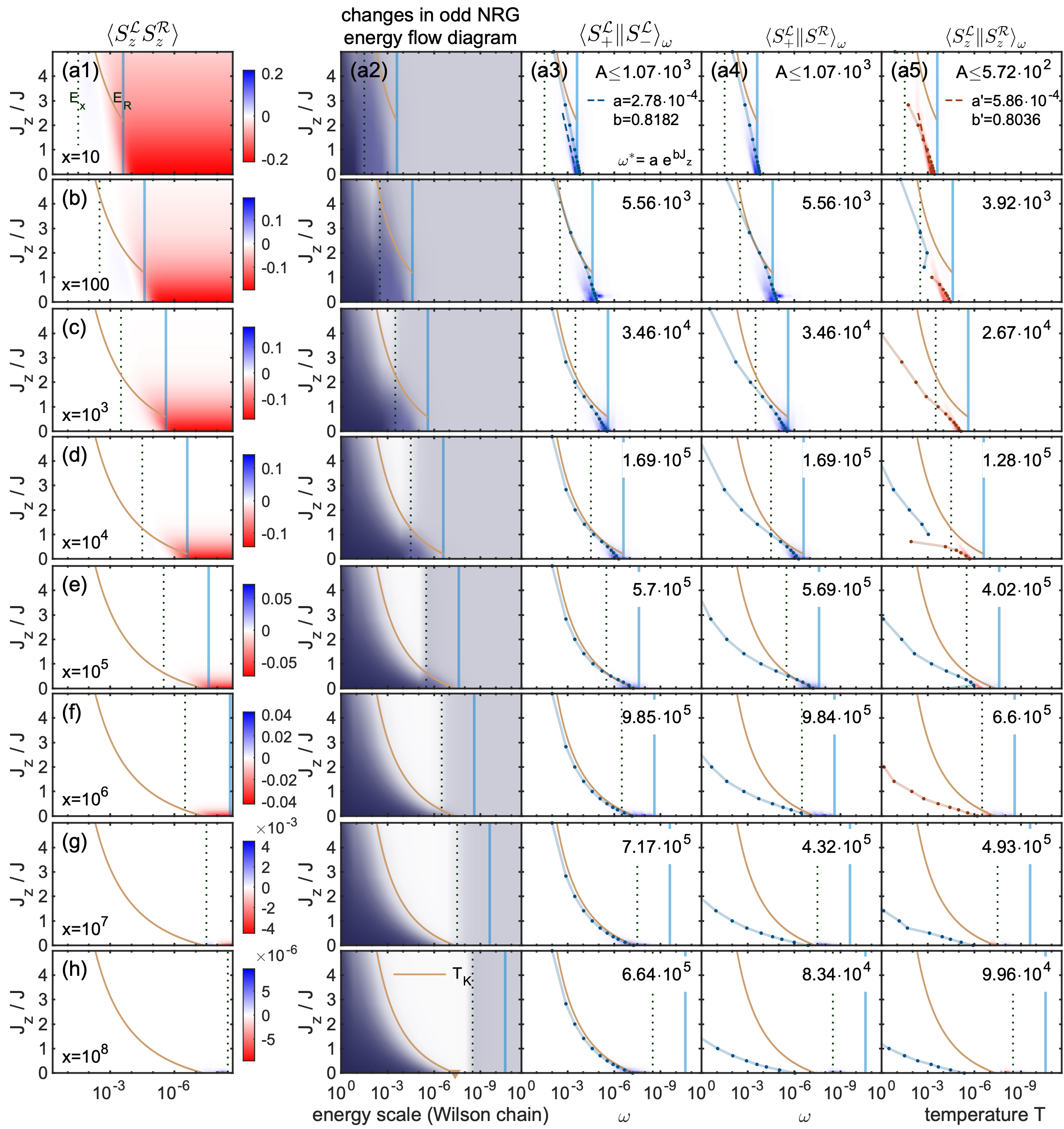}
\end{center}
\caption{NRG analysis of the 2HKM for $J=0.1$
   (i.e., $j_0 =0.05$) --
   Each row corresponds to a different integer impurity distance
   $x$ as indicated with the left panel, which increases
   exponentially from top to bottom. The vertical axis is the
   same for all panels, namely $J_z/J \equiv j_z/j_0$, whereas
   the horizontal axis represents energy in various forms. Here
   we adopt the NRG energy scale point of view, throughout,
   which starts at large energies at the left and then proceeds
   towards exponentially smaller energy scales towards the
   right. Rows (columns) are labelled by letters (numbers),
   respectively, e.g., having (a1) for the upper left panel.
   The vertical black dotted [solid blue] marker replicated in
   all panels indicate the coherence scale \Ex{} \eqref{eq:Ex}
   [RKKY scale \ER \eqref{H-RKKY}], respectively, as labelled in
   panel (a1), having $\Ex/\ER = 1/\pi j_0^2 = 127.3$,
   throughout. Similarly, the brown solid curved line shows the
   analytical single-impurity Kondo temperature $\TK(J,Jz)$
   [\App{App:TK-aniso}] for reference. This curve \TK is
   visually cutoff at \ER, because \TK is irrelevant at lower
   temperatures. The brown marker in panel (h2) shows the finite
   intercept $\TK(J=0.1,J_z)=4.31{\cdot}10^{-8}$ at $J_z=0$.
   Each column shows the quantity indicated above the top panel:
   Panels (:,1) [first column] show the static (equal-time)
   inter-impurity correlations $\langle S_z^{\iL}
   S_z^{\iR}\rangle$ vs. temperature and $J_z$. Panels (:,2)
   [second column] gives a visual impression of the changes
   along the NRG energy flow diagram vs. energy scale $\omega_n
   \sim \Lambda^{-n/2}$ (for precise prefactors, see
   \cite{Wb12_FDM}). This encodes cumulative effective thermal
   weights in three arbitrary but fixed low-energy symmetry
   sectors into red-green-blue (RGB) colors for all odd Wilson
   shells $n$ (to avoid even/odd effects) based on an effective
   inverse temperature $\beta_n \equiv 4 \omega_n$.
   The symmetry sectors chosen for this were $q \equiv
   (Q;S_z^\mathrm{tot}) \in (0; 0,1,-1)$ with $Q$ the total
   charge relative to half-filling.
   The remainder of the columns shows dynamical spin-spin
   correlation functions $\langle \hat{S}^\eta \Vert
   \hat{S}^{\eta'\dagger}\rangle_\omega$ as indicated at the top
   of each column at zero temperature ($T=1.8\, 10^{-11}$).
   The solid-dotted lines shows the respective derived inverse
   static susceptibility $T^{\eta\eta'}_{S} =
   1/4\chi^{\eta\eta'}_{S}$ [cf. \Eq{eq:TS:gen}]. Blue (red)
   indicates positive (negative) value, respectively. The number
   at the top right of each panel indicates the maximum absolute
   value the spectral data $A$ of the broadened NRG spectral
   data, and hence gives an impression of numerical range. Blue
   (red) shading in all panels except for the second column
   indicates positive (negative) values, respectively.
   The dashed lines in (a,c;3) and (a,c;5) represent exponential
   fits as indicated with (a5) of the maximum of the spectral
   data for $J/J_z \leq 0.5$. NRG parameters (e.g., see
   \cite{Wb12_FDM} for detailed definitions): $\Lambda=4$,
   truncation energy $E_\mathrm{trunc}=8$; $z$-averaged over
   $n_z=4$, with log-Gauss broadening $\sigma=0.3$ of the
   spectral data after $z$-averaging.
}
\label{fig:NRGspectra}
\end{figure*}

\section{NRG analysis of the 2HKM \label{sec:NRG:results}}

We proceed by presenting the NRG results for the 2HKM in this
section, before we explain in detail how we setup the
two-impurity helical setup within the NRG framework in the
subsequent \Sec{sec:NRG:setup}.
An exhaustive NRG parameter scan for the 2HKM at $J=0.1$ is
shown in \Fig{fig:NRGspectra}. All panels have the anisotropy
parameter $J_z/J$ on their vertical axis whereas the horizontal
axis shows energy in various forms [temperature [first column,
i.e., \Figs{fig:NRGspectra}(:,1)], NRG energy scale [second
column \Figs{fig:NRGspectra}(:,2)], or frequency (remainder of
columns, \Figs{fig:NRGspectra}(:,3-5))]. Throughout, energy is
decreasing towards the right, as motivated by the NRG approach
(second column), where large energy scales come first, followed
by a zoom into exponentially small energy scales towards the
right. Each row labelled by a letter shows data for a fixed
impurity distance $x$ as indicated in the left panel.
This distance is always chosen integer, i.e., on the grid
\eqref{eq:helical:xgrid}, and increases exponentially towards
lower panels, with the value specified in the left panels.

The inverse time to travel in between the impurities naturally
gives rise to its own energy scale
\begin{eqnarray}
   \Ex \equiv \tfrac{v}{x}
\label{eq:Ex}
\end{eqnarray}
which may be interpreted as the `Thouless energy' of a quantum dot
confined by the magnetic impurities. This energy scale in
itself is independent of any impurity properties other than
their distance [see also \Eq{eq:domega}]. Now since the
impurity distance explicitly enters the construction of the
Wilson ladder \eqref{eq:block-tridiag:2}, there is always a
clear qualitative change in the NRG finite size spectra, aka.
energy flow diagram at the energy scale \Ex, as visualized
in a condensed graphic way in \Fig{fig:NRGspectra}(:,2), i.e.,
the second column. There one observes a distinct change
(whitish to gray transition) right at \Ex (vertical dotted line).
This `curtain' is opened, i.e., moves towards the right as the
impurity distance is increased from the top to the bottom
panels. As such this `unveils' the underlying Kondo physics.
At largest distance shown in the bottom row of panels,
\Fig{fig:NRGspectra}(h,:), \ER has already dropped below the
smallest \TK reached at $J_z=0$ [brown marker at $\TK(J,0)
\simeq 4{\cdot}10^{-8}$]. Hence in this case the impurities are
fully Kondo screened individually, and RKKY physics plays no
role any longer, and hence is absent for all $J_z\geq 0$.

At larger energies above the coherence scale, $E \gtrsim \Ex$,
i.e., to the left of the vertical dotted black line, the system
is described by effectively independent impurities. Since
information cannot travel faster than $v$ between the impurities,
the impurities do not yet `see' each other at energy scales
$E > \Ex$. In this sense, the finite size spectra in the NRG
look identical in \Fig{fig:NRGspectra}(:,2) for large energies
$E \gtrsim \Ex$, i.e., to the left of vertical dotted line
across all panels in \Figs{fig:NRGspectra}(:,2).

For low energies, \Fig{fig:NRGspectra} shows that the smallest
energy scale from the point of the impurity is $\max(\ER,\TK)$.
Therefore \ER serves as a low-energy cutoff. For example, the
energy flow diagram is converged below \ER [uniform gray area to
the right of \ER (blue vertical marker) in \Figs{fig:NRGspectra}(:,2)].
On explicit physical grounds this is replicated in terms of static
inter-impurity correlations for $T<\Ex$ in \Figs{fig:NRGspectra}(:,1),
or by having no particular structure in the spectral density for
$|\omega|<\ER$ in the right panels. For this reason the brown
line which depicts the Kondo scale, is only shown above \ER and
hence terminated at the blue vertical marker line. For $J_z$
above this crossing point, Kondo screening sets in and eventually
fully dominates. This is seen in the brightening of the dark red
inter-impurity spin correlations in \Fig{fig:NRGspectra}(:,1)
towards white (no correlations) when increasing $J_z$ (vertical
direction).

The bare RKKY regime concerns the energy window $E \in
[\ER,\Ex]$. With $\rho_0 = 1/2D$ [\Eq{eq:lattice:a}], the
dimensionless Kondo coupling in \Fig{fig:NRGspectra} is small,
having $j_0 \equiv \rho_0 J = 0.05 \ll 1$, and thus
$\tfrac{\Ex}{\ER} = \tfrac{1}{\pi j_0^2} = 127.3$.
Therefore the bare RKKY regime spans about two orders of magnitude
in energy scale. It competes with Kondo physics for the case
$\TK(J,J_z) < \Ex$, i.e., when the Kondo scale is small enough
that the screening clouds of the two impurities overlap. We
refer to this intersect as the intermediate energy regime.

This intermediate regime  becomes visible as the
lighter-blue shaded area in \Fig{fig:NRGspectra}(a-f,2)
in between the two vertical makers below the brown solid line
which represents the single-impurity Kondo scale \TK.
The latter scale \TK is cut off at \ER, also visually so by
terminating its line towards the right at the blue vertical
marker. Hence the intermediate regime is present
(i) if there is a brown line segment in between
the two vertical markers (black dashed and blue),
in which case (ii) the intermediate regime occurs below it.
For example, with no brown line segment in the bare RKKY regime
in the lowest two rows in \Fig{fig:NRGspectra}, $\TK> \Ex$
dominates the low-energy regime throughout, and the system
consists of two individually Kondo screened impurities.
However, once $\TK<\Ex$, the buildup of the Kondo screening
cloud is affected by the presence of the other impurity. This
leads to characteristic changes in the NRG energy flow diagrams,
as seen in \Fig{fig:NRGspectra}(:,2).

\subsection{Spin-spin correlation functions and low-energy scales}

Data for dynamical spin-spin correlation function
$\mathcal{A}(\omega)$ [as defined in \Eq{eq:Aspec} in
\App{sec:complex-spectral}] is shown as color-plots in the
right panels of \Fig{fig:NRGspectra}. The frequency of the
maximum for fixed parameters sets the relevant low-energy scale
for the spins. It is well traced by the inverse static
susceptibility (solid-dotted lines)
\begin{eqnarray}
  \omega^\ast &\cong& T^{\eta\eta'}_{S}
  \equiv \tfrac{1}{4\chi^{\eta\eta'}_{S}}
\label{eq:TS:gen}
\end{eqnarray}
as derived from the dynamical susceptibility
\begin{eqnarray}
   \chi^{\eta\eta'}_{S}
   &\equiv& \chi^{\eta\eta'}_{0;S}
   \equiv \lim_{\omega\to 0}
   \langle \hat{S}^\eta\Vert \hat{S}^{\eta'\dagger\,}\rangle_\omega
   \ \ \in \mathbb{R}
\text{ .}\label{eq:chi0}
\end{eqnarray}
with $S \in \{S_z, S_\pm\}$, and the normalization convention
for $S_\pm \equiv \tfrac{1}{\sqrt{2}} (S_x \pm \iu S_y)$ chosen
such that $|S_\pm|^2 = |S_z|^2$ have the same Frobenius norm.

In case that $\eta=\eta'$, i.e., inter-impurity correlations,
only a single label may be shown, e.g., $T^{(\eta)}_{S} \equiv
T^{\eta\eta}_{S}$ or $\chi^{(\eta)}_{S} = \chi^{\eta\eta}_{S}$
(the $\eta$ label may be skipped altogether then, since the
impurities are considered identical, hence by symmetry, e.g.,
$\chi_{S} \equiv \chi^{\iL}_{S} = \chi^{\iR}_{S}$).
Strictly speaking, the interpretation as an energy scale is only
justified for `diagonal' correlations, (here intra-impurity
$\eta=\eta'$), as this guarantees positive spectral data and
hence a respective positive energy scale. But it is useful to
also include $\eta\neq\eta'$ here for the sake of the argument
and presentation. We will also refer to $\chi^{\eta\eta'}_{S_\pm}$
which involves a spin-flip, as the transverse susceptibility,
and $\chi^{\eta\eta'}_{S_z}$ which involves a phase flip, as
the longitudinal susceptibility.
The choice of the prefactor ($1/4$) is motivated by the standard
definition of the Kondo temperature in the plain single-impurity
Kondo model, $T_{K}^\mathrm{NRG} \equiv \tfrac{1}{4\chi_0}$
\cite{Hewson93,Bulla08}. For isolated RKKY impurities with the
energy spectrum as in \Fig{fig:Spectrum}, the spectral functions
have $\delta$-peaks at energies $\omega = \pm \ER$ for the
transverse, and $\pm 2\ER$ for the longitudinal correlation
function, thus resulting in $T_{S_\pm}^{\eta\eta';0} \equiv
\ER/2$ and $T_{S_z}^{\eta\eta';0} \equiv \mathrm{sgn}(\eta\eta') \ER$,
respectively [note the normalization convention of the spin
operators as indicated with \Eq{eq:chi0}]. Up to a sign, these
are independent of the choice of $\eta$ and $\eta'$, i.e., they
become the same for inter- and intra-impurity susceptibilities.
The latter is a direct consequence of the RKKY low-energy regime
where the pair of impurities, even though spatially separated,
act like a nearly-decouled microscopic unit governed by the
RKKY Hamiltonian.
In the following we will thus scale energies in the
numerical data by the smaller energy scale (where subscript `S'
denotes `spin'),
\begin{eqnarray}
   \TS \equiv T_{S_\pm}^{\eta\eta} \sim \ER/2
\label{eq:TS}
\end{eqnarray}
We prefer \TS over \ER, since \ER only represents a lowest order
estimate, whereas \TS includes the full many-body aspects of the
problem and is thus also self-contained and thus consistent
within the NRG.

The energy scales $T^{\eta\eta'}_{S}$ in \Eq{eq:TS:gen} capture
the low-energy scale of the impurity spins, as seen, e.g., in
the lowest panels \Figs{fig:NRGspectra}(f-h,3), [with the
impurity operators $S$ as well as their location $\eta$ and
$\eta'$ specified at the top of each of the right columns
\Figs{fig:NRGspectra}(:,3-5)]. There the inverse susceptibility
$T_{S_\pm}$ from the {\it intra}-impurity spin-spin correlation
(solid-dotted line) [\Figs{fig:NRGspectra}(:,3)] follows closely
the analytical Kondo scale $\TK(J,J_z)$ (brown line) up to a
constant prefactor of order one. This generally holds for
$\omega > \Ex$, i.e., to the left of the black dashed line in
\Figs{fig:NRGspectra}(:,3).

The two right-most columns of \Fig{fig:NRGspectra}, in contrast,
show {\it inter}-impurity correlations. These can only be due
to RKKY interactions, and hence diminish with increasing
impurity distance. Once this distance exceeds the size scale of
the Kondo screening cloud, i.e., $\TK > \Ex$, the inter-impurity
susceptibility becomes much smaller than the onsite
susceptibility, such that its inverse is orders of magnitude
larger in energy [e.g. compare solid-dotted line in
\Figs{fig:NRGspectra}(gh;4,5) to \TK (brown line)].
Its physical interpretation is that inter-impurity correlations
start to play a role {\it relative} to Kondo correlations only
once the latter are sufficiently suppressed, e.g., by a large
temperature scale $T \gtrsim T^{\iL\iR}_{S_\pm} \gg \TK$.

In the intermediate regime, where the low-energy physics is cut
off by RKKY, the inter- and intra-spin correlations start to
look identical when applying the same operators [e.g., compare
\Fig{fig:NRGspectra}(a,3) to \Fig{fig:NRGspectra}(a,4)].
This holds quantitatively as also seen by the overall scale
(see maximum spectral weight $A$ indicated with the panels).
This can be understood based on the RKKY impurity state
$|t\rangle$ in  \Eq{t-state} that (nearly) decouples as a
product state from the bath channel [cf. \App{App:finiteD}],
thus having
\begin{eqnarray}
  \langle \hat{S}^{\eta}_+ \hat{S}^{\eta'}_- \rangle \simeq
  \langle t | \hat{S}^{\eta}_+ \hat{S}^{\eta'}_- | t \rangle = \tfrac{1}{2}
\text{ ,}\label{eq:SpSm:correl:pm}
\end{eqnarray}
which holds for both, $\eta=\eta'$ (intra-impurity) as well as
$\eta\neq\eta'$ (inter-impurity), while bearing in mind that the
equal-time correlator above is identical to the integrated
spectral data over frequency [cf. sum rules].

The longitudinal inter-impurity correlations are shown in the
last column of \Fig{fig:NRGspectra}. Deep in the RKKY, the
dominant spectral weight of this non-diagonal dynamical
correlation functions is expected to be negative,
\begin{eqnarray}
  \langle \hat{S}^{\iL}_z \hat{S}^{\iR}_z \rangle
  \overset{\eqref{t-state}}{\simeq}
  \langle t | \hat{S}^{\iL}_z \hat{S}^{\iR}_z | t \rangle
  = -\tfrac{1}{4}
\text{ ,}\label{eq:SpSm:correl:zz}
\end{eqnarray}
and half the absolute value, in agreement with the red shading
(which indicates negative) and overall scale of the spectral
data, e.g., in \Fig{fig:NRGspectra}(a5) [see a detailed analysis
of the effects of finite bandwidth on the precise value of the
l.h.s. in \Eq{eq:SpSm:correl:zz} in \App{App:finiteD}, and in
particular \Fig{fig:RKKY-wband} therein]. Via spectral sum
rule, the frequency integrated data yields the data in
\Fig{fig:NRGspectra}(:,1) at given temperature (in the present
case, at the lowest temperature shown). Consistently, this
appears in a deep red in the RKKY state, indicating that the
spin-orientations of the two impurities are
antiferromagnetically (AF) correlated.

However, when crossing over into the Kondo regime, the weakened
inter-impurity longitudinal correlations can even change sign
and turn ferromagnetic (FM) [e.g., blue solid-dotted lines in
\Figs{fig:NRGspectra}(:,5) in the inverse susceptibility; this
necessarily has to originate from corresponding positive
spectral data, but its weight is too small, though, so it is not
visible in shading in the spectral data as shown]. The weak
ferromagnetic correlation can also be partly seen in the static
equal-time spin correlation [e.g., the very faint blue hue in
the intermediate regime just left of the red area in
\Fig{fig:NRGspectra}(b,1) or (e,1), and hence at significantly
elevated temperatures]. The sign change towards weak
ferromagnetic correlation observed in the longitudinal data is
not systematic, though. For example, not all inverse
susceptibilities in \Figs{fig:NRGspectra}(:,5) show a sign
change. Furthermore, deep in the Kondo regime, the system can
feature weak ferromagnetic (FM) inter-impurity spin correlations
and together with a sign change depending on the frequency
[e.g., faint red and blue shadings in
\Figs{fig:NRGspectra}(g-h,5) just above $J_z=0$ vs. $\omega$,
which is partly also visibly replicated vs. temperature e.g., in
\Figs{fig:NRGspectra}(g,1)]. For energies below \Ex this should
be well resolved by NRG, and not related to coherence affects
averaged out by NRG above \Ex due to coarse graining [cf.
discussion following \Eq{eq:domega}]. The appearance of weak AF
as well as FM correlations across the impurities may be related
to the fact, that for finite $J_z$ and finite bandwidth,
subleading terms can generate an effective small longitudinal
inter-impurity interaction $\mathcal{J}_z \hat{S}_z^\iL
\hat{S}_z^\iR$ which due to its oscillatory behavior vs.
distance may be ferromagnetic as well as antiferromagnetic [cf.
\App{App:finiteD}, and also the discussion around
\Eq{eq:Ilambda:zz:1}]. Consequently then, $z$-averaging of the
spectral data may lead to apparent non-systematic behavior in
the longitudinal correlations as a numerical artifact. In the
present case, however, we do not dwell on this any longer as
this concerns subleading effects.

\subsection{RKKY scale in spectral data}

The inverse transverse susceptibility deep in the RKKY regime
resembles a straight line on a semilog plot
[\Fig{fig:NRGspectra}(a,3-4)]. This also reflects the behavior
of the maximum in the actual spectral data. Generally, from its
very definition via the Kramers-Kronig integral relations, the
inverse susceptibility is also sensitive to the precise line
shape of the spectral data. Yet by having the maxima and inverse
susceptibility run in parallel for small $J/J_z$, this suggests
similar line shapes. Tracking and fitting the peak maxima in the
spectral data by the exponential fit specified with
\Fig{fig:NRGspectra}(a5) for $J/J_z \leq 0.5$, we obtain
$\omega^\ast \simeq 2.78{\cdot}10^{-4}\, e^{0.82 J_z}$ (blue
dashed line). It is lower-bound at $J_z=0$ by the analytically
obtained RKKY scale \ER [\Eq{H-RKKY}; blue vertical marker in
\Fig{fig:NRGspectra}], having $\ER = (\rho_0 J)^2/x =
2.50{\cdot}10^{-4}$ for $x=10$ [\Fig{fig:NRGspectra}(a,:)].
The difference of about 10\% is due to finite bandwidth
[cf.  \App{App:finiteD}].

As seen from the fits in \Fig{fig:NRGspectra}(a;3,5) the
low-energy scale in the intermediate regime diminishes
exponentially with decreasing $J_z$, as in $T_S \sim \omega^\ast
\simeq a e^{b J_z}$ with $b>0$. This is qualitatively similar
to the one-impurity Kondo temperature in the anisotropic
1-impurity case (cf. \App{App:TK-aniso}) yet with different
renormalized coefficients. For one, this shows that the RG /
poor-man's scaling for a single impurity needs to be stopped at
the RKKY scale. Moreover, the relative slope in the exponents
of the low-energy scale as seen in the semilog plots in
\Figs{fig:NRGspectra}(:,4) changes in the intermediate regime
with increasing impurity distance. The slope $1/b$ of the peak
vs. $\omega^\ast$ is larger than for \TK (brown line) in
\Fig{fig:NRGspectra}(a;3,4), about comparable in
\Fig{fig:NRGspectra}(e,4), and smaller in
\Fig{fig:NRGspectra}(f-h;4) where RKKY is absent. This clearly
underlines the continuous crossover from RKKY to Kondo.

A similar exponential fit on the maximum of the spectral data
was also obtained for the longitudinal spin correlations in
\Fig{fig:NRGspectra}(a5). The peak in the spectral data occurs
at a larger $a'=5.86{\cdot}10^{-4}$ at $J_z=0$ when compared to
the corresponding fit in \Fig{fig:NRGspectra}(a3). The slopes
$b$ are comparable, though ($b' = 0.80$ vs. $b=0.82$).
Therefore the maximum in the longitudical spectral data $\langle
\hat{S}^{\iL}_z \Vert \hat{S}^{\iR}_z \rangle_\omega$ is
systematically shifted at small $J_z$ by a factor of $5.86 /
2.78 = 2.11$ towards larger energies as compared to the maximum
in the spectral data that requires a spin-flip, $\langle
\hat{S}^{\iL}_+ \Vert \hat{S}^{\iR}_- \rangle_\omega$. Thus in
the RKKY regime at $x=10$ [\Figs{fig:NRGspectra}(a,:)],
\begin{eqnarray}
  \omega^\mathrm{max}_{S^{\iL}_z S^{\iR}_z}
  \simeq 2\
  (\omega^\mathrm{max}_{S^{\iL}_+ S^{\iR}_-}
   \simeq \ER)
\text{ ,}\label{eq:peak-position}
\end{eqnarray}
and consistent with the explicit analytical expression for \ER
in \Eq{H-RKKY} This is in agreement with the effective
two-impurity level-spectrum in \Fig{fig:Spectrum}, where with
reference to the Lehmann representation for spectral data in
\Eq{eq:Aspec}, one needs to pay an energy \ER for a spin-flip,
whereas one needs to pay an energy $2\ER$ for a sign-flip
(triplet $|t\rangle$ to singlet $|s\rangle$ transition).
The deviation of about 5\% from the expected factor of $2$ in
the fitted values is within the spectral resolution of NRG,
and thus likely due to z-averaging and broadening of the
spectral data.

The parameter scans in \Fig{fig:NRGspectra} give important hints:
they show that \ER (blue vertical marker in \Fig{fig:NRGspectra})
as obtained from a second order perturbative approach
[cf. \Eq{E-RKKY}], does not always describe the
RKKY low-energy scale correctly, as it can get renormalized
by the presence of single-impurity Kondo correlations.
While for $x=10$ this well
coincides with the $J_z=0$ low-energy scale [e.g., compare to
solid lines with symbols in \Fig{fig:NRGspectra}(a,4)], when
increasing the distance, the peak in the NRG data shifts towards
the left of the vertical blue marker, i.e., towards values that
are larger than \ER. That is, the fit value for $a$ shown with
\Fig{fig:NRGspectra}(a;3,5) effectively increases relative to
\ER when exponentially increasing the distance $x$ (no
additional fits shown, though, as this is a qualitative
argument). Thus even if at $J_z=0$ the one-impurity Kondo scale
may still be many orders of magnitude smaller than \ER, the
value of \ER already gets weakly affected (likely acquires
logarithmic corrections) due to the underlying Kondo
correlations, even if $\TK \ll \ER$. If at the same time it
also holds $\ER \ll D = 1 $, then integrating out the helical
edge mode starting from the initial band edge $D$ towards zero
energy, e.g., in a poor-man's scaling sense, this can be
expected to introduce an RG flow also for \ER. This suggests
that for a consistent interpretation of the NRG data with
analytics, the NRG data should be scaled by $T^{(\eta)}_{S_z}$
[cf. \Eq{eq:TS:gen}], rather than the lowest order analytical
estimate \ER in \Eq{E-RKKY}.

Moreover, the peaks seen in the spectral data are rather narrow,
i.e., within the resolution limit of the presented NRG data.
Based on the chosen NRG parameters $\Lambda=4$ with $n_z=4$
z-shifts (e.g., see \Sec{sec:NRGdisc}, or also \cite{Wb12_FDM}
for more detailed definitions), we expect a best possible
relative spectral resolution $\delta\omega/ \omega >
\Lambda^{1/2n_z} -1 = 0.19$, hence the value used for the
broadening of $\sigma=0.3$ on the discrete NRG raw data.
Based on perturbative approach, however, one may suspect
significantly narrower features as shown in \Fig{fig:NRGspectra}.
With this in mind, the spectral features seen in data in
\Fig{fig:NRGspectra} are likely overbroadened.

\begin{figure}[!]
\begin{center}
\includegraphics[width=1\linewidth]{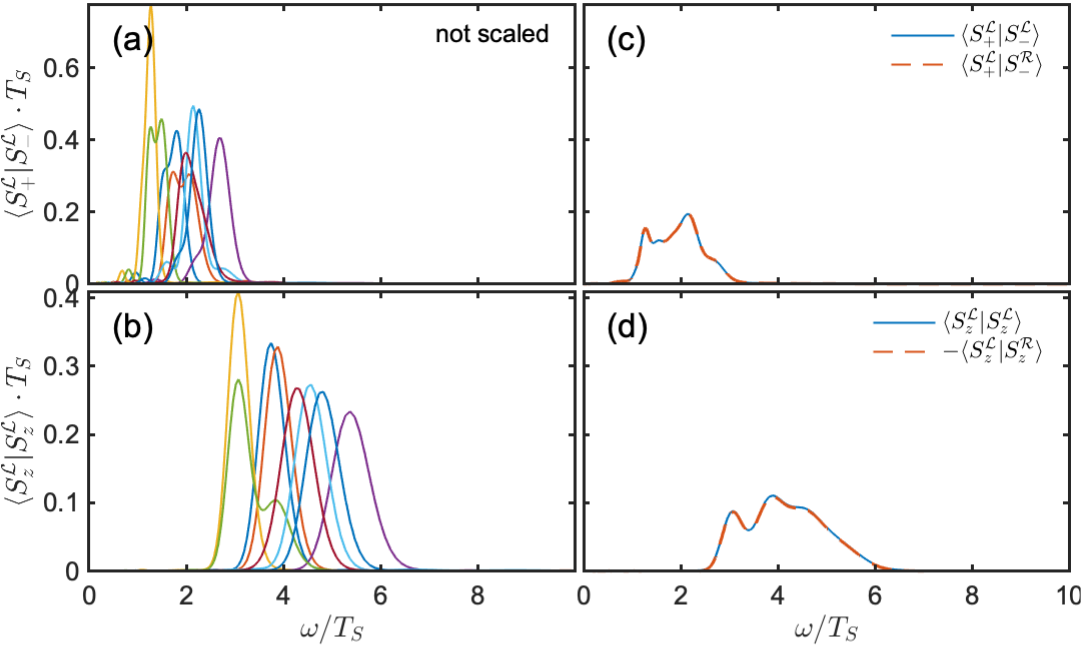}
\end{center}
\caption{
   NRG spectral data of the 2HKM for $J=0.1$, $J_z=0$, $x=10$
   for the dynamical correlations functions as partly indicated
   with the left axis [transverse in the upper panels,
   longitudinal in the lower panels; intra-impurity for the
   left panels, whereas in the right panels also includes the
   inter-impurity correlation for comparison; note the sign in
   the legend with panel (d)] -- Left panels: individual
   spectral data for z-shifted logarithmic discretization
   ($n_z=8$ curves with $z\in [0,1[$, having $\Lambda=4$,
   broadening $\sigma=0.1$), and the corresponding z-averaged
   data in the right panel. The horizontal and vertical axis
   are globally scaled by the z-averaged low-energy scale
   $T_{S_z} = 1.4{\cdot}10^{-4} = 0.556\, \ER$ only, so
   with spectral sum rules in mind, the
   data shown is of order one.
}
\label{fig:NRG-zspec00}
%\end{figure}
%
%\begin{figure}[!]
\begin{center}
\includegraphics[width=1\linewidth]{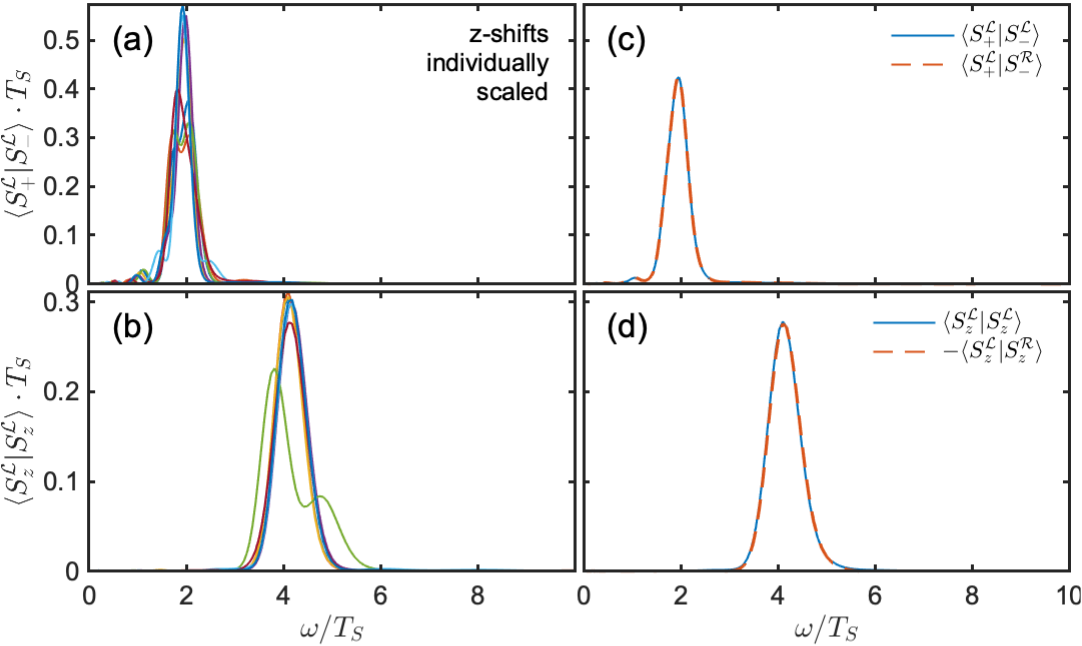}
\end{center}
\caption{
  Exactly the same bare data as in \Fig{fig:NRG-zspec00}, except
  that for each z-shift the frequency scale is first scaled by
  $T^{\eta\eta'}_{S_\alpha} / T^{\eta\eta'}_{S_\alpha}(z)$ (with
  $T^{\eta\eta'}_{S_\alpha} \equiv \langle
  T^{\eta\eta'}_{S_\alpha}(z)\rangle_z$ the z-averaged value,
  and $\alpha \in \{z,\pm\}$ chosen, respectively, for each
  curve), and then combined as in \Fig{fig:NRG-zspec00}.
  The z-shift specific $T_{S_z}(z)$ is also determined within
  the NRG in the same calculation as part of the post analysis
  [cf. \Eq{eq:TS:gen}]. Overall, this procedure leads to a
  significantly improved quality of peak shape that is
  significantly narrower, as compared to the spurious spread of
  spectral peaks in the right panels in \Fig{fig:NRG-zspec00}.
}
\label{fig:NRG-zspec00s}
\end{figure}

\begin{figure}[!]
\begin{center}
\includegraphics[width=1\linewidth]{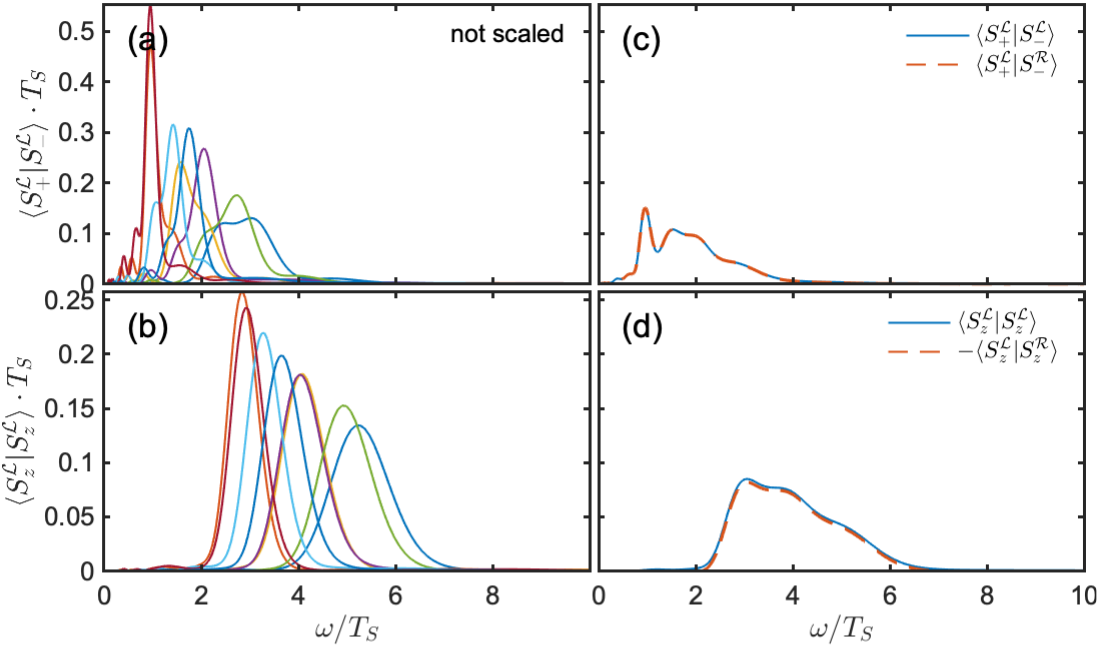}
\end{center}
\caption{
   Same analysis as in \Fig{fig:NRG-zspec03} except for an
   increased impurity distance $x=1\,000$, leading to the
   reduced value for $T_{S_z}= 3.0{\cdot}10^{-6} = 1.19\, \ER$.
   The broadening was also increase to $\sigma=0.15$.
}
\label{fig:NRG-zspec03}
%\end{figure}
%
%\begin{figure}[!]
\begin{center}
\includegraphics[width=1\linewidth]{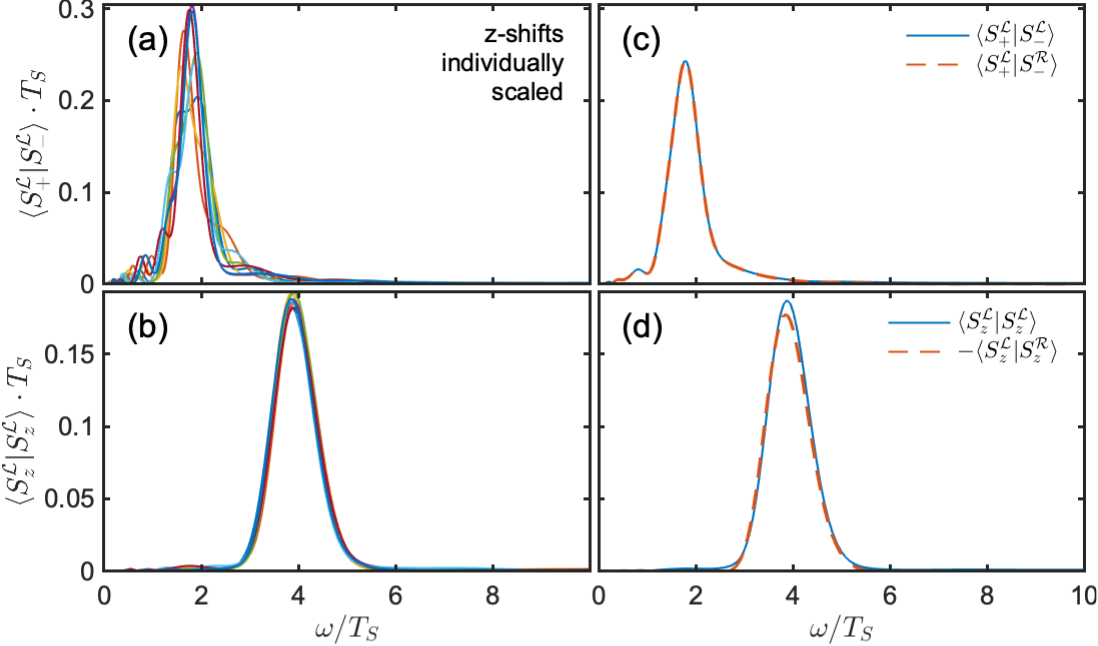}
\end{center}
\caption{
  Exactly the same bare data as in \Fig{fig:NRG-zspec03}, except
  that for each z-shift the frequency scale is first scaled by
  $T^{\eta\eta'}_{S_\alpha} / T^{\eta\eta'}_{S_\alpha}(z)$
  (similar to \Fig{fig:NRG-zspec00s}).
}
\label{fig:NRG-zspec03s}
\end{figure}

\subsection{More detailed spectral analysis (line shapes)}
\label{sec:lineshapes}

The RKKY `resonances' in the spectral data are expected to be
potentially very narrow. As it turns out, though, while sharp
peaks (near delta spikes) can be found in the bare discrete
spectral NRG data, the precise location in frequency of these
peaks is sensitive on the $z$-shift in the logarithmic
discretization of the NRG setup (e.g., see \Sec{sec:NRGdisc} or
also \cite{Wb12_FDM} for detailed definitions). Specifically, as
the z-shifts can shift energies by a factor of $\sqrt{\Lambda}$
in the discrete setup, in the present case, also the respective
`response' of the system in terms of the precise location of a
narrow spectral resonance for the 2HKM model can also vary
within a factor of $\sqrt{\Lambda}$. Therefore blind z-averaging
of the NRG data leads to artificial broadening and somewhat
irregular z-averaged data, as demonstrated in
\Fig{fig:NRG-zspec00} or \Fig{fig:NRG-zspec03} for two different
impurity distances, $x=10$ and $x=1\,000$, respectively.
Instead, when scaling the spectral data for each individual
z-shift by its respective $T_{S_\alpha}(z)$ and averaging that
resulting data, peak shapes are significantly improved, and in
particular also narrower (e.g., compare
\mbox{\Fig{fig:NRG-zspec00} $\to$ \ref{fig:NRG-zspec00s}}, or
\mbox{\Fig{fig:NRG-zspec03} $\to$ \ref{fig:NRG-zspec03s}}).

For the analysis here, the rather large $\Lambda=4$
is useful to emphasize how discretization manifests itself in
the spectral data which is more subtle here, as it also leads to
shifts. A major motivation for the larger $\Lambda=4$ is,
of course, that this also results in faster NRG calculation
or, conversely, in better converged NRG spectral data.
By careful $z$-averaging, we can obtain good spectral resolution
for $\Lambda=4$, nevertheless.  However, given that RKKY
peaks can be expected to become much narrower in terms of width
vs. location, this ultimately will be also challenging for
smaller $\Lambda$ in any case. Overall, the present analysis in
terms of  $\Lambda=4$ already clearly supports narrow features.
More importantly, the location of the peaks, and hence the
corresponding energy scale, can be considered reliable and
significantly more accurate than the width, given also that
these are directly related to inverse static susceptibilities.

The data in the right panels of \Fig{fig:NRG-zspec00s} gives a
good impression for the spectral data in the RKKY regime. As
compared to \Eq{eq:TS} with $\TS = 0.50\, \ER$, the actual data
in \Figs{fig:NRG-zspec00} \& \ref{fig:NRG-zspec00s} gives $\TS =
0.556\, \ER$ which is reasonably close. Furthermore, the peak
location in the spectral data is expected at $\omega^\ast = \ER
\simeq 2 \TS$ for the transverse spectral data, and at
$\omega^\ast = 2\ER \simeq 4 \TS$ in the longitudinal data,
again consistent with the data in \Fig{fig:NRG-zspec00s}. The
peak width in the longitudinal data related to a phase flip when
transitioning to the singlet state [cf. \Fig{fig:Spectrum}] is
found to be comparable as in the transverse data
\cite{VIYu_2022}. However, this width is limited by the
broadening $\sigma$ as seen with the individual data in the left
panels. While some structure can be observed with
\Fig{fig:NRG-zspec00s}(a) resolved by the spread with
$z$-shifts, the data in \Fig{fig:NRG-zspec00s}(b) is more smooth
this way, just showing the broadening $\sigma=0.10$ used. This
suggests that the data in (b) is still likely overbroadened by
given $\sigma=0.10$.
Finally, as already argued with \Fig{fig:NRGspectra}, the inter-
and intra-impurity correlations are identical to each other deep
in the RKKY regime. Here this is seen by having the dashed
lines in the right panels of \Fig{fig:NRG-zspec00s} on top of
the solid lines [note the sign change, though, as indicated with
legend in \Fig{fig:NRG-zspec00s}(d)], which again is rooted in
the triplet ground state in \Eq{t-state}.

\begin{figure}[tb!]
\begin{center}
\includegraphics[width=1\linewidth]{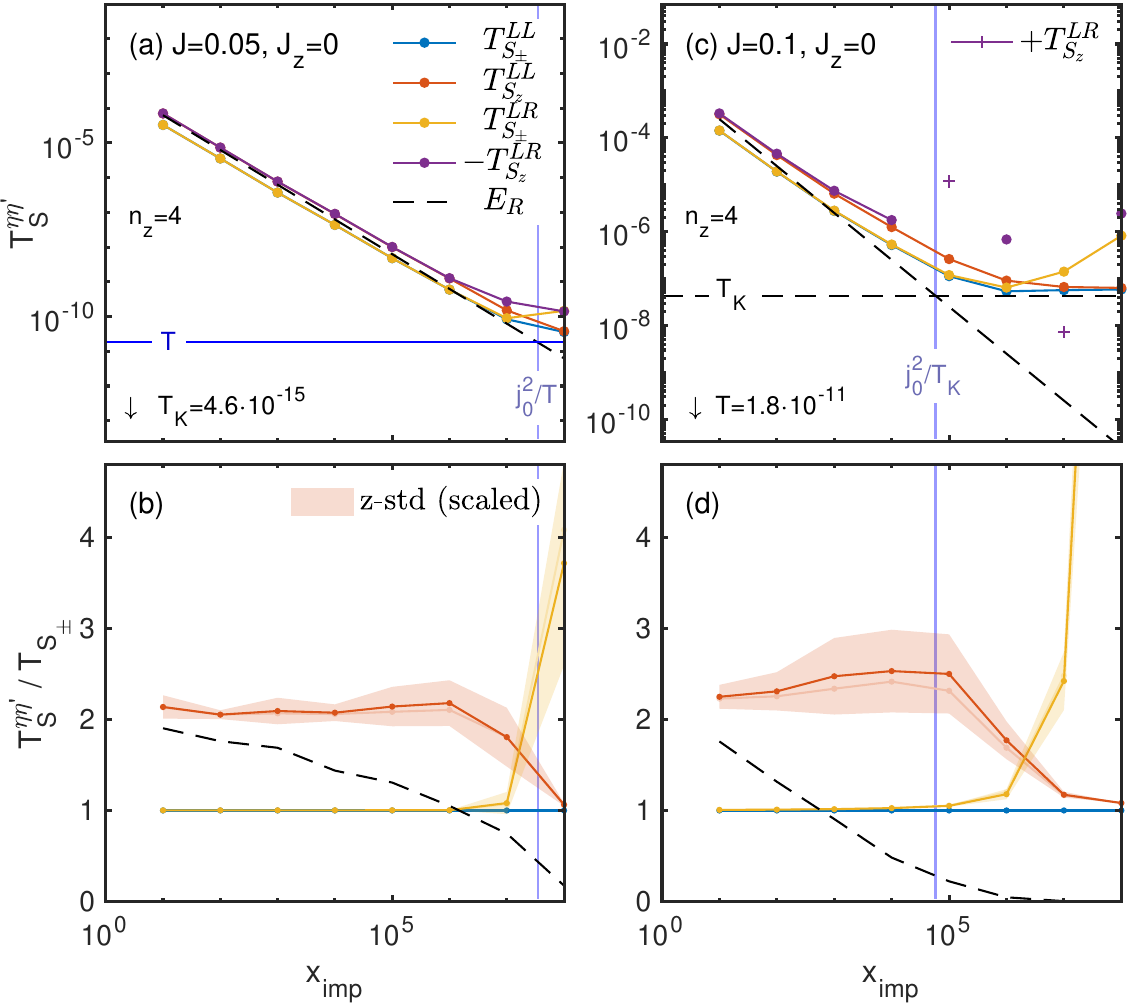}
\end{center}
\caption{
  Low-energy scales vs. distance from inverse static
  susceptibilities [cf. \Eq{eq:TS:gen}] as indicated in the
  legend of panel (a) for $J_z=0$, throughout. Left panels
  ($J{=}0.05$) [right panels ($J{=}0.10$)] have temperature
  $T<10^{-10}$ [the Kondo temperature $\TK \sim
  4{\cdot}10^{-8}$] as low-energy cutoff, respectively. The blue
  vertical markers translate the low-energy scale to distance,
  i.e., $x_0 \equiv j_0^2/\mathrm{max}(T,\TK)$ which also
  corresponds to the crossing point of \ER (black dashed) with
  $\mathrm{max}(T,\TK)$ (horizontal marker). The lower panels
  redraw the data in the upper panels, but vertically scaled by
  \TS. The shading indicates the standard deviation of the color
  matched energy scale due to $z$ shifts in the NRG
  discretization.
}
\label{fig:Tscale}
\end{figure}

Repeating the same analysis for the increased impurity distance
$x = 1\to1\,000$, the data in \Fig{fig:NRG-zspec03s} starts to
show a qualitative change in the spectral line shape. While
having increased the broadening to $\sigma=0.15$, as reflected
by the individual peaks in the left panels, the overall
lineshape in \Fig{fig:NRG-zspec03s}(d) is still largely
comparable. The inter- and intra-impurity correlations do lie
nearly exactly on top of each other, thus also suggesting a
clear RKKY low-energy state still. In particular, peak position
are still at the expected $\omega^\ast = 2 T_{S_z}$ or $4
T_{S_z}$ for the transverse or longitudinal data. The numerical
value for $T_{S_z}$, however, further deviates from the plain
second-order perturbative RKKY scale, having $T_{S_z}=
3.0{\cdot}10^{-6} = 1.19\, \ER$ which by now is clearly unequal
from the naive expected value of $0.5\,\ER$ (see also
\Fig{fig:Tscale} for a more detailed analysis in this regard).
Yet when scaling the data consistently fully within the NRG
framework, the schematic picture in \Fig{fig:Spectrum} still
works well when substituting $\ER \to 2\,T_{S}^\mathrm{NRG}$.

To conclude this section, we reemphasize that the NRG approach
correctly describes the position of the peaks in the response
functions as well as the integrated spectral weight, but yields
resolution-limited information about the peak width and thus its
overall shape. The latter unavoidably results from the
coarse-graining in energy space intrinsic to the NRG approach.

\subsection{RKKY vs. Kondo Energy scales}

An analysis of the low-energy scales vs. impurity distance is
shown in \Fig{fig:Tscale}. The longitudinal inter-impurity
susceptibilities are negative due to the antiferromagnetic RKKY
correlations; note the sign with $T_{S_z}^{\iL\iR}$ in the
legend. In the crossover region when leaving the RKKY regime,
this can become positive, though [cf. legend with
\Fig{fig:Tscale}(c)]. As seen in all panels, the energy scale
\TS `drifts' away from the bare RKKY scale \ER (black dashed
line) already many orders above the Kondo scale \TK (indicated
or specified with the upper panels). If the distance exceeds
the temperature length scale $v/T$ (left panels) or the inverse
Kondo scale (right panels), the impurities effectively become
independent of each other, such that RKKY is cutoff by
$\mathrm{max}(T,\TK)$. In \Fig{fig:Tscale}, the end of the RKKY
regime is seen where inter-impurity correlations start to
deviate significantly from their intra-impurity counter part
(e.g., compare blue vs. yellow in the lower panels).
By approaching the wide-band limit (or equivalently, by reducing
$J$, as in right vs. left panels in \Fig{fig:Tscale}), one can
observe in the lower panels that $T_{S_z} \simeq 2 (T_{S_\pm}
\equiv \TS)$ is obeyed over a wide distance (energy) window.
Also deviations from the relative factor of 2 diminish towards
the wide-band limit. The shading in the lower panels of
\Fig{fig:Tscale} indicates variations (i.e., the standard
deviation) due to by $z$-shifts in the logarithmic
discretization. These also become smaller towards the wide-band
limit [e.g., compare \Fig{fig:Tscale}(d $\to$ b)].

\section{Summary and outlook}

We presented a broad parameter scan of the two-impurity helical
Kondo model (2HKM) vs. impurity distance, coupling anisotropy,
temperature, and finite bandwidth within the NRG framework. We
emphasize that this setup substantially differs from a {\it
chiral} edge mode, where both spins propagate in the same
direction [for a recent NRG study on the latter, see e.g.,
\citet{Lotem22}; the chiral model is different from the helical
edge mode discussed here, both in the setup as well as in the
physics]. With the NRG being non-perturbative in character, our
presented results are reliable quantitatively in the full
parameter regime, with the only real constraint being energy
resolution in spectral data. We have established a plain
crossover from RKKY to Kondo with increasing impurity distance
or, conversely, increasing Kondo coupling or its anisotropy. The
Kondo screening of the impurities individually by the helical
edge mode is tuned continuously into an effective mutual `RKKY'
screening of the impurities themselves. In this sense, the
Kondo renormalization flow is cutoff by the RKKY energy \ER once
it exceeds the Kondo scale. The RKKY occurs in the energy
window determined by the inverse time scale required for the
impurity to travel in between the impurities (Thouless energy
\Ex), and the actual RRKY energy $\ER \ll \Ex$. If the Kondo
scale falls within this window one comes across a
continuous crossover.

The low-energy effective RKKY Hamiltonian gives rise to narrow
resonances with the helical edge. While the presented NRG
analysis, the location and overall spectral weight is reliable,
the spectral width, however, is likely (much) below the energy
resolution of our NRG data deep within the RKKY far from any
Kondo screening. Hence the precise linewidth e.g., in the
dynamical spin-spin correlation data for the fully interacting
model is left for future analytical and numerical studies.

\begin{acknowledgments}

We gratefully acknowledge collaboration with our long-standing
and esteemed colleague Vladimir I. Yudson. We further
acknowledge discussions with Matan Lotem und Moshe Goldstein
(University of Tel Aviv), Jan von Delft (LMU Munich), Seung-Sup
Lee (Seoul National University). O.\,M.\,Ye. acknowledges support
from the DFG through the grant YE 157/3. A.W. was supported by
the U.S. Department of Energy, Office of Science, Basic Energy
Sciences, Materials Sciences and Engineering Division.

\end{acknowledgments}

\ \\[1ex]
{\it Note added} -- At the time of submission
we learned about the work by Lotem et al.,
Ref.\,\cite{Lotem23}, which addresses the NRG approach for
the chiral systems as a follow-up to Ref.\,\cite{Lotem22}.

%%%%%%%%%%%%%%%%%%%%%%%%%%%%%%%%%%%%%%%%%%%%%%%%%%%%%%%%%%%%%%%%%%%%%%%%
\appendix
%%%%%%%%%%%%%%%%%%%%%%%%%%%%%%%%%%%%%%%%%%%%%%%%%%%%%%%%%%%%%%%%%%%%%%%%

\section{NRG setup for helical bath mode \label{sec:NRG:setup}}

The NRG coarse-grains in energy space from the point of view
of the impurities. As such, akin to the analytical treatment,
the NRG approach is also perfectly well-suited to directly
tackle a single helical edge mode in energy-momentum space.
The helical energy dispersion was introduced in \Sec{sec:H0}
as $\varepsilon_{k\sigma} = \sigma v k \equiv \sigma
\varepsilon_{k}$ with $k \in \tfrac{1}{v}[-D,D] \equiv
[\tfrac{-\pi}{a},\tfrac{\pi}{a}]$.
The helical modes propagate in opposite velocities for
different~$\etaB$.  With focus on the low-energy regime,
they are assumed to have simple linear dispersion
$|\tfrac{d}{dk}\varepsilon_{k\sigma}|=v$ over the entire
bandwidth. By working in momentum space, translational
invariance is implied, such that the one-particle bath
states are described by simple plain waves.
Therefore assuming for simplicity $N$ spin-dependent
one-particle modes $k$ over the entire bandwidth,
$\hat{c}_{k\sigma}(x) = \tfrac{1}{\sqrt{N}} e^{\iu k x} \,
\hat{c}_{k\sigma}$ such that for the one-particle state
\begin{eqnarray}
   \langle 0|\hat{c}^{\phantom{\dagger}}_{k}(x) \,
   \hat{c}^{\dagger}_{k}(0) |0\rangle
   \equiv \langle x| k\rangle
 = \tfrac{1}{\sqrt{N}} e^{\iu k x}
\label{eq:helical:xcor}
\end{eqnarray}
acquires a positive phase factor according to standard
conventions, when it travels from the left to the right
($k,x>0$).

\subsection{Hybridization function
\label{seq:hybridiziation}}

The helical 1D edge mode above is assumed to constitute a
fermionic macroscopic bath that  interacts equally with two
impurities $\hat{d}_{\etaI\sigma}$ symmetrically located at
positions $x_\etaI = \tfrac{\etaI }{2}\, x$ with $\etaI=\pm 1$
[\Eq{def:etaI}]. The NRG approach first focuses
on Anderson type impurities with explicit hybridization of
the impurities with the bath. The switch to Kondo-type
impurities can be taken as a subsequent step, e.g.,
via Schrieffer-Wolff transformation.
Thus the starting point is the hybridization Hamiltonian,
\begin{eqnarray}
   \hat{H}_\mathrm{hyb}
 &=& \sum_{\etaI\sigma k} \Bigl(
   \tfrac{V_k}{\sqrt{N}} \hat{d}^\dagger_{\etaI\sigma} \hat{c}^{\,}_{k\sigma}
   e^{\iu k x \frac{\etaI}{2}} + \mathrm{H.c.} \Bigr)
\notag \\ &\equiv&
   \sqrt{\tfrac{2D \Gamma}{\pi}} \sum_{\etaI\sigma}\Bigl(
      \hat{d}^\dagger_{\etaI\sigma}
      \tilde{f}^{\,}_{0\etaI\sigma}
    + \mathrm{H.c.} \Bigr)
\text{ .}\label{eq:H:hyb}
\end{eqnarray}
This defines the bath states $\tilde{f}^{\,}_{0\etaI\sigma}
{\equiv} \sqrt{\tfrac{\pi}{2 D \Gamma}}
\sum_k \tfrac{V_k}{\sqrt{N}} e^{\iu k x\frac{\etaI}{2}}
\hat{c}_{k\sigma}$ that the impurities couple to, which permits
the interpretation that these are the bath states at the
respective location of the impurities.  The states
$\tilde{f}^{\,}_{0\etaI\sigma}$ are normalized, but not
necessarily orthogonal yet, hence the tilde (cf.
\Sec{sec:ftilde}).

In \Eq{eq:H:hyb}, an electron with spin $\sigma$ can hop on and
off the corresponding helical branch, with the spin $\sigma$
preserved in the process. The hopping amplitude $V_k$ is assumed
independent and thus symmetric in the spin and impurity indices.
For simplicity, the bath is also assumed featureless,
characterized by two parameters only: a hybridization strength
$\Gamma$ and a finite half-bandwidth $D$. In the continuum
limit, the hybridization function for each impurity individually
is given by $\Gamma(\varepsilon) = \pi \rho_\varepsilon
|V_\varepsilon|^2 \equiv \Gamma \, \vartheta(D-|\varepsilon|)$
(aka. the default NRG box distribution), with $\rho_\varepsilon
= \rho_0\, \vartheta(D-|\varepsilon|)$ the one-particle density
of states, assuming constant $\rho_0 =\tfrac{1}{2D}$ without
restricting case.  The hybridization is cut off sharply in
energy, as depicted in \Fig{fig:helical}.
The integrated (norm-squared) hybridization strength is $\sum_k
|V_k|^2 \cong \int \rho_\varepsilon |V_\varepsilon|^2 = 2D\cdot\Gamma/\pi$,
which yields the split-off normalization factor in the second
line in \Eq{eq:H:hyb}.

From the perspective of the impurities, the full hybridization
function becomes a $2 \times 2$ matrix
\begin{eqnarray}
  \Delta_{\etaI\etaI'}^{[\etaB]}(\omega)
  &\equiv& \tfrac{2D\Gamma}{\pi}
  \langle \tilde{f}_{0\etaI\etaB}
    \Vert \tilde{f}_{0\etaI'\etaB}^\dagger \rangle_\omega
\notag \\
&=& \tfrac{1}{N} \sum_k \tfrac{V_{k}^{\phantom{\ast}} V_{k}^\ast}{
  \omega^+ - \varepsilon_k} \
  e^{\iu \etaB \frac{\varepsilon_k}{v} \frac{\etaI-\etaI'}{2} x}
\label{eq:Delta:eta'}
\end{eqnarray}
with $\etaI \in \{\iR,\iL \}$ indexing the impurities [\Eq{def:etaI}].
This includes a non-zero off-diagonal hybridization for
$\etaI\neq \etaI'$.  In the wide-band limit $D\to \infty$,
the hybridization becomes
\begin{subequations}
\begin{eqnarray}
  \boldsymbol{\Delta}(\omega) &\equiv& -\iu \Gamma \begin{pmatrix}
      1 & 2 e^{\iu \omega \tau} \,\vartheta(\tau) \\
      2 e^{-\iu \omega\tau}\, \vartheta(-\tau) & 1
  \end{pmatrix}
\label{eq:delta:tau}
\end{eqnarray}
where $\tau \equiv \tfrac{\etaB x}{v}$ is the time required
for a particle to travel from one impurity to the other,
since for example from \Eq{eq:Delta:eta'} for
a particle to travel from $\iL \to \iR$ (thus
with $x\neq0$, also $\tau \neq 0$),
\begin{eqnarray}
 \Delta_{12}(\omega) &\equiv& \Delta_{+-}(\omega)
 \equiv \tfrac{\Gamma}{\pi} \!\!
   \int\limits_{-D}^D d\varepsilon\,
   \tfrac{e^{+\iu \varepsilon \tau }}
   {\omega^+ - \varepsilon}
\label{eq:Delta12:tau} \\
 && \overset{D\to\infty}{\longrightarrow}
  -2\iu \Gamma \, e^{\iu \omega \tau }
  \vartheta(\tau)
\text{ .} \notag
\end{eqnarray}
\end{subequations}
This gives a non-zero contribution only for $\tau>0$. Assuming
$v, x > 0$, for example, this requires $\etaB>0$. Indeed, in
given helical setting, a spin up electron travels to the right,
that is the particle needs to be created at the left impurity
($\etaI'=-1$) which then can propagate to the right impurity
($\etaI=+1$). For spin down, the situation is vice versa.

We emphasize that the spectral representation $\Gamma(\omega)$
of the hybridization function cannot be simply written as
$-\mathrm{Im}\Delta(\omega)$ in the present case because the
matrix elements in \Eq{eq:Delta:eta'} are complex. Instead,
the imaginary part needs to be taken from the propagator only,
i.e., with $-\im \tfrac{1}{\omega^+ - \varepsilon}
= \pi \delta(\omega - \varepsilon)$
[see \Sec{sec:complex-spectral} for more details],
\begin{eqnarray}
  &&\Gamma_{\etaI\etaI'}^{[\etaB]}(\omega)
  \equiv
  \tfrac{1}{N} \sum_k V_{k}^{\phantom{\ast}} V_{k}^\ast
  \, \pi\delta(\omega - \varepsilon_k) \
  e^{\iu \frac{\etaI-\etaI'}{2} \tau \varepsilon}
\label{eq:Gamma:eta'} \\[-1ex]
  &&\cong
   \Gamma \int\limits_{-D}^D d\varepsilon\,
   \delta(\omega - \varepsilon) \
  e^{\iu \frac{\etaI-\etaI'}{2} \tau \varepsilon}
= \Gamma \vartheta( D - |\omega|) \,
  e^{\iu \frac{\etaI-\etaI'}{2} \tau \omega}
\text{ ,} \notag
\end{eqnarray}
where the spectral function $\Gamma_{\etaI\etaI'}(\omega)
\equiv [\boldsymbol{\Gamma}(\omega)]_{\etaI\etaI'}$
needs to be differentiated from the constant $\Gamma$.
Correspondingly, in matrix notation,
\begin{eqnarray}
  \boldsymbol{\Gamma}(\omega) = \Gamma \,
  \vartheta( D - |\omega|) \
  \begin{pmatrix}
      1 & e^{\iu \omega\tau} \\
      e^{-\iu \omega\tau} & 1
  \end{pmatrix}
\text{ ,}\label{eq:Gamma:hel}
\end{eqnarray}
which is non-zero now for both off-diagonal entries
$\etaI \neq \etaI'$ for either spin $\sigma$.
Only by taking the spectral data as above, this can be
simply completed to the full hybridization function using
Kramers-Kronig transform on the complex spectral data,
i.e., by folding the above possibly complex spectral
function with $1/(\omega^+ - \varepsilon)$.

Within the wide-band limit and in the absence of interactions,
the Green's function for the impurities with onsite energies
$\varepsilon_{d\eta}$ becomes, e.g., in the
spin-up channel ($\etaB=+1$, and hence $\tau>0$),
\begin{eqnarray}
  {\bf G}_\uparrow(\omega)
  &=& \bigl[ \omega^+ - \varepsilon_d - \boldsymbol{\Delta} \bigr]^{-1}
\label{eq:Gimp:tau} \\
  &\overset{\eqref{eq:delta:tau}}{=}&
  \begin{pmatrix}
      \omega - \varepsilon_{d+} + \iu\Gamma &
      2\iu\Gamma e^{\iu \omega \tau}  \\
      0 & \omega^+ - \varepsilon_{d-} +\iu\Gamma
  \end{pmatrix}^{-1}
\notag \\
 && \hspace{-.66in}
 = \tfrac{1}{
      (\omega - \varepsilon_{d+} + \iu\Gamma)
      (\omega - \varepsilon_{d-} + \iu\Gamma)}
  \begin{pmatrix}
      \omega^+ - \varepsilon_{d-}  + \iu\Gamma &
     -2 \iu\Gamma e^{\iu \omega \tau}  \\
      0 & \hspace{-.15in}\omega^+ - \varepsilon_{d+}  + \iu\Gamma
  \end{pmatrix}
\,\text{.}\notag
\end{eqnarray}
At particle hole symmetry with $\varepsilon_{d\etaI}=0$
on obtains for $\omega=0$,
\begin{eqnarray}
   -\Gamma \im {\bf G}_\uparrow(0)
  = \begin{pmatrix}
     1 & -2  \\
     0 & \phantom{+} 1
  \end{pmatrix}
= \pi\Gamma\, {\bf A}_\uparrow(0)
\text{ ,}
\end{eqnarray}
where the diagonal entries reflect half-filling based on the
Friedel sum rule. Due to the helicity, the diagonal terms in
\Eq{eq:Gimp:tau} are just the Green's function of decoupled
impurities, $G_{\etaI\etaI}(\omega)
= \tfrac{1}{\omega - \varepsilon_{d\etaI} + \iu \Gamma}$.
The non-zero off-diagonal term maintains the same matrix
position as in $\boldsymbol{\Delta}(\omega)$ in \Eq{eq:delta:tau},
consistent with the directedness of propagation. Similar to
\Eq{eq:delta:tau}, the off-diagonal term also shows oscillatory
behavior in energy with period
\begin{eqnarray}
    \delta\omega = \tfrac{2\pi}{|\tau|}
  = \tfrac{2\pi v}{x} \equiv 2\pi \Ex
\text{ .}\label{eq:domega}
\end{eqnarray}
This period is fixed by the energy scale $\Ex = v/x$
[\Eq{eq:Ex}] set by the inverse time $\tau$ required for an
electron to travel from one impurity to the other. In
particular, the period $\delta\omega$ is independent of the
energy scale $\omega$ in $G(\omega)$. Given that the NRG
discretizes logarithmically in energy space, assuming $\Ex \ll D
= 1$, there is no way these oscillations can be resolved to
orders of magnitude higher in energy all the way up to $D$.
On the other hand, on physical grounds at high energies,
e.g., at temperatures $T>\Ex$, the impurities effectively
no longer see each other. In this sense it appears reasonable to
expect for most physical quantities such as spin-spin
interactions that do not explicitly resolve one-particle phases
of propagation. In the presence of relaxation processes due to
interactions, these rapid oscillations likely average out at
large frequencies even for small temperature, and thus become
less important in detail. Yet this needs to be verified in
practice on a case-by-case basis by tracking the stability of
the data with respect to the level of course graining.

\subsection{Normalization of bath states and effects of
finite bandwidth \label{sec:ftilde}}

For either spin, the hybridization term \Eq{eq:H:hyb} in the
Hamiltonian defines two bath states
$\tilde{f}^{\phantom{\dagger}}_{0\etaI\sigma}$ with
$\etaI \in \{\iR,\iL\}$.  These are normalized, but not strictly
orthogonal, hence the tildes with the $f$ operators in
\Eq{eq:H:hyb} as a reminder. These bath states at the location
of the two impurities have finite overlap that can be determined
from the fermionic anticommutator relation,
\begin{subequations} \label{eq:f0:S}
\begin{eqnarray}
 S^{[\sigma]}_{\etaI \etaI'} &\equiv&
   \langle \tilde{0}_\etaI | \tilde{0}_{\etaI'} \rangle
\label{eq:f0:S:1} \\
 & \equiv &
   \langle 0| \tilde{f}_{0\etaI\sigma}^{\,}
   \tilde{f}_{0\etaI'\sigma}^\dagger | 0\rangle
 =\{ \tilde{f}^{\,}_{0\etaI\sigma},
      \tilde{f}^{\dagger}_{0\etaI'\sigma} \}
\notag \\
 &=& \tfrac{1}{2D} \int\limits_{-D}^D d\varepsilon\,
   e^{\iu  \sigma \frac{\etaI-\etaI'}{2v} \varepsilon x}
 = \left\{
     \begin{array}[c]{ll}
       1 & \etaI=\etaI' \\
       \tfrac{\sin(\pi x/a)}{\pi x/a} & \etaI \neq \etaI'
     \end{array}
  \right.
\text{ ,}\notag
\end{eqnarray}
with $| 0\rangle$ the vacuum state, and
$a\equiv \tfrac{\pi v}{D}$ as in \Eq{eq:lattice:a},
and therefore $\tfrac{\pi x}{a} = \tfrac{D x}{v} = |\tau D|$.
Rewritten in matrix notation
\begin{eqnarray}
   {\bf S}_0 = 1 + \underbrace{
     \tfrac{\sin(\pi x/a)}{\pi x/a}}_{
       \equiv r_0 % \sinc(\tfrac{\pi x}{a})
   }\tau_x
\text{ ,}\label{eq:helical:a}
\end{eqnarray}
it can be diagonalized, with eigenvalue matrix
\begin{eqnarray}
   s_0 \equiv 1 + r_0 \tau_z
\text{ ,}\label{eq:helical:r0}
\end{eqnarray}
\end{subequations}
with eigenvalues sorted as $s_{0\etaI} = 1 + \etaI r_0$, with
$\etaI \in \{+1,-1\}$, i.e., with the larger eigenvalue $s_{0+}
\geq 1$ coming first. The off-diagonal term in \Eq{eq:helical:a},
$r_0 \equiv a\, \delta_a(x)$, represents a $\delta$-function of
width $a$.  In this sense, the UV cut-off $D$ again directly
gives rise to the lattice constant already introduced with
\Eq{eq:lattice:a}. It has unit of length and resembles the
resolution in real space based on the given one-particle density
of states encoded into the box distribution. By having a finite
bandwidth, this translates into a cutoff in spatial resolution,
which in the present case naturally gives rise to a well-defined
lattice spacing: the overlap in \Eq{eq:f0:S} between the two
bath states becomes exactly zero for
\begin{eqnarray}
  x \in \{ x_n \equiv n a \  |\  n \in \mathbb{Z} \}
\label{eq:helical:xgrid}
\end{eqnarray}
when $n\neq 0$ which naturally suggests a discrete grid with
lattice spacing $a$. As pointed out with \Eq{eq:lattice:a},
choosing $D=1$ and $a=1$ then fixes the velocity to $v=1/\pi$.
For $n=0$, i.e., $x\to 0$, the off-diagonal overlap in
\Eq{eq:f0:S:1} becomes $1$, and thus identical to the diagonal
case $\etaI=\etaI'$.  In this case the two locations $\etaI$ and
$\etaI'$ approach the same `site' and thus become identical.

As an aside, we note that the argument here namely that a finite
bandwidth naturally gives rise to a discrete lattice spacing,
can also be straightforwardly carried over to a plain spinless
tight binding chain, with the minor difference that there due to
the structured density of states, the Fermi velocity $v$ differs
from the value above by a factor of order $1$.  Furthermore, the
Fourier transform of any quadratic Hamiltonian in momentum space
yields its full (long-range) hopping structure in real space.
Though a purely 1D lattice model
is unable to describe the topologically nontrivial phase of the
2D topological insulator, the one-particle dispersion
may nevertheless also be Fourier transformed for the isolated
single helical edge as in \Eq{eq:H0:k}. By
starting from momentum space, however, in the
present case this would mandate periodic (or infinite) BCs with
spin-dependent complex hopping coefficients
$t_{n\sigma} \equiv \sigma t_n$, where
with $\varepsilon_{k\sigma} \equiv \sigma\varepsilon_k$,
\begin{eqnarray}
   t_n &=& \tfrac{1}{N} \sum_k \varepsilon_k e^{\iu k x_n}
   \ \overset{N\to\infty}{\simeq}\
   a\!\!\int\limits_{-\pi/a}^{\pi/a} \tfrac{dk}{2\pi}\, v k\,
     e^{\iu k x}
\label{eq:tb:helical} \\[-2ex]
  &=& -\tfrac{\iu a v}{2\pi} \tfrac{d}{dx}
   \!\!\int\limits_{-\pi/a}^{\pi/a} dk\, e^{\iu k x}
   = -\tfrac{\iu a v}{\pi} \tfrac{d}{dx}
    \tfrac{\sin{\pi x/a}}{x}
 = - \iu v \tfrac{(-1)^n}{x_n}
\notag
\end{eqnarray}
for a chain of $N$ sites, with distance $x=x_n = n a>0$, having
$t_0=0$.  These hopping amplitudes are long range, decaying like
$1/x$, 
and by construction break inversion symmetry, $t_n = t_{-n}^\ast
\neq t_{-n}$. The long-range hoppings {\it cannot} be
eliminated by permitting deviations from the linear dispersion
close to the band edge.  For example, the altered dispersion
$\tilde{\varepsilon}_{k} = \frac{2v}{a} \sin \frac{ka}{2}$ which
still obeys $\tilde{\varepsilon}_{k} \simeq \varepsilon_{k} =
vk$ for small $k$, also results in long-range hopping. This is
due to the fact that the one-particle dispersion is
discontinuous across the boundary of the Brillouin zone.
Given this complications for numerical lattice simulations,
the starting point in energy-momentum space in \Eq{eq:H0:k} is
more natural and convenient.  Nevertheless, from the above it is
clear that the notion of lattice spacing and Brillouin zone are
perfectly valid also for a single helical edge mode even if the
one-particle dispersion is discontinuous across the boundary of
the Brillouin zone.

The analytic structure of the overlap in \Eq{eq:f0:S} is closely
related to static fermionic correlations versus distance.
For example, consider a filled helical Fermi sea for energies
$\varepsilon \in [-D,0]$.  Then at zero temperature with
$\tau = \frac{\sigma x}{v}$ [\Eq{eq:delta:tau}],
\begin{eqnarray}
 \langle \hat{c}_{x_0}^\dagger \hat{c}_{x_0 + x}^{\phantom{\dagger}}
 \rangle
 &=&
  \tfrac{1}{2D}\!\! \int\limits_{-D}^0\!\! d\varepsilon\,
   e^{\iu \tau \varepsilon}
 = e^{-\iu\frac{D \tau}{2}} \tfrac{\sin(\frac{D \tau}{2})}{D \tau}
\notag \\
 &\equiv&
 e^{-\iu \sigma k_f x} \ \tfrac{\sin(k_f x)}{2 k_f x}
\text{ .}\label{eq:helical:fcor}
\end{eqnarray}
the integral only includes the filled Fermi sea [whereas
\Eq{eq:f0:S:1} integrated over the entire `Brillouine zone'].
By comparison with the corresponding Fermionic correlations for
plain 1D tight-binding chain, this suggests the Fermi wave
vector $k_f$ as half the extent of the filled Fermi sea in
momentum space, i.e., given half-filling,
\begin{eqnarray}
  k_f \equiv \tfrac{D}{2v}
  \overset{\eqref{eq:helical:a}}{=}
  \tfrac{\pi}{2 a}
\text{ .}\label{eq:def:kf}
\end{eqnarray}
The leading phase factor in \Eq{eq:helical:fcor} has subtle
consequences when computing charge correlations, and results in
features that qualitatively differ from a plain tight binding
chain.

Assuming a continuous, i.e., non-discretized 1D edge mode in
terms of lattice sites spaced by $a$, then based on \Eq{eq:f0:S},
the overlap diminishes to zero for $x \gg a$. In particular,
this includes the wide band limit where $a\to0$. However, for
the sake of orthogonality of the fermionic states, the vanishing
of the overlap at finite bandwidth can be simply guaranteed by
adhering to the discrete grid in \Eq{eq:helical:xgrid} with
lattice spacing $a$ in complete analogy to a tight-binding
chain. Hence, in order to avoid complications based on
non-orthogonal $\tilde{f}^{\,}_{0\etaI\sigma}$ states, henceforth
distances will be chosen on the grid \eqref{eq:helical:xgrid},
i.e., with $a:=1$ having $x \in \mathbb{Z}$. With this
$\tilde{f}^{\,}_{0\etaI\sigma} \to \hat{f}^{\,}_{0\etaI\sigma}$
become well-defined orthonormalized local bath modes
at the location of the impurities, denoted by using
hats now instead of tildes.

While finite bandwidth is physically meaningful when having
particular 2D lattice models in mind, for a helical edge mode
this cutoff is peculiar in that the helical branches merge with
a continuum of bulk states. Therefore a sharp ultraviolet cutoff
for an isolated 1D helical edge mode can have potentially
artificial consequences. Lack of orthogonality of local bath
modes discussed with \Eq{eq:helical:a} above is one example.
The latter complication can be simply eliminated, though, by
adhering to the effective discrete lattice in \Eq{eq:helical:xgrid}.
On a related footing, the hybridization function in \Eq{eq:delta:tau},
which is closely related to the dynamical one-particle
propagation in between the impurities, reflects the directedness
of motion via the step functions $\theta(\tau)$.  This step
function, however, is strict for infinite bandwidth only.  For
finite bandwidth it also contains, in particular, a non-zero
oscillatory tail for $\tau<0$.  That is, for $|\omega| \ll D$
and $\tau<0$, rather than strictly being zero, the amplitude for
this enhanced backscattering probability decays like
$\frac{1}{D}\,e^{\iu(\omega\pm D)\tau} \sim \frac{1}{D}\,e^{\iu
\omega\tau}$ [see also \Eq{eq:Gamma:hel}] where the oscillatory
behavior with phase $D \tau = \tfrac{\pi x}{a}$ is similar to
\Eq{eq:helical:a}, thus having $e^{\iu\tau D} = \pm 1$ on the
grid \Eq{eq:helical:xgrid}.
However, this backscattering probability decays with increasing
bandwidth $D$, which eventually enforces strict directionality.
Yet for the above reasons finite bandwidth can generate a weak
subleading contribution $\mathcal{J}_z S_L^z S_R^z$ to the RKKY
Hamiltonian \eqref{H-RKKY} in the helical system
[cf.\,\App{App:RKKY:PT}].

\subsection{Coarse graining \label{sec:NRGdisc}}

For the sake of a numerical treatment, the continuum of the bath
needs to be discretized.  Here we use the NRG which, by
construction, always discretizes in energy-momentum space.
This allows us to target a single edge mode with plain linear
dispersion. To be specific, the NRG coarse-grains on a
logarithmic grid $D/\Lambda^{-(n+z)}$ in energy space with
$\Lambda>1$ (typically $\Lambda \gtrsim 2$) a dimensionless
discretization parameter, $n \in \mathbb{N}$, and $z\in [0,1[$
a plain `$z$-shift' of the logarithmic discretization
\cite{Oliveira91,Zitko09}.

Consider therefore some arbitrary but fixed energy interval $I_l
\equiv [\epsilon_{l}, \epsilon_{l+1}]$ of width
\mbox{$\Delta\epsilon_l \equiv \epsilon_{l+1}
{-}\epsilon_{l}>0$} and average energy $\bar{\epsilon}_l \equiv
\tfrac{1}{2}(\epsilon_l + \epsilon_{l+1})$ within the continuum
of the bandwidth.
Here $l>0$ will refer to energy intervals at positive energies,
with energy increasing with increasing $l$ (this is contrary to
the NRG, hence $l \sim N - n$ with $N$ the number of levels with
$\epsilon_l > 0$).  Since the helical mode in \Fig{fig:helical}
is symmetric around $\epsilon = 0$, the coarse graining for
positive and negative energies is also chosen symmetrically
around $\epsilon = 0$, having $\epsilon_{-l} = -\epsilon_{l}$
such that $l<0$ corresponds to negative energies.  The index
$l=0$ is generally considered excluded here, as it is typically
used to refer to the entire bandwidth, e.g., as with $r_0$ in
\Eq{eq:helical:a}.  Having $\epsilon_{-l} = -\epsilon_{l}$, the
index $l$ thus resembles momentum, in that the simultaneous
inversion of momentum and energy for a given spin flavor leaves
the Hamiltonian of the edge mode invariant.

The energy $\bar{\epsilon}_l$ is differentiated here from
$\varepsilon_l$ (note the different font) with the latter
eventually used for the effective level position for the full
interval $l$, typically having $\varepsilon_l \simeq \bar{\epsilon}_l$
similar but not exactly the same \cite{Zitko09}. When coarse
graining the bath, the integral for the hybridization is split
up into intervals,
\begin{eqnarray}
   \hat{H}_\mathrm{hyb}
   &\overset{\eqref{eq:H:hyb}}{=}&
    \sum_{\etaI \sigma} \Bigl( \hat{d}^\dagger_{\etaI\sigma}
    \sum_l \hspace{-.1in}\underbrace{
    \int\limits_{\epsilon_{l}} ^{\epsilon_{l+1}}
    \sqrt{\tfrac{\Gamma}{\pi} } \ d\varepsilon  \
   \hat{c}^{\,}_{\varepsilon\sigma}\
   e^{\iu \frac{\sigma }{v}\varepsilon \,
    \frac{\etaI x}{2}}
   }_{\equiv \ \sqrt{\frac{\Gamma\Delta\epsilon_l}{\pi}}\,
      \tilde{c}_{l \etaI\sigma} \ \equiv\
      \sum\limits_{\etaI'} T_{l,\etaI\etaI'} \hat{c}_{l\sigma\etaI'}
   }
   \!\!+\  \mathrm{H.c.} \Bigr)
\text{ ,}\notag\\[-6ex]
\label{eq:H:hyb:l} \\ \notag
\end{eqnarray}
with $\{ \hat{c}^{\,}_{\varepsilon\sigma},
         \hat{c}^{\dagger}_{\varepsilon\sigma} \}
{=}\delta_{\sigma \sigma'}\, \delta(\varepsilon {-}
\varepsilon')$, such that
$[\hat{c}^{\,}_{\varepsilon\sigma}] = \text{energy}^{-1/2}$.
The coarse-grained discrete and thus dimensionless bath modes
$\tilde{c}_{l \etaI\sigma}$ are normalized and orthogonal with
respect to spin $\sigma$, but not yet with respect to the
impurity location $\etaI$, as emphasized by using tildes. The
states $\tilde{c}_{l \etaI\sigma}$ for each individual interval
need to be orthonormalized in any case even if the distance $x$
is chosen on the grid in \Eq{eq:helical:xgrid}.
This orthonormalization has to occur prior to the subsequent
mapping of the so-called star geometry in \Eq{eq:H:hyb:l}
between the impurity and the bath states to an effective
one-dimensional (1D) chain geometry, as the latter requires
properly orthonormalized Fermionic levels.

Orthonormalization of the pair of bath states within each
interval $l$ for given spin can be achieved starting from their
overlap which is again simply related to the fermionic
anticommutator similar to \Eq{eq:f0:S},
\begin{eqnarray}
 \{ \tilde{c}_{l\sigma\etaI},  \tilde{c}^\dagger_{l'\sigma'\etaI'} \}
 =
  \delta_{l l'} \delta_{\sigma\sigma'}  \
  S_{\etaI \etaI'}^{l\sigma}
\label{eq:ctilde:overlap}
\end{eqnarray}
with the hermitian dimensionless $2\times 2$ overlap matrix
${\bf S}$ indexed by $\etaI \in \{\iR,\iL\}$ [\Eq{def:etaI}],
\begin{subequations}\label{eq:Scpl}
\begin{eqnarray}
S_{\etaI \etaI'}^{l\sigma} \equiv
  \tfrac{1}{\Delta\epsilon_l} \!\!
  \int\limits _{I_l}\!\!
     \ d\varepsilon  \,
   e^{\iu \frac{(\etaI-\etaI') \sigma x }{2 v}\varepsilon }
   \  \Rightarrow \
   {\bf S}^{l\sigma} {=}
\begin{pmatrix}
    1 & S^{l\sigma}_{+-} \\
    S^{l\sigma\ast}_{+-} & 1
\end{pmatrix}
\qquad\mbox{\ } \label{eq:fortho}
\end{eqnarray}
having ${\bf S}^{l\uparrow} = ( {\bf S}^{l\downarrow})^\ast$,
where with $\tau \equiv\tfrac{ \sigma x}{v} \neq 0$
[\Eq{eq:delta:tau}],
\begin{eqnarray}
   S^{l\sigma}_{+-} =
    \tfrac{1}{\Delta\epsilon_l} \!\!
    \int\limits _{I_l}
     \!\! d\varepsilon  \,
   e^{\iu \tau\varepsilon }
  = e^{\iu \tau\bar{\epsilon}_l} \
    \tfrac{\sin(\frac{\tau \Delta\epsilon_l}{2})}{
      \frac{\tau \Delta\epsilon_l}{2}
    }
   \equiv r_l \, e^{\iu 2\varphi_l}
\text{ .}\qquad
\label{eq:Scpl:1}
\end{eqnarray}
\end{subequations}
Here $r_l \in [0, 1]$ represents the absolute value, and
$2\varphi_l$ the complex phase (including the possible minus
sign from the sine factor). When the interval width is
sufficiently narrow, $|\tau \Delta\epsilon_l| \ll 1$, then $r_l
\to 1$ and one can resolve the phase in a single mode for given
distance, having $x \Delta k_l = |\tau \Delta\epsilon_l| \ll 1$.
This applies to the continuum limit, or also for NRG
discretization intervals at very low energies. In the latter
case it also holds $|\tau \bar{\epsilon}_l| \ll 1$, such that
\begin{eqnarray}
  S^{l\sigma}_{+-} \to 1
  \quad \Rightarrow \quad
  (r_l,\varphi_l) \to (1,0)
\text{ .} \label{eq:Scpl:contlim}
\end{eqnarray}
In this case, the overlap matrix singles out at the symmetric
state as the one with the dominant eigenvalue.
The impurities effectively couple symmetrically to a single bath
state in interval $l$ only, where the phase information
can no longer resolve the distance between the impurities.

As our model Hamiltonian conserves spin, the discretization
and subsequent mapping of the bath can proceed for each spin
individually. By time reversal symmetry,
the inversion and simultaneous spin flip leaves
the Hamiltonian invariant. Therefore the resulting structure
of the bath will be exactly the same for the opposite spin,
except that impurities are coupled in reverse order, i.e,
taking $\iL\leftrightarrow \iR$ [e.g., $x \to -x$ with the
overlap matrix in \Eq{eq:fortho}]. Therefore in what follows,
the coarse-graining of the bath proceeds for spin-up only,
skipping the spin index for readability, while occasionally
indicating it in brackets as a reminder. Similarly, the
interval index $l$ will be lowered, denoting $S_l \equiv
S_l^{(\uparrow)} \equiv {\bf S}^{l\uparrow}$.

By construction, the overlap matrix $S_l$
is hermitian and positive, with eigendecomposition
\begin{eqnarray}
   S_l = U_l s_l\, U_l^\dagger
\text { , }\
   s_{l\etaI} \equiv 1 + \etaI r_l \in [0,2]
  \quad (\etaI \in \{+,-\})\quad
\label{eq:Sl:eig}
\end{eqnarray}
and $s_{l\etaI}$ its eigenvalues. In matrix notation $s_l = 1 +
r_l\, \tau_z$ with $r_l$ as in \Eq{eq:Scpl}.  By convention, the
index order in the symmetric/antisymmetric space is $\etaI \in
\{+1,-1\}\equiv \{+,-\} \equiv \{1,2\}$, with the dominating
symmetric eigenstate ($\etaI=+1$) always listed first (which is
in contrast to the impurity location, where the first index
entry $\etaI=-1 \equiv \iL$ refers to the left impurity).

Orthogonality is ensured by taking symmetric and antisymmetric
combinations up to phase factors (skipping subscripts~$l$ for
readability),
\begin{eqnarray}
S_l & \overset{\eqref{eq:Scpl}} {=} &
  \begin{pmatrix}
    1 & \!\!\! r e^{\iu2\varphi} \\
    r e^{-\iu2\varphi} & 1
  \end{pmatrix} \cr
 & = & \underbrace{\begin{pmatrix}
    e^{\iu\varphi} & 0\\
    0 & \!\!\!e^{-\iu\varphi}
  \end{pmatrix}}_{\equiv \Phi_l}
  \underbrace{\begin{pmatrix} 1 & r \\ r & 1 \end{pmatrix}}_{
    = U_H^{\phantom{\dagger}} s_l\, U_H^{(\dagger)}
  }
  \begin{pmatrix}
    e^{-\iu\varphi} & 0\\
    0 & \!\!\!e^{\iu\varphi}
  \end{pmatrix}
\notag \\[-3ex]
\label{eq:S:Udef}
\end{eqnarray}
such that $U_l = \Phi_l U_H$ with $U_H \equiv \tfrac{1}{\sqrt{2}}
[1 \ 1; 1\ \text{-}1]$ the Hadamard matrix that switches to
symmetric/antisymmetric combinations.  Diagonalizing $S_l$ thus
leads to the fully orthonormal bath modes (with hat now),
\begin{eqnarray}
   \hat{c}_{l\etaI}
   =
   \sum_{\etaI''} P_{l,\etaI\etaI''} \Bigl(
   \tfrac{1}{\sqrt{s_{l\etaI}}} \
   \sum_{\etaI'} U_{l,\etaI'\etaI''}^\ast \,\tilde{c}_{l\etaI'}
   \Bigr)
\text{ ,}\label{eq:fortho:1}
\end{eqnarray}
where an additional unitary $P_l$ was left-multiplied as an
arbitrary rotation or phases to obtain the final bath modes
$\hat{c}_{l\etaI}$. It obviously leaves the fermionic levels
orthonormal, i.e., canonical. \EQ{eq:fortho:1} can be written
more compactly using spinor notation in $\etaI \in \{+,-\}$
(all for spin-up here),
\begin{eqnarray}
   \hat{c}_{l} &\equiv &
   \begin{pmatrix}
      \hat{c}_{l +} \\
      \hat{c}_{l -}
   \end{pmatrix}
   \text{, }
      \tilde{c}_{l} \equiv
   \begin{pmatrix}
      \tilde{c}_{l +} \\
      \tilde{c}_{l -}
   \end{pmatrix}
\ \Rightarrow\
   \tilde{c}_{l}
   = \underbrace{
   U_l \, \sqrt{s_l} \, P_l^\dagger}_{
      \overset{\eqref{eq:H:hyb:l}}{\equiv}
      \sqrt{\tfrac{\pi}{\Gamma \Delta\epsilon_l}} \, T_l
   }
   \hat{c}_{l}
\text{ ,} \notag \\[-4ex]
\label{eq:fortho:1b}
\end{eqnarray}
thus having,
\begin{eqnarray}
   T_l = \sqrt{\tfrac{\Gamma \Delta\epsilon_l}{\pi}}\,
   U_l \, \sqrt{s_l} P_l^\dagger
\text{ .}\label{eq:fortho:1c}
\end{eqnarray}
The overlap of the
$\tilde{c}$ states in \Eq{eq:fortho:1b} is consistent
with \Eq{eq:ctilde:overlap}, since
$\{ \tilde{c}_{l\etaI},  \tilde{c}^\dagger_{l'\etaI'} \}
 = \tfrac{\pi}{\Gamma \Delta\epsilon_l}
 (T_l^{\,} T_l^\dagger)_{\etaI\etaI'}$
with (all for spin up here)
\begin{eqnarray}
  \tfrac{\pi}{\Gamma \Delta\epsilon_l}
  T_l^{\,} T_l^\dagger
  = U_l^{\,} s_l^{\,} U_l^\dagger = S_l
\label{eq:ctilde:overlap:2}
\end{eqnarray}
Consequently, also the overlap of the bath states
at the location of the impurity,
$\tilde{f}_0 \equiv \sum_l \sqrt{\frac{\Delta\epsilon_l}{2D}}\,
\tilde{c}_{l \etaI\sigma}$
remains precisely the same in the discrete setting
as compared to the continuum in \Eq{eq:f0:S:1},
\begin{eqnarray}
 \{ \tilde{f}_{0+} , \tilde{f}_{0-}^\dagger \}
 &=& \tfrac{\pi}{2D\Gamma} \sum_{l} (T_l^{\,} T_l^\dagger)_{+-}
 = \sum_{l}
    \tfrac{\Delta\epsilon_l}{2D} \, S^l_{+-}
 \notag \\
&\overset{\eqref{eq:Scpl}}{=}&
 \tfrac{1}{2D}\int _{-D} ^{D}
    \!\! d\varepsilon \ e^{\iu \tau\varepsilon }
 \overset{\eqref{eq:f0:S}}{=}
   \tfrac{\sin(\pi x/a)}{x/a}
\text{ .}\label{eq:f0:olap:discrete}
\end{eqnarray}

\subsubsection{Bath representation in real numbers}

For $P_l$ in \Eq{eq:fortho:1b}, in principle,
one could have chosen $P_l = U_l$. This then approximately
returns to the original left\,/\,right association of levels
w.r.t. the impurities. However, it turns out more beneficial
to choose $P_l$ differently and in particular, independently
of the interval index $l$.  Based on the hybridization term,
the rotation to symmetric / antisymmetric states may be
temporarily also carried over to the impurity levels
themselves.  Consider the contribution to the hybridization
in \Eq{eq:H:hyb:l} from a particular energy intervall $l$
written in spinor notation
\begin{eqnarray}
   \hat{d}^\dagger \cdot
   T_{l} \, \hat{c}_{l}
 = (\underbrace{U_0\, \hat{d}}_{
     (\, \equiv \hat{f}_{-1} \, )
   })^\dagger \cdot
   \underbrace{(U_0 T_{l} )}_{
      \equiv \mathcal{T}_{l}
   } \hat{c}_{l}
\text{ ,}\label{eq:H:hyb:disc:3}
\end{eqnarray}
where $U_0 \sim U_H$ [\Eq{eq:S:Udef}] to be determined,
and where $\hat{f}_{-1}$ now refers to the impurity states
in the symmetric / antisymmetric basis. Then with
\begin{eqnarray*}
  \mathcal{T}_l = U_0 T_l
    \overset{\eqref{eq:fortho:1b}}{=}
    \sqrt{\tfrac{\Gamma \Delta\epsilon_l}{\pi}} \,
     U_0\!\! \underbrace{U_l}_{
     \overset{\eqref{eq:S:Udef}}{=} \Phi_l U_H}
     \sqrt{s_l}\, P_l^\dagger
\end{eqnarray*}
and by choosing $P_l$ diagonal (i.e., just complex phases)
it commutes with the diagonal matrix $\sqrt{s_l}$,
\begin{eqnarray}
 \mathcal{T}_l = \sqrt{\tfrac{\Gamma \Delta\epsilon_l}{\pi}} \
     U_0 \, \Phi_l
     \underbrace{U_H^{(\dagger)} P_l^\dagger}_{
       \overset{!}{=} U_0^\dagger
     } \sqrt{s_l}
\text{ ,}\label{eq:Ttilde:up:0}
\end{eqnarray}
which suggests the choice $U_0 := P_l U_H$ as indicated.
Now by also fixing $P_l$,
\begin{eqnarray}
   P_l := \mathrm{diag}([1, -\iu]) \equiv P_0 = \mathrm{const}
\end{eqnarray}
this separates real and imaginary part in the phases $\Phi_l$
such that $U_0 \Phi_l U_0^\dagger$ becomes purely real,
since with $U_0 = P_0 U_H$,
\begin{eqnarray}
   U_0 \tau_x U_0^\dagger &=& \tau_z \notag \\
   U_0 \tau_y U_0^\dagger &=& -\tau_x \\
   U_0 \tau_z U_0^\dagger &=& -\tau_y \notag
\label{eq:U0:tau}
\end{eqnarray}
and therefore
\begin{eqnarray}
   U_0  \hspace{-.3in}\underbrace{\Phi_l}_{
   =\cos\varphi + \iu\tau_z \sin\varphi }\hspace{-.3in}
   U_\dagger
 = \cos\varphi - \iu\tau_y \sin\varphi
 = \begin{pmatrix}
     \cos\varphi & \!\!-\sin\varphi \\
     \sin\varphi &  \cos\varphi
   \end{pmatrix}
 \equiv R(\varphi_l)
\notag\\[-3ex]\label{eq:R:def}
\end{eqnarray}
which represents a plain rotation by an angle $\varphi$.
In summary, with $R_l \equiv R(\varphi_l)$
\begin{eqnarray}
 \mathcal{T}_l = \sqrt{\tfrac{\Gamma \Delta\epsilon_l}{\pi}} \
     R_l \sqrt{s_l}
  \quad \in\ \mathbb{R}^{2\times 2}
\text{ ,}\label{eq:Ttilde:up}
\end{eqnarray}
with the coupling term in the hybridization now represented
by purely real coefficients. Since $s_{l\etaI} = 1 + \etaI
r_l$ depends on the particular discretization interval $l$,
$s_l$ is generally not proportional to the identity matrix.
The full hybridization becomes (so far all for the spin-up
sector),
\begin{eqnarray}
   \hat{H}_\mathrm{hyb}^{(\uparrow)} =
    (U_0 \hat{d})^\dagger \cdot
    \underbrace{\sum_{l} \mathcal{T}_{l} \,
    \hat{c}_{l}}_{
     \equiv \ \beta_{0} \hat{f}_{0}}
   \ +\ \mathrm{H.c.}
\label{eq:H:hyb:disc}
\end{eqnarray}

\subsubsection{Block tridiagonalization}

\EQ{eq:H:hyb:disc} is now the starting point for the Lanczos
block-tridiagonalization of the bath, given the initial pair
of orthonormal states $\hat{f}_{0}$ for the zeroth site with
real coefficients in the symmetric / antisymmetric basis
$\hat{c}_l$ of the bath. The bath itself is represented in
diagonal form just by the energies of $\hat{c}_l$, and hence
is clearly also real.  The resulting Wilson chain then
consists of two intercoupled chains,
\begin{eqnarray}
   \hat{H}_\mathrm{bath}^{(\uparrow)} =
   \sum_{n=0}^\infty
     \hat{f}_{n}^\dagger \cdot \alpha_n \hat{f}_{n}^{\,}
    +\bigl( \hat{f}_{n}^\dagger \cdot \beta_{n+1}^{\,} \hat{f}_{n+1}^{\,}
   \ +\ \mathrm{H.c.} \bigr)
\text{ ,} \quad
\label{eq:block-tridiag}
\end{eqnarray}
with $\alpha_n, \beta_n \in \mathbb{R}^{2\times 2}$ and
$\beta_0$ reserved for the coupling to $\hat{f}_{-1} \equiv
U_0 \hat{d}$, i.e., the impurities. By choosing the
symmetric\,/\,antisymmetric basis above, incidentally,
it turns out on numerical grounds, that within numerical
double-precision accuracy, there are no Creutz-couplings
\cite{Creutz99}, i.e., in between $\hat{f}_{n\etaI}$ and
$\hat{f}_{n+1,-\etaI}$. That is, $\beta_n$ turns out
diagonal in $\etaI$, and the Wilson chain becomes a pure
ladder (to be referred to as Wilson ladder \footnote{private
communication with Matan Lotem and Moshe Goldstein;
\cite{Lotem22}}).
Similarly, with the bath setup being symmetric in energy
and thus at half-filling, there are also no onsite energies
along the Wilson sites. All $\alpha_n$ are thus purely
off-diagonal, and encode the rung couplings of the
Wilson ladder. In summary, the block entries $\alpha_n$
and $\beta_n$ in \Eq{eq:block-tridiag} have the structure,
\begin{eqnarray}
  \alpha_n &=& a_n \tau_x \notag \\
  \beta_n &=& b_n (1 + \tfrac{\delta_n}{2} \tau_z)
\text{ .}\label{eq:block-tridiag:2}
\end{eqnarray}
As argued with \Eq{eq:Scpl:contlim}, the coupling to the
impurities is dominated at low energies by the symmetric
channel. Therefore the coupling strengths along the symmetric
and antisymmetric channel can differ, as quantified by the
relative difference $\delta_n$. This splitting scales like
$\delta_n \propto 1/x \ll 1$, and therefore is small for
early Wilson shells close to the impurity for large impurity
distance $x \gg 1$. The case $n=0$ is special, as it refers
to the coupling of the impurities to the $\hat{f}_0$ states.
In the spirit of symmetrically coupled Anderson impurities
where there is actual hybridization of the impurities with
the helical bath, it holds that $\delta_0=0$ if $x=x_n$
is chosen on the discrete lattice \Eq{eq:helical:xgrid}.
In this case, the $\tilde{f}_0$ states themselves are
orthogonal already, and so there are strictly no impurity
cross-coupling due to non-orthogonality of the $\tilde{f}_0$
states.

The fact that the block-tridiagonalization can proceed in
real arithmetic has several advantages, in practice.
The switch to the symmetric\,/\,antisymmetric basis prevents
certain numerical errors from piling up during the
block-tridiagonalization that are related to slowly drifting
complex phases. Also by dealing with real arithmetic, phase
conventions on basis states only refer to signs. The pair of
states within every tridiagonalization step represent
symmetric / antisymmetric states, and hence are already
orthogonal, by construction, so there is no immediate explicit
need for a Schmid decomposition within the pair of states
$\hat{f}_n$.

The total combined hybridization strength of the two
impurities for given spin(-up) is [cf. \Eq{eq:H:hyb:l}]
\begin{eqnarray}
   \sum_{l} \mathrm{tr}(T_{l}^{\,} T_{l}^{\dagger})
   \overset{\eqref{eq:fortho:1c}}{=}
    \tfrac{2D\Gamma}{\pi}
    \sum_l \tfrac{\Delta\epsilon_l}{2D}
    \underbrace{\mathrm{tr}(S_{l})
    }_{=2}
  = 2\times \tfrac{2D\Gamma}{\pi}
\label{eq:V2:helical:1}
\end{eqnarray}
where the prefactor of $2$ derives from the two impurity
channels. This total combined hybridization strength is
the same, on average, for the impurities $\hat{f}_{-1}$
in the symmetric / antisymmetric basis since
$ \sum_{l} \mathrm{tr}(
  \mathcal{T}_{l}^{\,} \mathcal{T}_{l}^{\dagger})
  \overset{\eqref{eq:H:hyb:disc:3}}{=}
  \sum_{l} \mathrm{tr}(T_{l}^{\,} T_{l}^{\dagger})
$
In the symmetric basis,
however, the contribution of the individual channels
can differ from each other, as reflected also
in $\delta_0 \neq 0$ if $x \neq x_n$
in \Eq{eq:block-tridiag:2}. With
\begin{eqnarray}
  \mathcal{T}_{l}^{\,} \mathcal{T}_{l}^{\dagger}
 &\overset{\eqref{eq:Ttilde:up}}{=}&
  \tfrac{\Gamma\Delta\epsilon_l}{\pi}\, R_l s_l R_l^\dagger
  \overset{\eqref{eq:H:hyb:disc:3}}{=}
  U_0 T_{l}^{\,} T_{l}^{\dagger} U_0^\dagger
= \tfrac{\Gamma\Delta\epsilon_l}{\pi}\,
  U_0 S_l U_0^\dagger
\notag \\
&\overset{\eqref{eq:Scpl}}{=}&
  \tfrac{\Gamma\Delta\epsilon_l}{\pi}\,
  U_0 \bigl[1 + r_l (
     \tau_x \cos2\varphi_l - \tau_y \sin2\varphi_l \bigr]
  U_0^\dagger
\notag \\
&\overset{\eqref{eq:U0:tau}}{=}&
  \tfrac{\Gamma\Delta\epsilon_l}{\pi}\,
  \bigl[1 + r_l (
    \tau_z \cos2\varphi_l + \tau_x \sin 2\varphi_l\bigr]
\label{eq:TlTldagger}
\end{eqnarray}
the diagonal elements for channel $\etaI$
are given by
\begin{eqnarray}
   &&\tfrac{\pi}{2D\Gamma} \sum_{l} \bigl(
    \mathcal{T}_{l}^{\,} \mathcal{T}_{l}^{\dagger}
    \bigr)_{\eta\eta}
  \overset{\eqref{eq:Ttilde:up}}{=}
    \sum_l \tfrac{\Delta\epsilon_l}{2D}
    (1 + \eta r_l \cos 2\varphi_l)
\notag \\
  &&\qquad \overset{\eqref{eq:Scpl}}{=}
    1 + \tfrac{\eta}{2D}\sum_l
     \Delta\epsilon_l \re\!\bigl( S_{l,+-} \bigr)
\notag \\
  &&\qquad \overset{\eqref{eq:helical:r0}}{=}
  1 + \eta r_0
  = s_{0\eta}
\text{ ,}\label{eq:V2:helical:2}
\end{eqnarray}
noting that $r_0$ in \Eq{eq:helical:a} has precisely the
same structure as $r_l$ in \Eq{eq:Scpl}, it just considers
the full bandwidth instead of the interval $\Delta\epsilon_l$,
i.e., $\Delta\epsilon_l \to 2D$. The expressions in
\Eq{eq:V2:helical:2} also describe the normalization of the
states $\hat{f}_0$ that initialize the block-tridiagonalization,
\begin{eqnarray}
   \hat{f}_{0}
   = \sqrt{\tfrac{\pi}{2D\Gamma} \tfrac{1}{s_{0}}}\
     \sum_{l} \mathcal{T}_{l} \,
     \hat{c}_{l}
   \overset{\eqref{eq:H:hyb:disc:3}}{=}
   \underbrace{\tfrac{1}{\sqrt{s_{0}}} U_0}_{\equiv \tilde{U}_0}
   \underbrace{
     \sum_{l} \sqrt{\tfrac{\Delta\epsilon_l}{2D}} \,
     \tilde{c}_{l}}_{= \tilde{f}_0}
\text{ ,}\label{eq:f0:explicit}
\end{eqnarray}
thus obtaining $\tilde{f}_0 = U_0^\dagger\sqrt{s_0} \hat{f}_0
\equiv \tilde{U}_0^{-1} \hat{f}_0$.
In the absence of the impurities, these levels are exactly
half-filled in the ground state (and also at finite temperature),
since with $r_{-l} = r_l$ and $\varphi_{-l} = -\varphi_{l}$,
\begin{eqnarray}
   \langle \hat{f}_{0\eta}^\dagger \hat{f}_{0\eta}^{\,} \rangle
   = \tfrac{\pi}{2D\Gamma} \tfrac{1}{s_{0\eta}}
     \underbrace{\sum_{l<0}}_{\to\ \frac{1}{2} \sum_l} \bigl(
    \mathcal{T}_{l}^{\,} \mathcal{T}_{l}^{\dagger}
    \bigr)_{\eta\eta}
   = \tfrac{1}{2}
\text{ .}
\label{eq:n0:occ}
\end{eqnarray}
For the off-diagonal expectation value one obtains,
\begin{eqnarray}
   \langle \hat{f}_{0+}^\dagger \hat{f}_{0-}^{\,} \rangle\!
   &=& \!\tfrac{1}{\sqrt{s_{0+} s_{0-}}}
     \sum_{l<0} \underbrace{ \tfrac{\pi}{2D\Gamma}
     \bigl(
      \mathcal{T}_{l}^{\,} \mathcal{T}_{l}^{\dagger}
     \bigr)_{-+}}_{
      = \frac{\Delta\epsilon_l}{2D}\im( r_l e^{\iu 2\varphi_l})
     }
   \overset{\eqref{eq:def:kf}}{=}
   \tfrac{-1}{\sqrt{1-r_{0}^2}} \tfrac{\sin^2 (k_f x)}{2 k_f x}
\notag\\[-3ex]
\label{eq:n0:+-}
\\[-4ex]\notag
\end{eqnarray}
since
\begin{eqnarray*}
  \sum_{l < 0}\!
    \tfrac{\Delta\epsilon_l}{2D} \, r_l \, e^{\iu 2\varphi_l}
  &\overset{\eqref{eq:Scpl}}{=}&
   \tfrac{1}{2D}\!\! \int\limits_{-D} ^{0}\!\!
     \ d\varepsilon \, e^{\iu \tau\varepsilon }
  = \tfrac{\sin \frac{\tau D}{2}}{\tau D}
    e^{-\iu \frac{\tau D}{2}}
\text{ ,}
\end{eqnarray*}
By construction, \Eq{eq:n0:+-} is real, and due to the
helical nature, it is antisymmetric under inversion $x \to
-x$. Yet since with \Eq{eq:helical:a} $\lim_{x\to 0} r_0 =
1$, \Eq{eq:n0:+-} becomes discontinuous across $x=0$.
When $x\neq 0$ is taken on the grid \eqref{eq:helical:xgrid},
$r_0=0$, and therefore the prefactor becomes $1$.
By construction, the local bath levels are clearly
also half-filled, $\langle \tilde{f}_{0\eta}^\dagger
\tilde{f}_{0\eta}^{\,} \rangle {=} \tfrac{1}{2}$ for $\eta
\in \{\iR,\iL\}$, but with the complex off-diagonal
expectation value
\begin{eqnarray}
  \langle \tilde{f}_{0L}^\dagger \tilde{f}_{0R}^{\,} \rangle
  \overset{\eqref{eq:helical:fcor}}{=}
  e^{-\iu \sigma k_f x} \ \tfrac{\sin(k_f x)}{2 k_f x}
\text{ .}
\label{eq:n0:LR}
\end{eqnarray}
This is identical with a plain tight-binding chain up to
the phase factor which gives rise to the somewhat different
expression in \Eq{eq:n0:+-}.

\subsubsection{Finalizing the Wilson setup}

Once the block-tridiagonalization is performed,
one can rotate the symmetric / antisymmetric impurity
space $\hat{f}_{-1} \equiv U_0 \hat{d}$ in
\Eq{eq:H:hyb:disc:3} back to the local representation,
for convenience,
\begin{eqnarray}
   H_\mathrm{hyb}^{(\uparrow)}
   &\overset{\eqref{eq:H:hyb:disc}}{=}&
   \hat{f}_{-1,\sigma}^\dagger \cdot \beta_0 \hat{f}_{0}^{\ }
   + \ \mathrm{H.c.}
  = \hat{d}_{\sigma}^\dagger \cdot
    \underbrace{U_0^\dagger \beta_0
    \hat{f}_{0\sigma}^{\ }}_{
      \overset{!}{=} \sqrt{\tfrac{2D \Gamma}{\pi}}
      \tilde{f}_{0}
    }
   + \ \mathrm{H.c.}
\notag \\[-4ex]
\label{eq:helical:Hcpl-0}
\end{eqnarray}
The last representation must reflect the original local
$\tilde{f}_0$ states as this was the very starting point,
namely that these couple symmetrically and diagonally with
the respective impurity only. Thus by identifying
\begin{eqnarray}
  \beta_0
  \overset{\eqref{eq:block-tridiag:2}}{\equiv}
  b_0 (1 + \tfrac{\delta_0}{2} \tau_z)
  \overset{\eqref{eq:f0:explicit}}{=}
  \sqrt{\tfrac{2D \Gamma}{\pi}}
  \hspace{-.15in}\underbrace{\sqrt{s_0}}_{
     \overset{\eqref{eq:helical:r0}}{\simeq}
     1 + \tfrac{1}{2} r_0 \tau_z
  }
\label{eq:helical:Hcpl:s}
\end{eqnarray}
this shows $b_0 \simeq \sqrt{\tfrac{2D \Gamma}{\pi}}$ and
$\delta_0 \simeq r_0$ for small $r_0$,
e.g., large distance $x$, or in particular $r_0=0$
if the impurity distance is chosen on the grid
\eqref{eq:helical:xgrid}. Therefore, in the case
of $\beta_0$ this permits an analytic identification
of the structure observed numerically for $\beta_n$
in \Eq{eq:block-tridiag:2}.
The hybridization in \Eq{eq:helical:Hcpl-0} thus becomes
\begin{eqnarray}
   H_\mathrm{hyb}^{(\uparrow)} &=&
    \sqrt{\tfrac{2D \Gamma}{\pi}}\,
    d_{\sigma}^\dagger
   \underbrace{U_0^\dagger \sqrt{s_0} \hat{f}_{0}^{\ }
    }_{
      = \tilde{f}_{0}
    }
   \ + \ \mathrm{H.c.}
\text{ .}\label{eq:helical:Hcpl-1}
\end{eqnarray}
consistent with \Eq{eq:f0:explicit}.
Similarly, the overlap of the states $\tilde{f}_{0\eta}$
agrees with \Eqs{eq:f0:S}, since  by denoting
$\tilde{U}_0^{-\dagger} \equiv (\tilde{U}_0^{-1})^\dagger$,
with
$ \{ \tilde{f}_{0\eta}^{\,}, \tilde{f}_{0\eta'}^\dagger \}
\overset{\eqref{eq:f0:explicit}}{=} \bigl(
  \tilde{U}_0^{-1} \tilde{U}_0^{-\dagger}
  \bigr)_{ \eta \eta'}$
and given that $P_0$ in $U_0$ commutes with $s_0$,
\begin{eqnarray}
   \tilde{U}_0^{-1} \tilde{U}_0^{-\dagger}
   &=& U_H^{(\dagger)}
     \hspace{-.1in}\underbrace{s_0}_{= \, 1 + r_0\tau _z}
     \hspace{-.1in}
     U_H
   = \begin{pmatrix}
      1 & r_0 \\
      r_0 & 1
   \end{pmatrix}
   \overset{\eqref{eq:f0:S}}{\equiv} S_0
\label{eq:ftilde:overlap:1} \\
   \tilde{U}_0^{-\dagger} \tilde{U}_0^{-1}
   &=& s_0
\label{eq:ftilde:overlap}
\text{ .}
\end{eqnarray}
Overall, the back-transformation towards $\hat{d}$ and $\tilde{f}_0$
alters the initial couplings of the Wilson ladder as
follows,
\begin{subequations}\label{eq:f0:backtrafo}
\begin{eqnarray}
   \beta_0 &\to& \mathfrak{b}_0 \equiv
   U_0^\dagger \,\beta_0\, \tilde{U}_0^{\,}
 = U_0^\dagger \,\sqrt{\tfrac{2D \Gamma}{\pi} s_0} \,
   \tilde{U}_0^{\,}
 = \sqrt{\tfrac{2D \Gamma}{\pi}}\, {\bf 1}
\label{eq:f0:backtrafo:b0}
\\[2ex]
   \beta_1 &\to& \mathfrak{b}_1 \equiv
   \tilde{U}_0^\dagger\beta_1
 =
   U_0^\dagger \tfrac{1}{\sqrt{s_0}} \, \beta_1
\label{eq:f0:backtrafo:b1} \\[1ex]
   \alpha_0 &\to& \mathfrak{a}_0 \equiv
   \tilde{U}_0^{\dagger} \,\alpha_0\, \tilde{U}_0^{\,}
   \overset{\eqref{eq:block-tridiag:2}}{=}
   U_0^\dagger \underbrace{\tfrac{1}{\sqrt{s}} (a_0 \tau_x)
       \tfrac{1}{\sqrt{s}}}_{
     = \tfrac{a_0}{\sqrt{1-r_0^2}} \tau_x
   } U_0
 = \tfrac{-a_0}{\sqrt{1-r_0^2}} \,\tau_y
\,\text{.}
\notag\\[-4ex]\label{eq:f0:backtrafo:a0}
\end{eqnarray}
\end{subequations}
This back-transformation generates (i) the only complex entries
in the Wilson ladder setup and (ii) also introduces Creutz
couplings in $\mathfrak{b}_1$ via $U_0$ which thus render the
Wilson setup non-bipartite. Both are important on physical
grounds. The latter prevents the model from having an SU(2)
particle-hole symmetry, which is absent in a helical system.
Rather, it is reduced to a discrete $\mathbb{Z}_2$ particle-hole
symmetry [cf. \Sec{sec:symmetries}].  This is specific to the
effective model of the helical edge used here, though, stressing
that such a particle-hole symmetry is absent in 2D time-reversal
invariant topological insulators \cite{schnyder_2008}.
The complex phases with $\mathfrak{a}_0$ and $\mathfrak{b}_1$
close to the impurity are important and hence cannot be gauged
away, since e.g. {\it within} a fixed spin flavor, the helical
Hamiltonian is not time-reversal invariant [if derived from a
real-space lattice the Hamiltonian necessarily would have to
include complex spin-orbit coupling terms; in the diagonal
eigenbasis, as in \Eq{eq:H0:k}, any quadratic Hamiltonian
becomes real of course].

The coupling $a_0$ is fully determined by $\langle 0|
\hat{f}_{0+}^{\,} \hat{H}_\mathrm{bath} \hat{f}_{0-}^\dagger
|0\rangle$ with $| 0\rangle$ the vacuum state, and hence
can be expressed analytically,
\begin{eqnarray}
  a_0
 &\overset{\eqref{eq:f0:explicit}}{=}&
  \tfrac{1}{\sqrt{s_{0+} s_{0-}}}
  \tfrac{\pi}{2D\Gamma}
  \sum_l \varepsilon_l \hspace{-.12in}\underbrace{
  (\mathcal{T}_{l}^{\,} \mathcal{T}_{l}^{\dagger})_{+-}}_{
    \overset{\eqref{eq:TlTldagger}}{=}
    \tfrac{\Gamma\Delta\epsilon_l}{\pi}\, r_l \sin 2\varphi_l
  }
\\
 &=&
  \tfrac{1}{\sqrt{1 - r_0^2}}\
  \im \tfrac{1}{2D} \sum_l
   \underbrace{\Delta \epsilon_l \,
   r_l \, e^{2\iu\varphi_l}}_{
     \overset{\eqref{eq:Scpl}}{=}
     \int\limits _{\epsilon_l} ^{\epsilon_{l+1}}
     d\varepsilon  \, e^{\iu \tau\varepsilon }
   } \varepsilon_l
 \simeq
  -\tfrac{1}{\sqrt{1 - r_0^2}} \
   \tfrac{d r_0}{d\tau}
\notag
\end{eqnarray}
The precise value for $a_0$ is thus sensitive on the
discretization, as seen with the second line. In the continuum
limit, $\varepsilon_l \to \varepsilon$ may be pulled inside the
integral, which yields the last expression having used
$\varepsilon e^{\iu \tau \varepsilon} = (-\iu \tfrac{d}{d\tau})
e^{\iu \tau \varepsilon}$. While $r_0=0$ on the grid
\eqref{eq:helical:xgrid}, the derivative is generally non-zero
with alternating signs and decaying like $1/\tau \sim 1/x$.
Therefore $a_0$ is non-zero for any $x=x_n$. That is, the two
impurities weakly see each other right away upon a single
application of $\hat{H}_\mathrm{bath}$. On intuitive grounds,
one can compare this to a tight-binding chain with long-range
hoppings that decay like $1/x$ [cf. \Eq{eq:tb:helical}] which
also gives rise to a similar behavior.

\begin{figure}[tb!]
\begin{center}
\includegraphics[width=.8\linewidth]{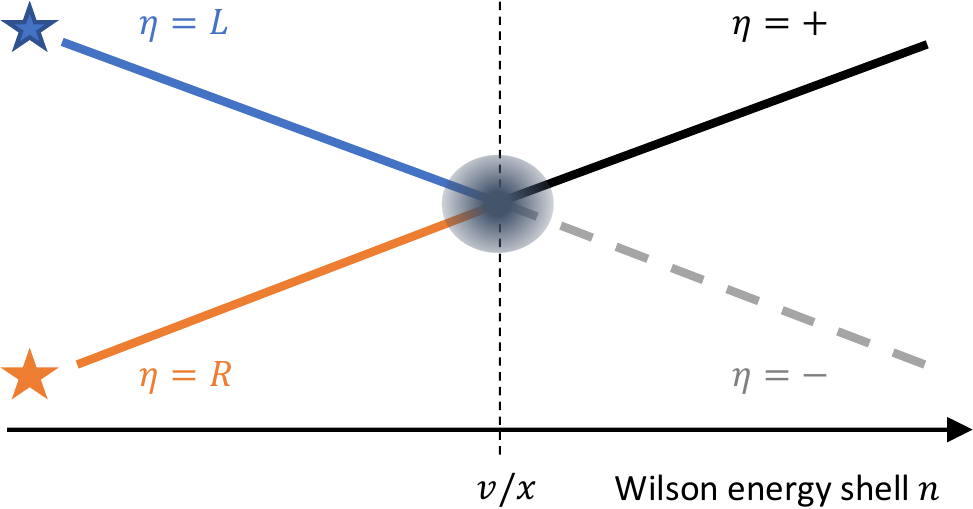}
\end{center}
\caption{Schematic representation of the 2-impurity
  Wilson setup of the bath giving rise to two intercoupled chains
  (for a quantitative example, see \Fig{fig:Wilson:ladder}).
  The system moves to small energies towards the right with
  the Wilson shells $n$ having energy $\varepsilon_n \sim
  \Lambda^{-n/2}$. At high energies $\varepsilon \gg \Ex
  \equiv \tfrac{v}{x}$ [\Eq{eq:domega}] one may think
  of the bath states coupled to their respective impurity.
  Therefore the physics represents two independent copies of
  one-impurity problems down to energies \Ex. Around the
  energy scale of \Ex the impurities start to coherently
  interact with each other (if the impurities are, for example,
  already Kondo screened above the energy \Ex, then the
  two impurities remain decoupled down to zero energy).
  The 2-impurity physics takes place at energies
  $\varepsilon \lesssim \Ex$ where the relevant description
  of the bath effectively changes from the local $(\iR,\iL)$
  to the symmetrized $(+,-)$ representation.
  The antisymmetric channel ($\eta=-$) starts to smoothly
  decouple, but stays in the system as a passive spectating
  bath space. The symmetric channel ($\eta=+$) is the one
  that remains fully coupled to the impurities. While the
  representation of the bath in the main text is always in
  the symmetric/antisymmetric configuration [cf.
  \Eq{eq:helical:Hcpl-0}, except for a final rotation
  of $\hat{f}_0$ back to the local representation].
  If the same unitary mixing of modes were to be applied
  for subsequent Wilson shells still, the property of
  having two independent copies of bath modes remains
  intact down to $E_x$.
}
\label{fig:Wilson}
\end{figure}

\subsubsection{Decoupling of anti-symmetric sector at
low energies}

\begin{figure}[tb!]
\begin{flushleft}
{(a)}\\[2ex]
\includegraphics[width=1\linewidth]{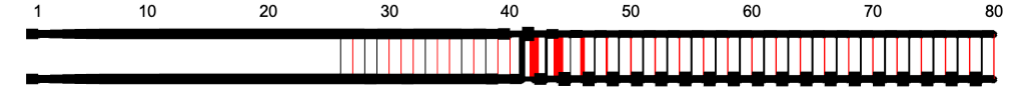} \\[3ex]
{(b)}\\[2ex]
\includegraphics[width=1\linewidth]{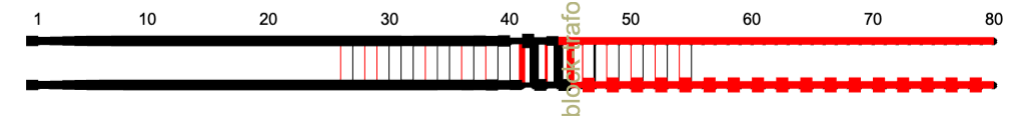}
\end{flushleft}
\caption{
   Typical Wilson ladder for the 2-impurity helical system
   as described by \Eq{eq:block-tridiag} for an
   impurity distance $x=10^6$. The labels on top
   indicate the Wilson shell $n$ (based on $\Lambda=2$).
   The widths of the bonds are proportional to the
   hopping amplitudes rescaled by $\Lambda^{-n/2}$.
   They are all real-valued,
   where black (red) color shows positive (negative) hopping
   amplitudes, respectively.
   Around the energy $1/x$ (shell $n\sim 40$),
   the structure of the Wilson ladder changes,
   as qualitatively already argued in \Fig{fig:Wilson}.
   The upper leg corresponds to the symmetric
   ($\eta=+$), and the lower leg to the antisymmetric  channel
   ($\eta=-$). The rescaled Creutz couplings have amplitudes
   below double precision accuracy, hence are absent.
(b) Same as in (a), except that starting from the
   position `blocktrafo' towards the right
   a nearest-rung shell-mixing numerically determined
   block transformation was performed on top of (a).
   This shows that at low energies
   the system can be exponentially decoupled towards
   later shells into two independent channels.
}
\label{fig:Wilson:ladder}
\end{figure}

The antisymmetric sector gets suppressed once energies are much
below $\varepsilon \ll \Ex \equiv \tfrac{v}{x}$ [\Eq{eq:domega}],
or more precisely in the limit $|\tau \Delta\epsilon_l \ll 1|$,
This occurs in practice at very low energies in the NRG context.
With $r_l \to 1^-$, while $s_{l+}$ approaches the finite value
of $2^-$, the smaller eigenvalue $s_{l-} \to 0^+$ approaches
zero as $1 - 1^-$ which is numerically ill-conditioned.  Hence
form a numerical perspective, it is computed via an expansion
around small $\xi \equiv \tau \Delta\epsilon_l /2 = x \Delta
k_l/2$, i.e., from \Eq{eq:Scpl:1},
\begin{eqnarray}
   s_{l-} &=&  1 - r_l = 1 - \tfrac{\sin\xi}{\xi}
   = \tfrac{1}{3!} \xi^2 + \mathcal{O}(\xi^3)
\label{eq:s-:asymptotic}
\end{eqnarray}
When setting $s_{l-}$ strictly to zero below some threshold
$s_{l-} < 10^{-16}$, the block tridiagionalization eventually
switches over to a hopping amplitude that decays twice as fast,
because the antisymmetric levels that actually couple have been
exhausted.  On the other hand, keeping the asymptotic dependence
in \Eq{eq:s-:asymptotic} down to the lowest energies considered,
the hopping amplitudes along the Wilson ladder always decay like
$\omega_n \sim \Lambda^{-n/2}$. That is, the antisymmetric channel
{\it remains} in the system, throughout. The reason for this is
that the decoupling occurs smoothly. So once the energy scale
(or more precisely, the energy resolution) drops below $v/x$,
the antisymmetric channel does not decouple in an instant, and
so it stays in the system, as schematically depicted in
\Fig{fig:Wilson}.  At energies much below $v/x$, however, one
can show in practice that the Wilson chain, indeed, switches
over to two fully decoupled chains [cf.  \Fig{fig:Wilson:ladder}(b)].
While the symmetric sector which remains coupled to the
impurities, shows a smooth decay of the hopping amplitudes,
in the antisymmetric channel the hopping amplitudes along its
corresponding leg in the Wilson ladder becomes increasingly
alternating (lower legs in \Fig{fig:Wilson:ladder}): namely the
paired up antisymmetric levels at energies $\pm \bar{\varepsilon}_l$.
They form strong bonds along the Wilson chain at zero energy,
where bonding and antibonding states reveal the original $\pm
\bar{\varepsilon}_l$ states in the star geometry.

Since the impurity distance directly enters the coarse graining
in the NRG, there will always be a qualitative change in the NRG
energy flow diagram around the energy scale $v/x$ [cf.
\Figs{fig:Wilson} and \ref{fig:Wilson:ladder}].  But this change
in the representation of the bath can become irrelevant for
static or dynamic properties from the perspective of the
impurity.  In this sense, the `apparent' energy scale strongly
visible in the standard NRG energy flow diagrams at the energy
scale $v/x$ may be irrelevant for the impurity.  Nevertheless,
this may leave minor artificial features (wiggles) in the
temperature or frequency dependence of physical properties
around the energy scale of $v/x$. This effect is expected to be
more pronounced for coarser discretization (larger $\Lambda$),
but to diminish for smaller $\Lambda$.

\subsubsection{Block-tridiagonal structure for opposite spin}

Switching the spin has the same effect as changing the sign of energy
or spatial inversion [cf. discussion after \Eq{eq:Scpl:contlim}].
This is also explicitly reflected in the the variable $\tau
\equiv \tfrac{\sigma x}{v}$ [\Eq{eq:delta:tau}] that appears in
much of the above treatment.  Therefore by construction of the
starting vectors for the block-tridiagonalization of the spin-up
channel above, from the point of view of the impurity, spin-down
couples to an identical Wilson ladder of its own, except that
the local $\tilde{f}_{0}$ modes couple to the impurity levels in
reverse order.  In this sense, the back-transformation to the
local representation of the impurity in \Eqs{eq:f0:backtrafo} is
useful as a prior step.  Then the Pauli matrix $\sigma_x$ below
accounts for the reversed order,
\begin{eqnarray}
   \beta_0^\downarrow \ \to\  \mathfrak{b}_0^\downarrow \equiv
   \sigma_x \mathfrak{b}_0^\uparrow
  = \sqrt{\tfrac{2D \Gamma}{\pi}} \sigma_x
\text{ ,} \label{eq:f0:backtrafo:4}
\end{eqnarray}
while everything else remains the same for spin down
as for spin up. The effect of the cross terms on
the impurity spectral function, for example, are therefore,
\begin{eqnarray}
   \langle \hat{d}_{\uparrow \eta}   \Vert
           \hat{d}_{\uparrow,-\eta}
   \rangle_\omega
 = \langle \hat{d}_{\downarrow,-\eta} \Vert
            \hat{d}_{\downarrow \eta}
   \rangle_\omega
\text{ .}\label{eq:helical:Aimp}
\end{eqnarray}
This makes intuitive sense, since motion in between the two
impurities occurs in opposite directions for different spins.

\subsection{Dynamical correlation functions
\label{sec:complex-spectral}}

Within the NRG approach, correlation functions are computed by
evaluating the Lehmann representation \cite{Oliveira91,Bulla08}
which can be carried out exactly at arbitrary temperature $T$
within the full density matrix approach (fdm-NRG \cite{Wb07}).
To be specific, a retarded dynamical correlation function
\begin{eqnarray}
    \mathcal{G}_{BC}^R(t)
  = -\iu \vartheta(t) \,\langle \hat{B}(t)  \hat{C}^\dagger\rangle_T
\label{eq:Gt}
\end{eqnarray}
for two local impurity operators $\hat{B}$ and $\hat{C}$, with
$\vartheta(t)$ the Heaviside step function, $\hat{B}(t) =
e^{\iu \hat{H} t} \hat{B} e^{-\iu \hat{H} t}$, and $\langle
\ldots\rangle_T$ the thermal average, is computed via its
spectral Lehmann representation
\begin{eqnarray}
   \mathcal{A}_{BC}(\omega) =
   \sum_{a b} \rho_a^{\,}
   B_{ab}^{\,} C_{ab}^\ast \, \delta(\omega - E_{ab})
   \ \ \in\  \mathbb{C}
\label{eq:Aspec}
\end{eqnarray}
with $a$ and $b$ complete sets of many-body eigenstates, having
$\hat{H} |a\rangle = E_a |a\rangle$ with $E_{ab} \equiv E_b {-}
E_a$, thermal weights $\rho_a {=} \tfrac{1}{Z}\,e^{-\beta
E_{a}}$, $\beta \equiv \tfrac{1}{T}$, and $Z \equiv \sum_a
e^{-\beta E_{a}}$ the partition function.  The iterative
diagonalization within the NRG approach explicitly generates the
full many-body state space $\hat{H} |a\rangle \simeq E_a
|a\rangle$ above.  In practice, this yields a tractable
symmetry-respecting, eigenstate decomposition of the entire
impurity Hamiltonian on the Wilson chain \cite{Anders05}
which while approximate, is well-controlled and complete.

Standard fermionic and bosonic
correlation functions have an (anti)commutator
in their Green's function,
\begin{eqnarray}
  \langle \hat{B} \Vert \hat{C}^\dagger\rangle_t
  &\equiv& G_{BC}^R(t)
= -\iu \vartheta(t) \langle [ \hat{B}(t), \hat{C}^\dagger]_s \rangle_T
\notag \\
  &\equiv& \mathcal{G}_{BC}(t) +
       s\, \mathcal{G}_{C^\dagger B^\dagger}(-t)
\text{ ,}
\end{eqnarray}
with the commutator ($s{=}-1$) for bosonic correlations such as
spin-spin correlation functions, or the anticommutator
($s{=}+1$) for fermionic correlation functions such as the
fermionic local density of states.  Equivalently, in frequency
space $G_{BC}^R(\omega) = \mathcal{G}_{BC}(\omega) + s\,
\mathcal{G}_{C^\dagger B^\dagger}(-\omega)$, where by
construction at zero temperature $\mathcal{A}_{BC}(\omega<0)=0$
since $E_{ab}\geq 0$ in \Eq{eq:Aspec}, such that the two
contributions in the (anti)commutator to the full correlation
function are separated in frequency space, since they
exclusively contribute to positive or negative frequencies only.

The important point with \Eq{eq:Aspec}, as already also pointed
out with the hybridization function in \Sec{seq:hybridiziation},
is that for off-diagonal correlations $\hat{B}\neq \hat{C}$, the
spectral data cannot only be negative, but fully complex, i.e.,
$\mathcal{A}_{BC}(\omega)  \in \mathbb{C}$, if the matrix elements
themselves are complex.  In this sense, one cannot simply write
the spectral data $\mathcal{A}(\omega)$ as $-\tfrac{1}{\pi} \im
\mathcal{G}(\omega)$.  Instead, by only taking $-\tfrac{1}{\pi}
\im (\ldots)$ of the propagator $\tfrac{1}{\omega^+ - E_{ab}}$,
the full Green's function in frequency space can be simply
obtained by standard means, i.e., via Kramers-Kronig
transformation, or by folding with the propagator,
\begin{eqnarray}
   \mathcal{G}_{BC}(\omega) =
   \int d\omega' \, \mathcal{A}_{BC}(\omega')\,
   \tfrac{1}{\omega^+ - \omega'}
\text{ .}\quad (\mathcal{A} \in \mathbb{C})
\end{eqnarray}
By construction, with \Eq{eq:Aspec} one still also has full
access to the well-known simple spectral sum rules,
\begin{eqnarray}
   \int d\omega \, \mathcal{A}_{BC}(\omega)
 = \sum_{a b} \rho_a^{\,} B_{ab}^{\,} C_{ab}^\ast
 = \langle B C^\dagger\rangle_T
    \  \in \mathbb{C} \quad
\label{eq:Aspec:C}
\end{eqnarray}
which may be complex for $B \neq C^\dagger$ in the present
helical setting. This is relevant, for example, when computing
the one-particle correlation function across the impurity
$\langle \hat{d}_{\sigma \iL} \Vert \hat{d}_{\sigma\iR}^\dagger
\rangle_\omega$ [e.g., see \Eqs{eq:Gamma:hel} or
\eqref{eq:Gimp:tau} for the non-interacting case].

\subsection{Symmetries \label{sec:symmetries}}

The global symmetries of the effective 1D model of the helical
edge in \Sec{sec:model} also manifest themselves in the
structure of correlation functions. Aside from the symmetry
U(1)$_\mathrm{charge} \otimes$ U(1)$_\mathrm{spin}$ already
discussed when introducing the helical model system in
\Sec{sec:model} and also explicitly exploited in our numerical
simulations, the isolated helical edge can support further
symmetries that are actually absent in the original full-fledged
2D topological system.  The original topological aspect is
reflected here in the fact that the isolated helical edge exists
as a valid physical model in the first place.  With this in
mind, the isolated helical edge with two Kondo impurities (2HKM)
located symmetrically around the origin also preserves
\begin{itemize}

\item

$\mathbb{Z}_2$ time reversal symmetry
($\mathbb{Z}_2^\mathrm{TRS}$): momentum together with spin
reversal is preserved by the helical edge. The local
impurity-bath Kondo interaction is also spin-reversal symmetric,
with the impurities themselves located symmetrically around the
origin [cf.~\Eq{def:etaI}]. Therefore $k \to -k$ together with
the reversal of the impurities, \mbox{$\eta \to -\eta$}, leaves
the local impurity-bath interaction invariant.  Now the combined
operation $k\to -k$ and $\eta \to -\eta$ is equivalent to
spatial inversion. Hence, for our model setup with an isolated
helical edge, TRS can be translated into spatial inversion with
simultaneous global spin reversal.

\item $\mathbb{Z}_2$ particle-hole symmetry
($\mathbb{Z}_2^\mathrm{p/h}$): the helical channel in
\Fig{fig:helical} was chosen such that for every level at
$\varepsilon_{ k\sigma}>0$ there is a level at
$\varepsilon_{-k\sigma}<0$, having $\varepsilon_{k\sigma} =
-\varepsilon_{-k\sigma}$.  By having half-filling, this converts
a particle to a hole or vice versa. Furthermore, by
construction, the Kondo interaction of the impurities with the
helical channel are also particle-hole symmetric.

\end{itemize}
The Hamiltonian in \Sec{sec:model} has two impurities located
symmetrically around the origin [cf. \Eq{def:etaI}]. From an
Anderson impurity point of view with explicit hybridization as
in \Eq{eq:H:hyb}, this is the only point where complex numbers
enter the total Hamiltonian.  From a numerical perspective, the
Fermionic operators $\hat{c}_{k\sigma}$ and $\hat{d}_{\eta\sigma}$
can be encoded by real matrix elements, such that the only
complex entry is the phase $e^{\iu kx \frac{\eta}{2}}$. This
also holds when switching from Anderson-type hybridization to
Kondo spin interactions when represented in terms of
$\hat{S}^\eta_\pm$ or $\hat{S}^\eta_z$, as their matrix elements
are also real.  With this perspective, it holds that complex
conjugation of the Hamiltonian, $\hat{H} \to \hat{H}^\ast$, is
equivalent to reversing the locations of the impurities
$\hat{d}_{\eta\sigma} \to \hat{d}_{-\eta,\sigma}$. Denoting the
latter by $R_I$, with $R_I H|a\rangle = H^\ast R_I |a\rangle$ it
holds that if $|a\rangle$ is an eigenstate of $H$ to eigenvalue
$E_a$, then so is $|a'\rangle \equiv [R_I |a\rangle]^\ast$.
Hence, with $K$ denoting complex conjugation, $\mathfrak{R}_I
\equiv R_I K$ is an anti-unitary symmetry of the system, with
$|a'\rangle = \mathfrak{R}_I |a\rangle$ also an eigenstate of
the Hamiltonian with the same eigenenergy.

As a consequence of the above symmetries, for example, the
spectral data of spin-spin correlations as in \Eq{eq:Aspec:C}
is real, after all.  Based on the Lehmann representation in
\Eq{eq:Aspec}, one encounters the matrix elements
$  \langle a|
     (\hat{d}_{\sigma}^\dagger \hat{d}_{\sigma'}^{\,})_\eta
   |b \rangle
   \langle b|
     (\hat{d}_{\sigma'}^\dagger \hat{d}_{\sigma }^{\,})_{\eta'}
   |a \rangle
$.
For the case that $\eta=\eta'$, i.e., intra-impurity spin
correlations, this product of matrix elements can be combined
into $|\langle a| .. |b \rangle|^2$, which is real.  For the case
of inter-impurity spin correlations, $\eta = -\eta'$, taking the
complex conjugate and inserting
$\mathfrak{R}_I^\dagger\mathfrak{R}_I^{\,}$ twice,
\begin{eqnarray}
   \langle a|
     (\hat{d}_{\sigma }^\dagger \hat{d}_{\sigma'}^{\,})_{\eta}
   |b \rangle^\ast
 &=& \langle b|
     (\hat{d}_{\sigma'}^\dagger \hat{d}_{\sigma }^{\,})_{ \eta}
   |a \rangle
\\
 &=& \underbrace{\langle b| (\mathfrak{R}_I^\dagger}_{ \equiv \langle b'|}
   \underbrace{\mathfrak{R}_I^{\,})
     (\hat{d}_{\sigma'}^\dagger \hat{d}_{\sigma }^{\,})_{\eta}
   (\mathfrak{R}_I^\dagger }_{
     =(\hat{d}_{\sigma'}^\dagger \hat{d}_{\sigma }^{\,})_{\eta'}
    }
   \underbrace{\mathfrak{R}_I^{\,}) |a \rangle}_{\equiv |a'\rangle}
\notag \\
 &=&
   \langle b'|
     (\hat{d}_{\sigma'}^\dagger \hat{d}_{\sigma}^{\,})_{\eta'}
   |a' \rangle
\notag
\end{eqnarray}
Now since $a$ and $b$ can be chosen to also be simultaneous
eigenstates of $\mathfrak{R}_I$, i.e., having $a=a'$ and $b=b'$
with the same eigenvalue w.r.t.  $\mathfrak{R}_I$, the Lehmann
sum of the spin-spin spectral functions, while having complex
matrix elements, yields a purely real result. In practice, $a$
and $b$ are not eigenstates of $\mathfrak{R}_I$.  Yet by
explicitly resorting to complete many-body basis sets within the
NRG \cite{Anders05} evaluated within fmd-NRG \cite{Wb07}, the
spectral data has an imaginary contribution with relative
strength comparable to numerical noise based on double precision
accuracy, and hence can be ignored.

\subsection{From Anderson to (anisotropic) Kondo type model}

The entire discussion above assumed Anderson-type impurities
that hybridize with the helical edge mode. Now if the local
Coulomb interaction $U$ with each impurity is large, charge
fluctuations get frozen out. Therefore in the low-energy regime,
charge fluctuations can be integrated out {\it locally} with
each impurity via Schrieffer Wolff transformation. Assuming
half-filling of each impurity with a single magnetic spin-half
moment, second order perturbation theory based on \Eq{eq:H:hyb}
yields the Kondo-type interaction
\begin{eqnarray}
   \hat{H}_\mathrm{K} \sim \sum_{\eta\in L,R}
   \tfrac{1}{-U}
   (\hat{H}_\mathrm{hyb}^\eta)^\dagger
   \hat{H}_\mathrm{hyb}^\eta
\label{eq:Kondo:0}
\end{eqnarray}
which is diagonal in $\eta$. When projecting the into the
low-energy Kondo regime of the Anderson model by scaling up
local interactions, this leaves the representation of the bath
untouched. Therefore from the NRG point of view the bath
remains completely unaffected by whether one resorts to an
Anderson-type or low-energy Kondo-type impurity setup.
Based on the coarse-grained version in \Eq{eq:H:hyb:disc} then,
\begin{eqnarray}
   \hat{H}_\mathrm{K} \sim 2J \sum_{\eta\in L,R}
   \hat{\bf{S}}_{\eta} \cdot \hat{\bf{S}}_{0\eta}^{\,}
\text{ ,}\label{eq:Kondo}
\end{eqnarray}
with $J$ the Kondo coupling, $\hat{\bf{S}}_{\eta} \equiv
 \hat{\bf{S}}_{d\eta}$ the spin operator of impurity $\eta$, and
$\hat{\bf{S}}_{0\eta} \equiv \sum_{\sigma\sigma'}
\tfrac{\boldsymbol{\tau}_{\sigma\sigma'}}{2}
\tilde{f}_{0 \eta\sigma}^\dagger
\tilde{f}_{0 \eta\sigma'}^{\phantom{\dagger}}$
the spin operator with respect to the bath site at the location
of impurity $\eta$. Therefore a single impurity interacting
with in a helical edge mode is identical to a regular 1-impurity
Anderson or Kondo model without a helical character. The
dynamically generated low-energy Kondo scale $T_\mathrm{K} \cong
\sqrt{\rho_0 J} e^{-\frac{1}{\rho_0 J}}$ \cite{Hewson93} for the
1-impurity problem with $\rho_0$ the one-particle density of
states around the Fermi level, however, also represents a
relevant energy scale for the 2-impurity problem.
Assuming the impurity distance on the grid
\eqref{eq:helical:xgrid}, the local bath operators $\tilde{f}_{0
\eta\sigma}$ are already properly orthonormalized, such that
there are no issues with crosstalk between the impurities.

For spin-independent hybridization between the Anderson
impurities and the bath, the resulting Kondo coupling in
\Eq{eq:Kondo} is SU(2) spin symmetric. While bearing in mind
that the helical bath mode itself already has the SU(2) symmetry
broken, the spin symmetry can also be broken at the level of the
Kondo Hamiltonian, giving rise to anisotropic local spin-spin
interactions. Assuming a single preferred direction ($z$) with
$J_x = J_y \equiv J \neq J_z$, $J_z > J$ ($J_z < J$ describes an
easy-axis (easy-plane) regime, respectively. The global U(1)
spin symmetry thus remains preserved.

\subsection{Effects of finite bandwidth with Kondo interaction
\label{App:finiteD}}

\begin{figure}[tb!]
\begin{center}
\includegraphics[width=1\linewidth]{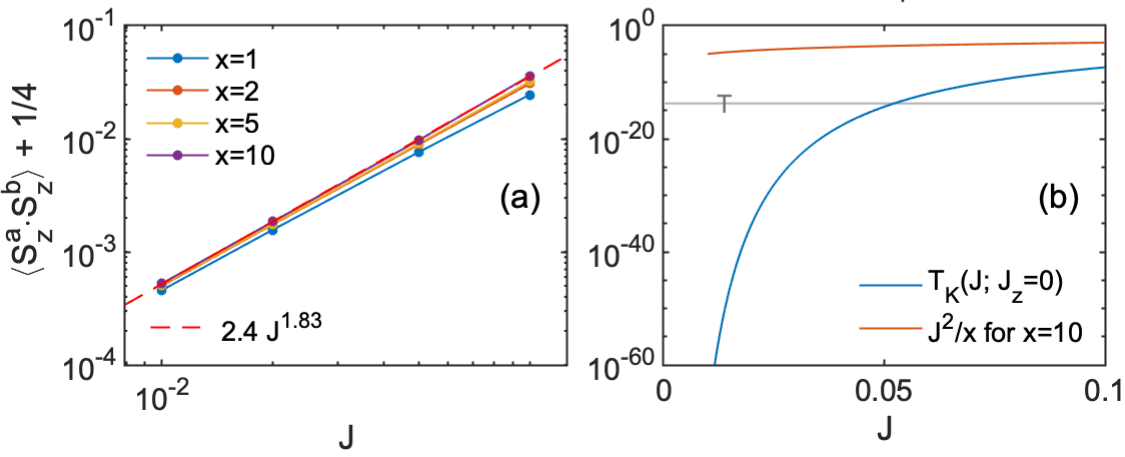}
\end{center}\vspace{-.1in}
\caption{
   Static $\langle S_L^z S_R^z \rangle$ impurity
   correlations for $\TK \ll \ER \sim \tfrac{J^2}{x}$ obtained
   by NRG towards the wide-band limit $D\gg J (=J_x=J_y)$
   for $J_z=0$.
(a) Deviations of the computed $\langle S_L^z S_R^z \rangle$
   from the expected RKKY value of $-1/4$ on a loglog plot.
   Data is shown for various impurity distances $x$, with
   the polynomial fit obtained for $x=10$ showing
   approximate quadratic behavior (red dashed line).
(b) Comparison of energy scales for the same parameter
   range as in (a) where the temperature $T\sim 10^{-9}$
   used in NRG simulations in (a) is indicated by the
   horizontal gray line. The data for the Kondo scale
   \TK was obtained via poor-man's scaling [cf.
   \App{App:TK-aniso}]. Red line shows a simple estimate
   for the RKKY energy at $x=10$, demonstrating that (a) is
   deeply in the RKKY regime, i.e., $\ER \gg \TK, T$.
}
\label{fig:RKKY-wband}
\end{figure}

When starting from the Anderson model, one may scale the
local Coulomb interactions properly to infinity in relation
to other parameters, with the result that also in the
numerical setting, one effectively arrives at the Kondo
model \cite{Hanl14}. For the Anderson model the effective
bandwidth relevant for the impurities is cut off by $U$ if
$U<D$. However, by taking $U\gg D$ when transitioning
towards the Kondo model, bandwidth keeps playing a
considerable role, and the universal wide-band limit is
approached rather slowly.

Here for the anisotropic 2HKM in the present case, if the Kondo
scale $\TK(J,J_z) \ll \ER \ll D=1$ for each impurity individually
is just several orders of magnitude smaller than all other
energy scales, effects of finite-bandwidth are still
considerable. In order to reach the wide-band scaling limit
(here $J\lesssim 0.02$), the single-impurity Kondo scale is
already {\it many many} orders of magnitude lower than the band
width $D=1$ ($\TK \lesssim 10^{-40}$), as demonstrated in
\Fig{fig:RKKY-wband}, and also consistent with the literature on
the single-impurity case \cite{Hanl14}. In the present case, we
see that similar arguments also carry over to the RKKY regime,
even if Kondo physics is irrelevant (in the sense that it sets
in at {\it much} smaller energy scales, even much smaller than
any temperature of practical interest).

In the RKKY regime and in the wide-band limit, the impurities
are expected to be well decoupled from the bath and described by
$\HR {=} -\ER \left( S^+_L S^-_R + \mathrm{H.c.} \right)$ [cf.
\App{App:RKKY:PT}] with the dynamically generated effective
direct RKKY impurity coupling \ER. The impurities are thus
expected in the $S_z^\mathrm{tot}{=}0$ triplet state $|0\rangle
\equiv \tfrac{1}{\sqrt{2}} |{\uparrow \downarrow} {+}
{\downarrow\uparrow} \rangle$ which yields the low-energy static
spin correlation $\langle S_L^z S_R^z \rangle = -1/4$. In
practice, however, one sees substantial deviations from this
expectation value up to nearly 20\% for $J=0.1$ in the NRG data
even using $J_z=0$ as shown in \Fig{fig:RKKY-wband}(a).
Hence these deviations must derive from higher-order processes
that go beyond second order PT [see \App{App:RKKY:PT}]. The
coupling to the bath remains finite in the low-energy regime,
thus inducing fluctuations in the impurity spin configuration.
These deviations can be reduced systematically by lowering the
Kondo coupling $J$, e.g., having a deviation already below 1\%
for $J \lesssim 0.02$. As shown in \Fig{fig:RKKY-wband}(a) the
static expectation value $\langle S_L^z S_R^z \rangle$
approaches $-1/4$ in a polynomial fashion as $J$ is lowered,
down to the smallest $J$ considered. Therefore, indeed, the
deviations seen in \Fig{fig:RKKY-wband}(a) are clearly due to
finite bandwidth. This demonstrates that finite bandwidth does
play an observable role in the Kondo setting \cite{Hanl14} when
comparing numerical to analytical results if the latter
strictly assumed the wideband limit. If one considers the
wide-band scaling limit reached within deviations in
observables of about $10^{-3}$, this suggests approximately $J
\lesssim 0.02$ in \Fig{fig:RKKY-wband}(a). On the single
impurity level, this already corresponds to astronomically
small Kondo scales $\TK \ll 10^{-40}$ in
\Fig{fig:RKKY-wband}(b) consistent with earlier NRG studies
\cite{Hanl14}. From a physics point of view, however, we do not
expect that the observed minor variations change the overall
physical picture in any significant qualitative manner.

\section{Poor-mans scaling for anisotropic Kondo
\label{App:TK-aniso}}

\begin{figure}[tb!]
\begin{center}
\includegraphics[width=1\linewidth]{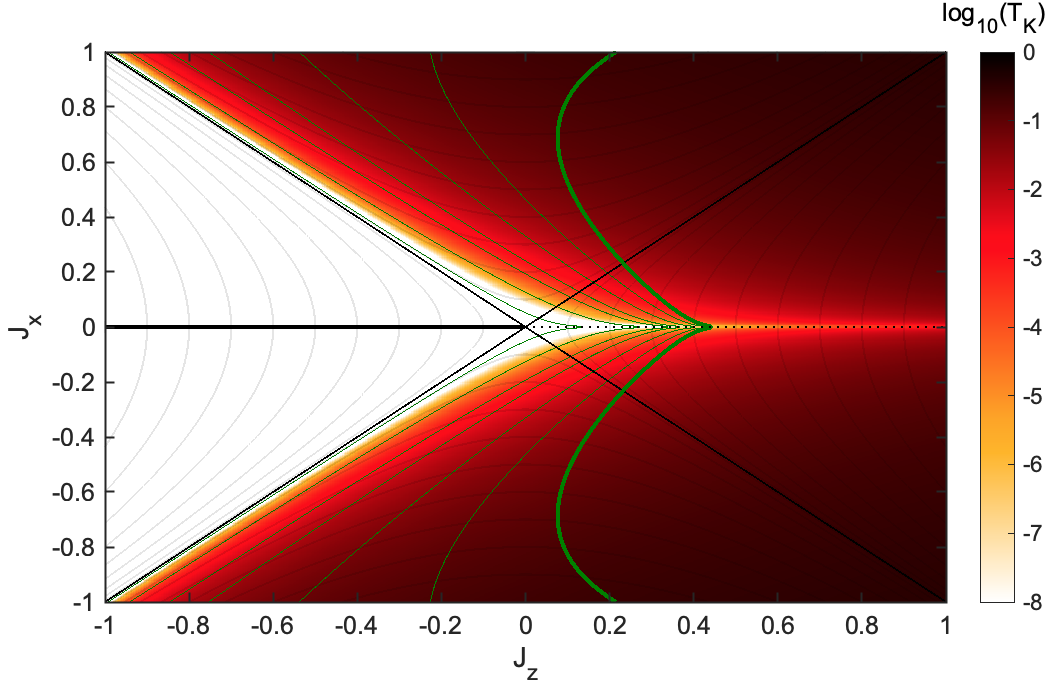}
\end{center}
\caption{Kondo scale of the anisotropic Kondo model from
  poor Man's scaling [\Eqs{eq:anisoK:z} and
  \eqref{eq:anisoK:xy}]. $J_x<0$ is physically equivalent
  to $J_x>0$, such that the lower half of the panel is a
  mirror image of the upper half. The thick black
  horizontal line ($J_z<0$ for $J_x=0$) represents a stable
  line of fixed points without a Kondo scale. The entire
  white region below the diagonal lines to the left flows
  towards it. The black dotted horizontal line ($J_z>0$ for
  $J_x=0$) represents a line of unstable fixed points that
  flow to strong coupling for any small but finite $J_x$
  (strictly at $J_x=0$ one has $\TK=0$, represented by white
  color). The gray lines represent RG flow contours as in
  \Eq{eq:anisoK:contours}. The Kondo temperature is defined
  by the starting point $(x_0,z_0)$ on such a contour.
  The Kondo scale increases monotonically from left to
  right, and also with increasing $|J_x|$.
  The green contours show lines
  of constant $T_K(J,J_z) - (\rho_0 J)^2 \leq 0$.
  The thick green contour to the right
  describes $T_K(J,J_z) = (\rho_0 J)^2 = \ER(x=1)$
  which represents the largest
  RKKY energy possible by putting two impurities
  right next to each other, with the closest
  distance being one `lattice spacing' [cf. \Eq{E-RKKY}].
}
\label{fig:TK-aniso}
\end{figure}

The poor man's scaling equation
for the renormalization group (RG) flow of
the anisotropic Kondo model
are given by \cite{Hewson93,Kogan18,KanaszNagy18}
\begin{eqnarray}
   \tfrac{\partial}{\partial \ln D} (\rho_0 J_x)
   = - (\rho_0 J_y) (\rho_0 J_z)
\text{ ,}
\end{eqnarray}
and similarly for the other components using cyclic permutations
of $(x,y,z)$.  In particular, with $x \equiv \rho_0 J_x = \rho_0
J_y$ and $z \equiv \rho_0 J_z$,
\begin{eqnarray}
   \tfrac{\partial}{\partial \ln D} \,x &=& - x z \notag \\
   \tfrac{\partial}{\partial \ln D} \,z &=& - x^2
\text{ ,}\label{eq:anisoK:flow}
\end{eqnarray}
It follows from the above that
\begin{eqnarray}
   -\tfrac{d z}{x^2} = d \ln D
 = -\tfrac{d x}{x z}
 \quad &\Rightarrow&
   z \, dz = x \, dx
\notag \\
&\Rightarrow&
   z^2 = x^2 \pm a^2
\label{eq:anisoK:contours}
\end{eqnarray}
where $\pm a^2$ with $a\geq 0$ is some constant of integration.
If the starting point has $|z_0|>|x_0|$ (easy-axis), then the
positive sign is chosen for $a^2$, whereas the regime
$|z_0|<|x_0|$ (easy-plane) has the negative sign. The contours
described by \Eq{eq:anisoK:contours} exactly reflect the RG
paths of the anisotropic Kondo (parabolas or hyperbolas
separated by $|x_0| = |z_0|$, as shown by the gray lines in
\Fig{fig:TK-aniso}). It simply also follows from
\Eq{eq:anisoK:flow} that for $x\to 0$, $(x,z)$ stops flowing.
The model is physically equivalent for $x_0 \to -x_0$ (vertical
flip in \Fig{fig:TK-aniso}) as this can be absorbed into a gauge
transformation of the spin basis.

For $z_0>0$ and $|x_0|>|z_0|$, the anisotropic Kondo model
flows to the isotropic strong coupling regime. That is
integrating out the bath starting from large $D$ needs to be
stopped at some $D^\ast > 0$ where $x$ and $z$ diverge,
which thus defines the dynamically generated low-energy
Kondo scale $\TK := D^\ast$.

In the easy-axis Kondo regime ($J_z > |J_x|$), with $z \ge a$,
the RG differential equations diverge at
\begin{eqnarray}
  \ln \tfrac{D^\ast}{D_0} = \tfrac{1}{2a} \ln
     \tfrac{z_0-a}{z_0+a}
  \ , \qquad a = \sqrt{z_0^2 - x_0^2}
\label{eq:anisoK:z}
\end{eqnarray}
The Kondo scale vanishes to exponentially small energy
scales for $J_x \to 0$, i.e., $a \to z_0^-$.

In the easy-plane regime ($|J_x| > |J_z|$), with $|x| \geq a$,
the RG differential equations diverge at
\begin{eqnarray}
  \ln \tfrac{D^\ast}{D_0} = \tfrac{1}{a} \bigl(
     \tan^{-1} \tfrac{z_0}{a} - \tfrac{\pi}{2}
  \bigr)
  \ , \quad a = \sqrt{x_0^2 - z_0^2}
\label{eq:anisoK:xy}
\end{eqnarray}
Both Kondo scales, \Eq{eq:anisoK:z} and \Eq{eq:anisoK:xy}
connect smoothly across $J_x = J_z > 0$, where
\begin{eqnarray}
  \ln \tfrac{D^\ast}{D_0} = -\tfrac{1}{z_0}
   \qquad (a \to 0)
\label{eq:isoK}
\end{eqnarray}
thus having $\TK^\mathrm{isotropic}\!\simeq D_0\,
e^{-\frac{1}{\rho_0 J}}$. Here the Kondo scale remains finite
if $J_z \to 0$,
\begin{eqnarray}
  \ln \tfrac{D^\ast}{D_0} = -\tfrac{\pi}{2 x_0}
   \qquad (a \to x_0)
\label{eq:anisoK:xy0}
\end{eqnarray}
thus having $\TK^{J_z=0} \simeq D_0\, e^{-\pi/2\rho_0 J_x}$.

For $J_z < -|J_x|$ the system flows to the ferromagnetic regime
with $J_x \to 0$ without any Kondo physics. That is $D$ can be
integrated out all the way to zero, which thus implies $D^\ast =
\TK = 0$.

\section{Basic comparison Kondo vs. RKKY energy
\label{App:TKvsER}}

The single-impurity Kondo scale from the previous section may be
compared to the RKKY energy \ER [cf. \Eq{E-RKKY}]. The latter
decays with distance, hence is largest for short distances, with
the shortest possible distance given by one `lattice spacing'
[cf. \Eq{eq:lattice:a}], hence $\ER \leq (\rho_0 J)^2$ in the
adopted units [\Sec{sec:H0}]. With this in mind, \Fig{fig:TK-aniso}
also includes contours of constant $T_K(J,J_z) - (\rho_0 J)^2
\leq 0$ (green lines), with the thick line representing
$T_K(J,J_z) = (\rho_0 J)^2$. Therefore the parameter regime to
the right of the thick green contour always has $T_K > \ER$ for
any $x\geq1$, meaning that in this regime Kondo is always
dominant over RKKY. This regime has been predicted in
Ref.\,\cite{Yevt_2018}. In particular, this includes the
isotropic Kondo for $J\gtrsim 0.25$, corresponding to a
dimensionless coupling $j_0 \equiv \rho_0 J = 0.125$. While
still clearly below 1., this is on the upper end of what may be
considered acceptable on physical grounds for the Kondo model:
Note that the poor-man's scaling approach for the Kondo model
only includes second order virtual processes by integrating out
the bath which assumes (and thus is justified only if)
$\tfrac{J^2}{D} \ll D$, i.e., $j_0 \ll 0.5$. Conversely, note
that coming from an Anderson model, one has $j_0^\mathrm{eff}
\simeq \tfrac{4\Gamma}{\pi U}$ with hybridization strength
$\Gamma$ and onsite interaction $U$. So in order for the Kondo
model to be justified in the first place, the onsite interaction
$U$ needs to be large enough so that charge fluctuations can be
integrated out via virtual second order processes (cotunneling)
in order to obtain a pure effective spin Hamiltonian. This
clearly requires $U\gg\Gamma$, and thus also $j_0 \ll 1$ on
physical grounds.

Conversely, to the left of the thick green line in
\Fig{fig:TK-aniso}, RKKY can dominate in the low-energy regime
if a pair of impurities is just brought close enough to each
other. In particular, RKKY also occurs within the entire white
region to the left where $\TK=0$. With $\ER > \TK=0$ then, the
impurity distance $x$ can be taken to infinity while still
seeing RKKY in the low-energy regime (note that since RKKY is
second order, with $\ER \propto J^2/x$, the sign of the Kondo
couplings is irrelevant).

\section{Second order perturbation processes entering RKKY
\label{App:RKKY:PT}}

Complimentary to the field theoretic approach in the main text,
the RKKY processes can also be analyzed from the point of view
second order perturbation theory (PT). Ultimately, this
generates the same effective Hamiltonian as in the field
theoretic approach, but purely within the Hamiltonian
setting, e.g. via the Feshbach formalism
\cite{Feshbach58,Fano61}.

Let the full Hamiltonian of two Kondo-coupled impurity spins
$\eta \in \{\iR,\iL\} \equiv \{ +1 , -1 \}$ at location
$\frac{\eta x}{2}$, i.e., at mutual distance $x$,
interacting with a one-dimensional helical bath encoded by
the Fermionic operators $\hat{c}_{k\sigma}$ be given by
\begin{subequations} \label{eq:RKKY:tb:H}
\begin{eqnarray}
   \hat{H} &=& \!\!\!
   \sum_{\eta a} 2J_a \
     (\hat{S}_\eta^{a})^\dagger \cdot \hat{s}_\eta^{a}
    + \sum_{k\sigma}
      \varepsilon_{k\sigma}^{\,}
      \hat{c}_{k\sigma}^\dagger \hat{c}_{k\sigma}^{\,}
\label{eq:RKKY:tb:H.1}
\end{eqnarray}
with $a \in x,y,z$, and
bath spins (with $\tau_a$ the Pauli matrices),
\begin{eqnarray}
   \hat{s}_\eta^{a}
 \equiv \tfrac{1}{2} \hat{\sigma}_\eta ^{a}
 = \tfrac{1}{N} \!\!\!
   \sum_{k\sigma, k'\sigma'} \!\!\!\bigl(
   \hat{c}_{k\sigma}^\dagger \,\tfrac{1}{2} \tau_a^{\sigma\sigma'}
   \hat{c}_{k'\sigma'}^{\,}
   \bigr)
   e^{\iu (k'-k)\frac{\eta x}{2}}
\text{,}\quad\label{eq:RKKY:tb:H.2}
\end{eqnarray}
assuming a total of $N$ levels $k$ for each spin. Furthermore,
the energies of the bath $\varepsilon_{k\sigma} = \sigma v k$
are constrained to within a finite half-bandwidth
$|\varepsilon_{k\sigma}| \leq D$.
Taking $J {\equiv} J_x {=} J_y$ and $\hat{\sigma}_\eta
^{\pm} \equiv \hat{s}_\eta ^{\pm} \equiv \hat{s}_\eta ^{x}
\pm \iu \hat{s}_\eta ^{y}$, the Kondo interaction can be
written from the point of view of the helical Fermions as,
\begin{eqnarray}
   2J_a (\hat{S}_\eta^a)^\dagger \cdot \hat{s}_\eta ^{a}
&\equiv& J\, \bigl(\hat{S}_\eta ^{+} \hat{\sigma}_\eta ^{-}
   + \mathrm{H.c.}\bigr)
 + J_z\ \hat{S}_\eta ^{z} \hat{\sigma}_\eta ^{z}
\notag\\
&\equiv& \tfrac{J}{N} \!\!\!\sum_{k\sigma, k'\sigma'}\!\!\! \bigl(
   \hat{c}_{k\sigma}^\dagger \hat{T}^\eta_{\sigma\sigma'} \,
   \hat{c}_{k'\sigma'}^{\,}\bigr)
   e^{\iu (k'-k)\frac{\eta x}{2}}
\notag
\end{eqnarray}
with the matrix notation in $\sigma {\in} \{\uparrow, \downarrow \}
{\equiv} \{ 1, 2\}$ for the operators of each impurity
[see also \Eq{eq:V:def}],
\begin{eqnarray*}
  \hat{T}^{\eta} &=&
  \begin{pmatrix}
     \Delta \hat{S}^{z}_{\eta} & \hat{S} ^{-}_{\eta} \\
     \hat{S} ^{+}_{\eta} & -\Delta \hat{S} ^{z}_{\eta}
  \end{pmatrix}
  \equiv \hat{S}^-_{\eta} {\otimes} \tau_+
  + \hat{S}^+_{\eta} {\otimes} \tau_- + \Delta \hat{S}^z_{\eta} {\otimes} \tau_z
\text{ ,}
\end{eqnarray*}
\end{subequations}
with anisotropy $\Delta {\equiv} \tfrac{J_z}{J}$ and
$\tau_\pm = \tfrac{1}{2}(\tau_x \pm \iu \tau_y)$. The
isotropic case $\Delta{=}1$ just reduces to $\hat{T}^\eta
= \hat{S}^{\dagger}_\eta \cdot \tau$.

The effective Hamiltonian in between the impurities, i.e.,
their direct interaction, is generated via second order PT
in the Kondo setting ($4^\mathrm{th}$ order in the hybridization
for an Anderson model). Then with the shortcuts $1 \equiv
(k_1\sigma_1)$, etc., a typical second order process is given by
\begin{widetext}
\begin{eqnarray}
   \delta \hat{H}(x)
 &\sim& P_0 \,
   \bigl(
     \hat{c}_{1}^\dagger  \hat{T}_{\sigma_1\sigma_2}^{\eta}
     \hat{c}_{2}^{\,}\,
     e^{\iu (k_2-k_1)\frac{\eta x}{2}} \bigr)\
   \tfrac{(J/N)^2}{\iu 0^+  - (\varepsilon_{3} - \varepsilon_{4})}\
   \bigl(
     \hat{c}_{3}^\dagger  \hat{T}_{\sigma_3\sigma_4}^{\eta'}
     \hat{c}_{4}^{\,}\,
     e^{\iu (k_4-k_3)\frac{\eta' x}{2}} \bigr)
   \,P_0
\label{eq:RKKY:dH}
\end{eqnarray}
\end{widetext}
with degrees of freedom such as impurity location $\eta
{\in} \{L,R\} {\equiv} \{-1,1\}$, or spin $\sigma$ or
momenta $k$ to be summed over eventually.
$P_0$ is a projector into the target low-energy regime of the
bath without acting at the impurities. This is in the spirit
of generating the low-energy effective Hamiltonian such as
the Feshbach-Fano partitioning \cite{Feshbach58,Fano61}.
There, by construction, one needs to start out of the
low-energy regime, say by considering some infrared cutoff
energy $D^\ast \ll D$. In the low-energy regime described by
$P_0$ then, all bath levels with energy $\varepsilon_{k\sigma}
< -D^\ast$ are strictly occupied and all energy levels
$\varepsilon_{k\sigma} > +D^\ast$ are strictly empty. Now in
\Eq{eq:RKKY:dH}, the perturbation $\hat{T}$ on the r.h.s.
generates a particle-hole (p/h) pair with particle energy
$\varepsilon_3 > 0$ and hole energy $\varepsilon_4 <0$.
Therefore as long as either $\varepsilon_3 > +D^\ast$ or
$\varepsilon_4 < -D^\ast$ (or both), this represents state
space outside the low-energy regime $P_0$.

If $D^\ast \ll D$, the overwhelming number of processes will
involve both, particle and hole at energy cost above
$D^\ast$. There in order to return to the low energy regime,
exactly the same p/h pair must be destroyed again. This
constrains the second processes above to $\delta_{1 4}
\delta_{2 3}$, with the shortcut notation $\delta_{i j}
\equiv \delta_{k_i k_j} \delta_{\sigma_i \sigma_j}$.
Actually, the processes $\delta_{1 4} \delta_{2 3}$ can be
included all the way down to zero energy, i.e., there is
apriori no need for an infrared energy cutoff $D^\ast$,
so one can assume $D^\ast \to 0^+$ as long as the sum is
well-defined (which it turns out it is). With the bath
projected via $P$ onto the exactly filled helical Fermi sea,
this generates the effective, purely local inter-impurity
Hamiltonian
\begin{widetext}
\begin{eqnarray}
   \hat{H}^\mathrm{eff}_\mathrm{imp} (x)
 &=&
   \sum_{\eta\eta'}
   \sum_{k_1 \sigma_1}^{\varepsilon_{1}<0}
   \sum_{k_2 \sigma_2}^{\varepsilon_{2}>0}
   \bigl(
     \hat{T}_{\sigma_1\sigma_2}^{\eta} \,
     e^{\iu (k_2-k_1) x_\eta} \bigr)\
   \tfrac{(J/N)^2}{\iu 0^+  - (\varepsilon_{2} - \varepsilon_{1})}\
   \bigl(
     \hat{T}_{\sigma_2\sigma_1}^{\eta'} \,
     e^{\iu (k_1-k_2) x_{\eta'}} \bigr)
\notag \\[-1ex]
 &=&
    -(\rho_0 J)^2 \sum_{\substack{\eta\eta' \\ \sigma_1\sigma_2 }}
     \hat{T}_{\sigma_1\sigma_2}^{\eta }
     \hat{T}_{\sigma_2\sigma_1}^{\eta'}
     \underbrace{
      \int\limits_{-D}^0 d\varepsilon_1
      \int\limits_{ 0}^D d\varepsilon_2\,
      \tfrac{e^{\iu (\sigma_2\varepsilon_2 - \sigma_1\varepsilon_1)
      \frac{1}{v} (x_\eta - x_{\eta'})}}%
         {-\iu 0^+  + (\varepsilon_{2} - \varepsilon_{1})}
   }_{\equiv I^{\eta\eta'}_{\sigma_1\sigma_2}}
\label{eq:HRKKY:0}
\end{eqnarray}
\end{widetext}
having taken the continuum limit $i \equiv (k_i\sigma_i) \to
(\varepsilon_i \sigma_i)$, with $\rho_0$ the constant one-particle
density of states of the helical edge for a given spin at the
Fermi level as in \Eq{eq:lattice:a}.
The phase factor in the second line
rewritten in terms of energies
is already also specific to the helical edge.
As an aside, we note that
the energy denominator of $I^{\eta\eta'}_{\sigma_1\sigma_2}$
can be rewritten into an integral over imaginary frequencies,
\begin{eqnarray}
  \int \tfrac{d\omega}{2\pi\iu}\,
  \tfrac{1}{\iu\omega - \varepsilon_1}
  \tfrac{1}{\iu\omega - \varepsilon_2}
= \tfrac{\theta(-\varepsilon_1 \varepsilon_2)}%
  {|\varepsilon_2 - \varepsilon_1|}
\text{ ,}
\end{eqnarray}
which resembles a Matsubara summation at zero temperature. Via
contour integral, the result is non-zero only if $\varepsilon_1$
and $\varepsilon_2$ have opposite sign as encoded into the
Heaviside step function on the r.h.s. This already reflects the
particle-hole nature of the excitations in \Eq{eq:HRKKY:0} where
$\varepsilon_1 < 0$ and $\varepsilon_2 > 0$. The case with
opposite signs is included in \Eq{eq:HRKKY:0} via the overall
sum. Assuming $\eta {\neq} \eta'$, then the case $\eta
\leftrightarrow \eta'$ gives the additional contribution above,
while also exchanging the labels $1 \leftrightarrow 2$, such
that the sign in the phase factor is properly restored, while
also having $[\hat{T}^\eta, \hat{T}^{\eta'}] {=} 0$ for $\eta
{\neq} \eta'$. Hence by summing over all second order
processes, one can connect the second order perturbative
approach here to the double-propagator structure in
\Eq{Spin-action-expanded} in the main text obtained from the
field-theoretic approach. Collecting phase factors in $k_1$
and $k_2$, the latter thus permits the interpretation that any
second order process in the effective impurity Hamiltonian
requires the free propagator of two particles shuttling back
and forth in between the impurities (one particle needs to
propagate the distance $+x$, and another the distance $-x$).
Here in the helical setting, however, the directions that
particles can move are constrained depending on their spin.
This manifests itself in the overall structure of the resulting
RKKY Hamiltonian \cite{Yevt_2018}.

The integral $I^{\eta\eta'}_{\sigma_1\sigma_2}$ in
\Eq{eq:HRKKY:0} is generally well defined when both,
$\varepsilon_1$ and $\varepsilon_2$ approach zero energy,
irrespective of the phase factor since
\begin{eqnarray}
   \bigl| I^{\eta\eta'}_{\sigma_1\sigma_2} \bigr|
 &\le&
   \int_{-D}^0 \!\!\! d\varepsilon_1
   \int_0^D    \!\!\! d\varepsilon_2\
   \bigl|\tfrac{1}{-\iu 0^+ + (\varepsilon_2 - \varepsilon_1)} \bigr|
 = 2D \ln 2
\text{ .}\qquad \label{eq:RKKY:Iabs}
\end{eqnarray}
The imaginary part form $\iu 0^+$ does not contribute, as by the
integral limits, it can only contribute at $\varepsilon =
\varepsilon' = 0$, where by the double integral the integrated
weight vanishes.

When $\eta {=} \eta'$, the complex phases drop out, and with
$S_\eta^2 \propto 1\!\!1$ the integral in \Eq{eq:HRKKY:0} just
generates a plain irrelevant shift in the global energy
reference as estimated above (the generation of the single
impurity Kondo couplings needs to consider finite $D^\ast$ and
relax the condition $\delta_{1 4} \delta_{2 3}$ above). Hence
the following discussion focusses on $\eta {\neq} \eta'$, i.e.,
$\eta' {=} -\eta$, in which case the phase factors in the
enumerator are non-trivial and reflect the underlying helical
physics. In contrast to the energy denominator, however, the
phase factors carry momenta as arguments. Converting these into
energies, in the helical setting with $\varepsilon_{k\sigma} =
\sigma v k$, this involves signs depending on the spin as
already indicated with the integral
$I^{\eta\eta'}_{\sigma_1\sigma_2}$ in \Eq{eq:HRKKY:0}.

Contributions that involve a spin flip ($\sigma_2 {=} -
\sigma_1$), will have opposite relative sign of $\varepsilon_1$
vs. $\varepsilon_2$ in the phase factor in \Eq{eq:HRKKY:0} as
compared to the energy denominator, whereas in the absence of a
spin flip the relative sign is the same.
To evaluate this integral including the phases, it is therefore
convenient to change variables to symmetric and antisymmetric
combinations
\begin{eqnarray}
   \begin{pmatrix} \varepsilon' \\ \varepsilon \end{pmatrix}
   \equiv
   \underbrace{
   \begin{pmatrix} \ 1 & 1 \\ \!\!-1 & 1 \end{pmatrix}}_{\equiv U}
   \begin{pmatrix} \varepsilon_1 \\ \varepsilon_2 \end{pmatrix}
\text{ .}
\end{eqnarray}
With $\varepsilon_1 < 0$ and $\varepsilon_2>0$, by construction,
$0 \le \varepsilon \le 2D$ and $\varepsilon' \in [-\varepsilon,
\varepsilon]$. The integral will converge with large $D$, such
that the upper integral limit can be taken more loosely by
deforming the integration area to $\varepsilon \lesssim
\tilde{D} =\frac{3D}{2}$, with the upper integral limit
$\tilde{D}$ assumed large, and eventually taken to infinity if
well-defined.  With this one obtains for the case including a
spin flip, with $\tilde{\tau} \equiv \frac{\sigma_2}{v} (x_\eta
- x_{\eta'})$, and thus $\tau \equiv |\tilde{\tau}| = \tfrac{x}{v}$
having $\eta\neq \eta'$, with $x = |x_\eta - x_{\eta'}|$,
\begin{eqnarray}
 &&I^{\eta\neq\eta'}_{\sigma_1\neq\sigma_2}
 =
    -\int\limits_{-D}^0 \!d\varepsilon_1
     \int\limits_{ 0}^D \!d\varepsilon_2\,
     \tfrac{e^{\iu \tilde{\tau} (\varepsilon_2 + \varepsilon_1) }}
      {\iu 0^+  - (\varepsilon_{2} - \varepsilon_{1})}
\label{eq:HRKKY:I1} \\
&&\simeq -\underbrace{\tfrac{1}{\det U}}_{=\frac{1}{2}}
  \int\limits_{0}^{\tilde{D}} \!d\varepsilon \underbrace{
  \int_{-\varepsilon}^{\varepsilon}   \!d\varepsilon'\,
  \tfrac{e^{\iu \tilde{\tau} \varepsilon'}}{\iu 0^+  - \varepsilon}
  }_{
   = \tfrac{1}{\iu\tilde{\tau}}
     \tfrac{2\iu \sin \tilde{\tau}\varepsilon}{\iu 0^+  - \varepsilon}
  }
= \tfrac{1}{\tilde{\tau}} \underbrace{
  \int\limits_{0}^{\tilde{D}} \!d\varepsilon \,
     \tfrac{\sin \tilde{\tau}\varepsilon}{\varepsilon}
  }_{\simeq\ \frac{\mathrm{sgn}\tilde{\tau}}{2} \pi}
\simeq \tfrac{\pi}{2\tau}
= \tfrac{\pi v}{2x}
\notag
\end{eqnarray}
where the contribution from the imaginary part $\iu 0^+$ in
the denominator vanishes, as this gives $\int d\varepsilon\,
\delta(\varepsilon) \sin \tau\varepsilon \to 0$. The remaining
integral is well-defined and converges for $|\tau D| =
\frac{xD}{v} \gg 1$ to $\pi/2$ as indicated. Combining all
$\sigma_1 =- \sigma_2$ terms adds Hermitian conjugate,
whereas summing over $\eta =- \eta'$ just duplicates
entries. Thus the RKKY Hamiltonian for the helical setting
generated by second order processes becomes
\begin{eqnarray}
   \HRh (x) &=& -\underbrace{
   (\underbrace{\rho_0 J}_{\equiv j_0})^2 \!\!
    \underbrace{\tfrac{\pi v}{x}}_{= \pi \Ex}}_{ = \ER}
   \bigl(
     \hat{S}^{+}_{\iL} \hat{S}^{-}_{\iR}
    +\hat{S}^{-}_{\iL} \hat{S}^{+}_{\iR}
   \bigr)
\label{eq:RKKY:PT}
\end{eqnarray}
with the dimensionless Kondo coupling strength $j_0$.
Overall, this generated direct impurity interaction is
ferromagnetic and non-oscillatory with a plain decay with
inverse distance $x$, having the RKKY energy scale
\begin{eqnarray}
   \ER = \pi j_0^2 \Ex
\text{ ,}\label{eq:ER:Ex}
\end{eqnarray}
as already encountered in \Eq{H-RKKY} in the main text. The
RKKY Hamiltonian is independent of the bandwidth $D$, and aside
from the dimensionless Kondo coupling strength $j_0$, only
references the coherence scale $\Ex \equiv \tfrac{1}{\tau}$ with
$\tau = x/v$ the time required for a helical particle to travel
from one impurity to the other. With the lattice spacing in
\Eq{eq:lattice:a}, nevertheless, this energy scale may be
rewritten as $\Ex = \tfrac{D/\pi}{x/a}$. That is when measuring
distance on the grid \eqref{eq:helical:xgrid}, by definition,
this involves a finite bandwidth, such that the bandwidth does
appear in the definition of $\Ex$. In the continuum wide-band
limit, however, the natural way to think about the coherence
scale is $\Ex = 1/\tau$ without any reference to bandwidth.

In the absence of a spin flip, i.e., $\sigma_1 = \sigma_2$
the integral can be similarly evaluated,
\begin{eqnarray}
   I^{\eta\neq\eta'}_{\sigma_1=\sigma_2}
   &=&-\int\limits_{-D}^0 \!d\varepsilon_1
     \int\limits_{ 0}^D \!d\varepsilon_2\,
     \tfrac{e^{\iu \tilde{\tau} (\varepsilon_2 - \varepsilon_1) }}
      {\iu 0^+  - (\varepsilon_{2} - \varepsilon_{1})}
\label{eq:HRKKY:I2} \\[-2ex]
&\simeq& -\underbrace{\tfrac{1}{\det U}}_{=1/2}
  \int_{0}^{\bar{D}} \!d\varepsilon \underbrace{
  \int_{-\varepsilon}^{\varepsilon}   \!d\varepsilon'\,
  }_{ = 2\varepsilon }
  \tfrac{e^{\iu \tilde{\tau} \varepsilon}}{\iu 0^+  - \varepsilon}
= \int\limits_{0}^{\bar{D}} \!d\varepsilon\,
  e^{\iu \tilde{\tau} \varepsilon}
\text{ ,} \notag
\end{eqnarray}
with some simple averaged upper bound $\bar{D} \sim
\tfrac{3D}{2}$, assuming that the integral is well-defined
in the wideband-limit.
Again the contribution from the imaginary part $\iu 0^+$
in the denominator vanishes, since $\int d\varepsilon\,
\delta(\varepsilon)\, \varepsilon e^{\iu\tilde{\tau}\varepsilon}
\to 0$, as already expected from \Eq{eq:RKKY:Iabs}.
By summing over spin or location, with $\tilde{\tau} \propto
\eta \sigma_2$, only the real part remains,
$ \sum_{\eta\sigma_1} I^{\eta=-\eta'}_{\sigma_1=\sigma_2}
  \simeq \tfrac{4}{\tau} \sin \tau \bar{D}$
which seems to suggest $1/x$ behavior similar in magnitude to
\Eq{eq:RKKY:PT}. In the present case, however, the integral
remains sensitive on the bandwidth $D$. Therefore the
assumption for introducting $\bar{D}$ above does not hold,
and the integral needs to be evaluated more carefully.
The exact representation of the integral in the first
line of \Eq{eq:HRKKY:I2} yields with
$\tilde{z} \equiv \tilde{\tau} \varepsilon$,
$\tilde{\lambda} \equiv \tilde{\tau} D$, and
$\lambda \equiv |\tilde{\lambda}| = \tau D$,
\begin{eqnarray}
  \sum_{\eta\sigma_1}
  I^{\eta\neq\eta'}_{\sigma_1=\sigma_2}
&=& 4 \re \Bigl( \int\limits_{0}^{D} \!d\varepsilon\,
  e^{\iu \tilde{\tau} \varepsilon}
+ \int\limits_{D}^{2D} \!d\varepsilon\,
  \tfrac{2D -\varepsilon}{\varepsilon}
  e^{\iu \tilde{\tau} \varepsilon}
  \Bigr)
\notag \\[-2ex]
&=& (1 - e^{\iu \tilde{\tau} D})
  \underbrace{ \tfrac{4}{\tau}
  \sin \tau D}_{\to\ 0}
  \,+\,
  8D \hspace{-1ex}\underbrace{\re\!\!
    \int\limits_{\tilde{\lambda}}^{2\tilde{\lambda}} \!\!
    \tfrac{d\tilde{z}}{\tilde{z}}\, e^{\iu \tilde{z}}
  }_{= {\rm Ci}(\lambda) - {\rm Ci}(2\lambda) }
\ \qquad \label{eq:Ilambda:zz:1}
\end{eqnarray}
with ${\rm Ci}()$ the {\it cosine integral} function.
The first term vanishes on the grid \eqref{eq:helical:xgrid}
having $|\lambda|=\tau D = \pi x / a$ a multiple of $\pi$.
As apparent from the oscillating averaging structure
in the second term, it also vanishes for large $\lambda$.
In the asymptotic form for large $\lambda$,
the leading term of the cosine integral
${\rm Ci}(\lambda) \sim \frac{\sin \lambda}{\lambda}$
again drops out on the grid \eqref{eq:helical:xgrid}.
Therefore the subleading term
${\rm Ci}(\lambda) \sim \frac{\cos \lambda}{\lambda^2}$
becomes the dominant one for large $\lambda$.
But with $D {\rm Ci}(\lambda) \sim \frac{D}{\lambda^2} \sim
\frac{1}{D x^2}$, this does not just decay faster over
distance as compared to RKKY for a normal 1D metallic mode,
but is also suppressed in the wide-band limit.
Therefore as expected the $ZZ$
contribution to the RKKY Hamiltonian properly vanishes for
$D\to\infty$ even for finite $J_z$.
At finite bandwidth, however, there is a finite return
probability, resulting in a small but finite $S_{\iL}^z
S_{\iR}^z$ interaction strength across the impurities.
This may be considered acceptable on physical grounds, bearing
in mind that for a true 2D model the spin-dependent directedness
of motion only concerns the edge but not the bulk states gapped
out to energies $>D$.

Finally, we point out that above integrals also appear in
the theory of a normal metallic edge when one assumes the same
linear dispersion and finite bandwidth as for the helical system
here. However, by having additional branches of the opposite
helicity in the dispersion for back-propagation, the same
integrals appear and are summed over all spin interactions XX,
YY, and ZZ.  This way the RKKY interaction becomes isotropic for
a normal metallic edge.  The above argument that the leading
oscillatory term vanishes is particular to the one-particle
dispersion chosen here, and would also apply to the normal
metallic edge with the dispersion indicated.  This differs
crucially from a system of free particles in a Fermi sea with a
quadratic dispersion and a finite Fermi energy.  In this case,
the analytic structure of the respective integrals is different.
As a consequence, this permits the leading $2k_f$ oscillatory
term $\propto 1/x$ to be present in a normal metallic 1D mode
\cite{Yafet87}.

\bibliography{refs}

\end{document}

%% file: shortcuts.tex
\newcommand{\citex}[1]{%
\begin{NoHyper}\citeauthor{#1} (\citeyear{#1})\,\end{NoHyper}%
\cite{#1}}

  \newcommand{\func}[1]{{\,\rm{#1}\,}}

  \newcommand{\Sec}[1]{Sec.~\ref{#1}}
  \newcommand{\App}[1]{App.~\ref{#1}}
  \newcommand{\Eq}[1]{Eq.~\eqref{#1}}
  \newcommand{\Eqs}[1]{Eqs.~\eqref{#1}}
  \newcommand{\EQ}[1]{Equation~\eqref{#1}}
  \newcommand{\Fig}[1]{Fig.~\ref{#1}}
  \newcommand{\Figs}[1]{Figs.~\ref{#1}}

  \newcommand{\rsym}[1]{\ensuremath{%
  \mathfrak{r}_{\mathrm{\ifx\@empty{}\else{#1}\fi}}}}

  \def\re{\ensuremath{{\func{Re}}\,}}
  \def\im{\ensuremath{{\func{Im}}\,}}
  \def\iu{\ensuremath{\mathrm{i}\mkern1mu}}

  \def\etaI{\eta}
  \def\etaB{\sigma}

  \def\iL{\ensuremath{\mathcal{L}}\xspace}
  \def\iR{\ensuremath{\mathcal{R}}\xspace}

  \def\TK{\ensuremath{T_{\rm K}}\xspace}
  \def\ER{\ensuremath{E_{\rm R}}\xspace}
  \def\Ex{\ensuremath{E_x}\xspace}
  \def\TS{\ensuremath{T_{\rm S}}\xspace}

  \def\HRh{\ensuremath{\hat{H}_{\rm R}}\xspace}
  \def\HR{\ensuremath{H_{\rm R}}\xspace}
  \def\SR{\ensuremath{{\cal S}_{\rm R}}\xspace}